\RequirePackage{rotating}
\documentclass[a4paper,fleqn,usenatbib]{mnras}

\usepackage{mathptmx}

\usepackage{graphicx}	
\usepackage{amsmath}	
\usepackage{amssymb}	
\usepackage{epsfig}
\usepackage{color}
\usepackage{longtable}
\usepackage{tabu}
\usepackage{threeparttable} 
\usepackage{threeparttablex} 
\usepackage{rotating}
\usepackage{xtab}
\usepackage{multicol}
\usepackage{pdflscape}

\def\aj{AJ}                   
\def\apj{ApJ}                 
\def\apjl{ApJ}                
\def\apjs{ApJS}               
\def\aap{A\&A}                
\def\aapr{A\&A~Rev.}          
\def\aaps{A\&AS}              
\def\mnras{MNRAS}             
\def\mnraslett{MNRAS Letters}             
\def\nat{Nature}              
\def\pasp{PASP}               
 
\def\memras{Mem. R. Astron. Soc.}
\def\ssr{Space Sci Rev}

\makeatletter
\newlength{\abovecaptionskip}%
\makeatother

\title[LOFAR observations of the Lockman Hole field]{The Lockman Hole project: LOFAR observations and spectral index properties of low-frequency radio sources }

\author[E. K. Mahony et al.]{E. K. Mahony$^{1,2,3}$\thanks{E-mail:elizabeth.mahony@sydney.edu.au}, R. Morganti$^{1,4}$, I. Prandoni$^{5}$, I. M. van Bemmel$^{6}$, T. W. Shimwell$^{7}$, 
\newauthor M. Brienza$^{1,4}$, P. N. Best$^{8}$, M. Br\"uggen$^{9}$, G. Calistro Rivera$^{7}$, F. de Gasperin$^{7}$, 
\newauthor M. J. Hardcastle$^{10}$, J. J. Harwood$^{1}$,  G. Heald$^{11,4}$, M. J. Jarvis$^{12,13}$, S. Mandal$^{7}$, G. K. Miley$^{7}$, 
\newauthor E. Retana-Montenegro$^{7}$, H. J. A. R\"ottgering$^{7}$, J. Sabater$^{8}$, C. Tasse$^{14,15}$, S. van Velzen$^{16}$, 
\newauthor R. J. van Weeren$^{17}$, W. L. Williams$^{10}$ and G. J. White$^{18,19}$ \\
$^{1}$ASTRON, the Netherlands Institute for Radio Astronomy, Postbus 2, 7990 AA, Dwingeloo, The Netherlands\\
$^{2}$Sydney Institute for Astronomy, School of Physics A28, The University of Sydney, NSW 2006, Australia \\
$^{3}$ARC Centre of Excellence for All-Sky Astrophysics (CAASTRO) \\
$^{4}$Kapteyn Astronomical Institute, University of Groningen, Postbus 800, 9700 AV Groningen, The Netherlands\\
$^{5}$INAF - Istituto di Radioastronomia, via Gobetti 101, 40129 Bologna, Italy \\
$^{6}$Joint Institute for VLBI in Europe, Dwingeloo, Postbus 2, 7990 AA, The Netherlands\\
$^{7}$Leiden Observatory, Leiden University, P.O. Box 9513, NL-2300 RA Leiden, The Netherlands \\
$^{8}$SUPA, Institute for Astronomy, Royal Observatory, Blackford Hill, Edinburgh, EH9 3HJ, UK\\
$^{9}$University of Hamburg, Hamburger Sternwarte, Gojenbergsweg 112, 21029 Hamburg, Germany \\
$^{10}$Centre for Astrophysics Research, School of Physics, Astronomy and Mathematics, University of Hertfordshire, College Lane, Hatfield AL10 9AB, UK\\
$^{11}$CSIRO Astronomy and Space Science, 26 Dick Perry Avenue, Kensington 6151 WA, Australia \\
$^{12}$Astrophysics, University of Oxford, Denys Wilkinson Building, Keble Road, Oxford, OX1 3RH, UK \\
$^{13}$Physics and Astronomy Department, University of the Western Cape, Bellville 7535, South Africa \\
$^{14}$GEPI, Observatoire de Paris, CNRS, Universite Paris Diderot, 5 place Jules Janssen, 92190 Meudon, France\\
$^{15}$Department of Physics \& Electronics, Rhodes University, PO Box 94, Grahamstown, 6140, South Africa\\
$^{16}$Department of Physics \& Astronomy, The Johns Hopkins University, Baltimore, MD 21218, USA \\
$^{17}$Harvard-Smithsonian Center for Astrophysics, 60 Garden Street, Cambridge, MA 02138, USA\\
$^{18}$RAL Space, The Rutherford Appleton Laboratory, Chilton, Didcot, Oxfordshire OX11 0NL, England\\
$^{19}$The Department of Physical Sciences, The Open University, Milton Keynes MK7 6AA, England
}

\date{Accepted XXX. Received YYY; in original form ZZZ}

\pubyear{2016}

\begin{document}
\label{firstpage}
\pagerange{\pageref{firstpage}--\pageref{lastpage}}
\maketitle

\begin{abstract}

The Lockman Hole is a well-studied extragalactic field with extensive multi-band ancillary data covering a wide range in frequency, essential for characterising the physical and evolutionary properties of the various source populations detected in deep radio fields (mainly star-forming galaxies and AGNs). In this paper we present new 150-MHz observations carried out with the LOw Frequency ARray (LOFAR), allowing us to explore a new spectral window for the faint radio source population. This 150-MHz image covers an area of 34.7 square degrees with a resolution of 18.6$\times$14.7\,arcsec and reaches an rms of 160\,$\mu$Jy\,beam$^{-1}$ at the centre of the field. 

As expected for a low-frequency selected sample, the vast majority of sources exhibit steep spectra, with a median spectral index of $\alpha_{150}^{1400}=-0.78\pm0.015$. The median spectral index becomes slightly flatter (increasing from $\alpha_{150}^{1400}=-0.84$ to $\alpha_{150}^{1400}=-0.75$) with decreasing flux density down to $S_{150} \sim$10\,mJy before flattening out and remaining constant below this flux level. For a bright subset of the 150-MHz selected sample we can trace the spectral properties down to lower frequencies using 60-MHz LOFAR observations, finding tentative evidence for sources to become flatter in spectrum between 60 and 150\,MHz. Using the deep, multi-frequency data available in the Lockman Hole, we identify a sample of 100 Ultra-steep spectrum (USS) sources and 13 peaked spectrum sources. We estimate that up to 21\,per\,cent of these could have $z>4$ and are candidate high-$z$ radio galaxies, but further follow-up observations are required to confirm the physical nature of these objects.

\end{abstract}


\section{Introduction} \label{intro}

Although the majority of the earliest radio surveys were carried out at very low frequencies (i.e. the 3rd, 6th and 7th Cambridge Surveys; \citealt{3c,3cr,6c,Hales2007} and the Mills, Slee and Hill (MSH) survey; \citealt{msh}), in more recent years large area radio surveys have primarily been performed at frequencies around 1\,GHz such as the NRAO VLA Sky Survey (NVSS; \citealt{nvss}), the Sydney University Molonglo Sky Survey (SUMSS: \citealt{sumss} and the Faint Images of the Radio Sky at Twenty Centimeters (FIRST) survey; \citealt{first}). With the advent of radio interferometers using aperture arrays such as the Low-Frequency ARray (LOFAR), we now have the ability to re-visit the low-frequency radio sky and survey large areas down to much fainter flux density levels, and at higher resolution, than these earlier surveys. 

LOFAR is a low-frequency radio interferometer based primarily in the Netherlands with stations spread across Europe \citep{lofar}. It consists of two different types of antenna which operate in two frequency bands: the Low-Band-Antennas (LBA) are formed from dipole arrays which operate from 10 to 90\,MHz, and the High-Band Antennas (HBA) are tile aperture arrays which can observe in the frequency range 110--240\,MHz. The long baselines of LOFAR allow us to probe this frequency regime at much higher spatial resolution than previously done; up to 5\,arcsec resolution for the longest dutch baseline, and up to 0.5\,arcsec resolution for the European baselines. The combination of LOFAR's large field-of-view, long baselines and large fractional bandwidth make it an ideal instrument for carrying out large surveys. 

The majority of sources detected in these surveys to date have been radio-loud AGN, with star-forming galaxies only beginning to come into the sample at lower flux densities ($S<5-10$\,mJy). Obtaining large samples of these objects allows us to study the source population in a statistically significant manner and investigate how the properties of radio galaxies evolve over cosmic time. In addition, large surveys allow us to search for rare, unusual objects in a systematic way. 

However, in order to maximise the scientific value of these large surveys, complementary multi-wavelength data are essential to obtain a comprehensive view of the source populations. One such field with extensive multi-band coverage is the Lockman Hole field. This field was first identified by \citet{lockman} who noted the region had a very low column density of Galactic H{\sc i}. This smaller amount of foreground H{\sc i} makes it an ideal field for deep observations of extragalactic sources, particularly in the IR due to the low IR background \citep{Lonsdale2003}. Because of this, there is extensive multi-band ancillary data available, including deep optical/NIR data from ground based telescopes (e.g. \citealt{Fotopoulou2012}), midIR/FIR/sub-mm data from the {\it Spitzer} and {\it Herschel} satellites \citep{Mauduit2012a,Oliver2012} and deep X-ray observations from {\it XMM-Newton} and {\it Chandra} \citep{Brunner2007,Polletta2006}. 

In addition, the Lockman Hole field has an extensive amount of radio data covering a wide range in frequency. This includes the 15-GHz 10C survey \citep{10c, Whittam2013}, deep 1.4-GHz observations over 7 square degrees observed with the Westerbork Synthesis Radio Telescope (WSRT; \citealt{Guglielmino2012}, Prandoni et al., 2016a, in preparation), 610-MHz GMRT observations \citep{Garn2008} and 345-MHz WSRT observations (Prandoni et al., 2016b, in preparation). In this paper we present LOFAR observations of the Lockman Hole field, which extends this multi-frequency information down to 150\,MHz, allowing us to study the low-frequency spectral properties of the faint radio source population. For a brighter sub-sample we were also able to perform a preliminary analysis of the spectral properties down to 60\,MHz.

Studying the spectral index properties of the radio source population can provide insight into a range of source properties. For example, the radio spectral index is often used to distinguish between source components in AGN (i.e. flat-spectrum cores vs. steep-spectrum lobes or ultra-steep spectrum relic emission). Spectral information can also be used to derive approximate ages of the radio source based on spectral ageing models (see e.g. \citealt{Harwood2013, Harwood2015} and references therein), providing insight into the average life-cycle of radio-loud AGN. 

Previous studies that have looked at the average spectral index properties of large samples have reported conflicting results as to whether the median spectral index changes as a function of flux density \citep{Prandoni2006, Ibar2010, Randall2012, Whittam2013}. These studies have generally been carried out at GHz frequencies, with studies of low-frequency selected sources typically showing evidence for the median spectral index to become flatter at fainter flux densities \citep{Ishwara-Chandra2010, Intema2011, Williams2013}. However, most of these latter studies have been biased against detecting steep-spectrum sources at the fainter end of the flux density distribution due to the flux limits imposed by the different surveys used.

The wide frequency coverage available in the Lockman Hole, along with the large area surveyed, also allows us to search for sources with more atypical spectral properties such as those with ultra-steep or peaked spectra.

In this paper we present 150-MHz LOFAR observations of the Lockman Hole field. In Section 2 we discuss the observational parameters, data reduction and source extraction of the 150-MHz LOFAR observations, followed by a brief overview of additional 60-MHz LOFAR observations that are used for the spectral analysis. In Section 3 we present an analysis of the source sizes and resolution bias which are used to derive the 150-MHz source counts and in Section 4 we investigate the spectral index properties of low-frequency selected radio sources. Section 5 presents a deeper look at sources that exhibit more unusual spectral properties (e.g. ultra steep spectrum or peaked spectrum sources), providing insight into how many of these sources we might expect to find in the completed LOFAR all-sky survey. We conclude in Section 6. Throughout this paper we use the convention $S_{\nu} \propto \nu^{\alpha}$. 

\section{Observations and data reduction}

\subsection{LOFAR HBA observations}

The Lockman Hole field (centred at $\alpha$=10h47m00.0s, $\delta$=$+$58d05m00s in J2000 coordinates) was observed on 18 March 2013 for 10 hrs using the LOFAR High Band Antenna (HBA) array. A total of 36 stations were used in these observations: 23 core and 13 remote stations. The `HBA\_DUAL\_INNER' array configuration was used, meaning that the two HBA substations of each core station were treated as separate stations, resulting in 46 core stations. In addition, only the inner 24 HBA tiles in the remote stations are used so that every station has the same beam\footnote{FWHM of $\sim3.5^\circ$ at 150\,MHz in the case of the Lockman Hole field}. Using this array configuration resulted in baselines ranging from 40\,m up to 120\,km. These observations used a bandwidth of 72\,MHz, covering the frequency range from 110 to 182\,MHz, which was split into 366 subbands of 0.195\,MHz. Each subband consists of 64 channels. In order to set the flux scale, primary flux calibrators 3C196 and 3C295 were observed for 10\,min on either side of the Lockman Hole observations with the same frequency setup. An overview of the LOFAR observation details of the Lockman Hole field is given in Table \ref{observationstab}.

\begin{table}
\caption{Observational parameters for the Lockman Hole field. \label{observationstab}}
\begin{tabular}{ll}
\hline
Observation IDs & L108796 (3C295) \\
 & L108798 (Lockman Hole)\\
 & L108799 (3C196) \\
Pointing centres (J2000) & 08h13m36.0s +48d13m03s (3C196) \\
& 10h47m00.0s +58d05m00s (Lockman Hole) \\
& 14h11m20.5s +52d12m10s (3C295) \\
Date of observation & 18 March 2013 \\
Total observing time & 9.6\,hrs (Lockman Hole) \\
 & 10\,min (3C196, 3C295) \\
Integration time & 1\,s \\
Correlations & XX, XY, YX, YY \\
Number of stations & 59 total \\
& 23 core (46 split) \\
& 13 remote \\
Frequency range & 110--182\,MHz \\
Total Bandwidth & 72\,MHz\\
Subbands (SB) & 366 \\
Bandwidth per SB & 195.3125 kHz \\
Channels per SB & 64 \\
  \hline  
\end{tabular}
\end{table}

\subsection{Data reduction} \label{data_reduction}

\subsubsection{Pre-processing pipeline}

The initial steps of the data reduction were carried out using the automated Radio Observatory pre-processing pipeline \citep{Heald2010}. This involves automatic flagging of RFI using {\sc AOflagger} \citep{Offringa2012,Offringa2013} and averaging in time and frequency down to 5\,s integration time and 4 channels per subband. Since LOFAR views such a large sky area the brightest radio sources at these frequencies (referred to as the `A-team' sources: Cygnus A, Cassiopeia A, Taurus A and Virgo A) often need to be subtracted from the visibilities following a process termed `demixing' \citep{Vandertol2007}. Using simulations of the predicted visibilities for these observations it was determined that there was no significant contribution of the A-team sources so no demixing was required for these data\footnote{The nearest of the A-team sources is Virgo A at an angular distance of 50$^{\circ}$.}. These averaged data are then stored in the LOFAR Long Term Archive (LTA).

\subsubsection{Calibration of the data} \label{calib}

To set the amplitude scale we used the primary flux calibrator 3C295. Antenna gains for each station were obtained using the Black Board Selfcalibration tool ({\sc BBS}; \citealt{Pandey2009}), solving for XX, YY and the rotation angle for each subband separately. Solving for the rotation angle allows us to simultaneously solve for the differential Faraday rotation and remove this effect from the amplitudes measured in XX and YY without directly solving for XY and YX, thereby speeding up the process. The amplitude and phase solutions were calculated for each station according to the model of 3C295 presented by \citet{scaife+heald} and transferred to the target field for every subband separately. Although the phases are expected to be quite different between the calibrator field and the target field, this was done as an approximate, first-order correction for the delays associated with the remote stations not being tied to the single clock of the core station. As such, the subsequent phase-only calibration solves for the phase difference between the target and calibrator fields rather than the intrinsic phases for that pointing.

To ensure there was enough signal in the target field for the initial phase calibration, the data were combined into groups of 10 subbands (corresponding to a bandwidth of $\sim2$\,MHz), but maintaining the 4 channel/subband frequency resolution. An initial phase calibration was performed on 10 subbands centred at 150\,MHz using a skymodel obtained from LOFAR commissioning observations of the Lockman Hole field \citep{Guglielmino2012}. These 10 subbands were then imaged using the LOFAR imager {\sc AWImager} \citep{Tasse2013} which performs both $w$-projection to account for non-coplanar effects \citep{Cornwell1992} and $A$-projection to properly account for the changing beam \citep{Bhatnagar2008} during the 10 hr observation. 

\subsubsection{Peeling 3C244.1}

\begin{figure}
\centering{\includegraphics[width=\linewidth]{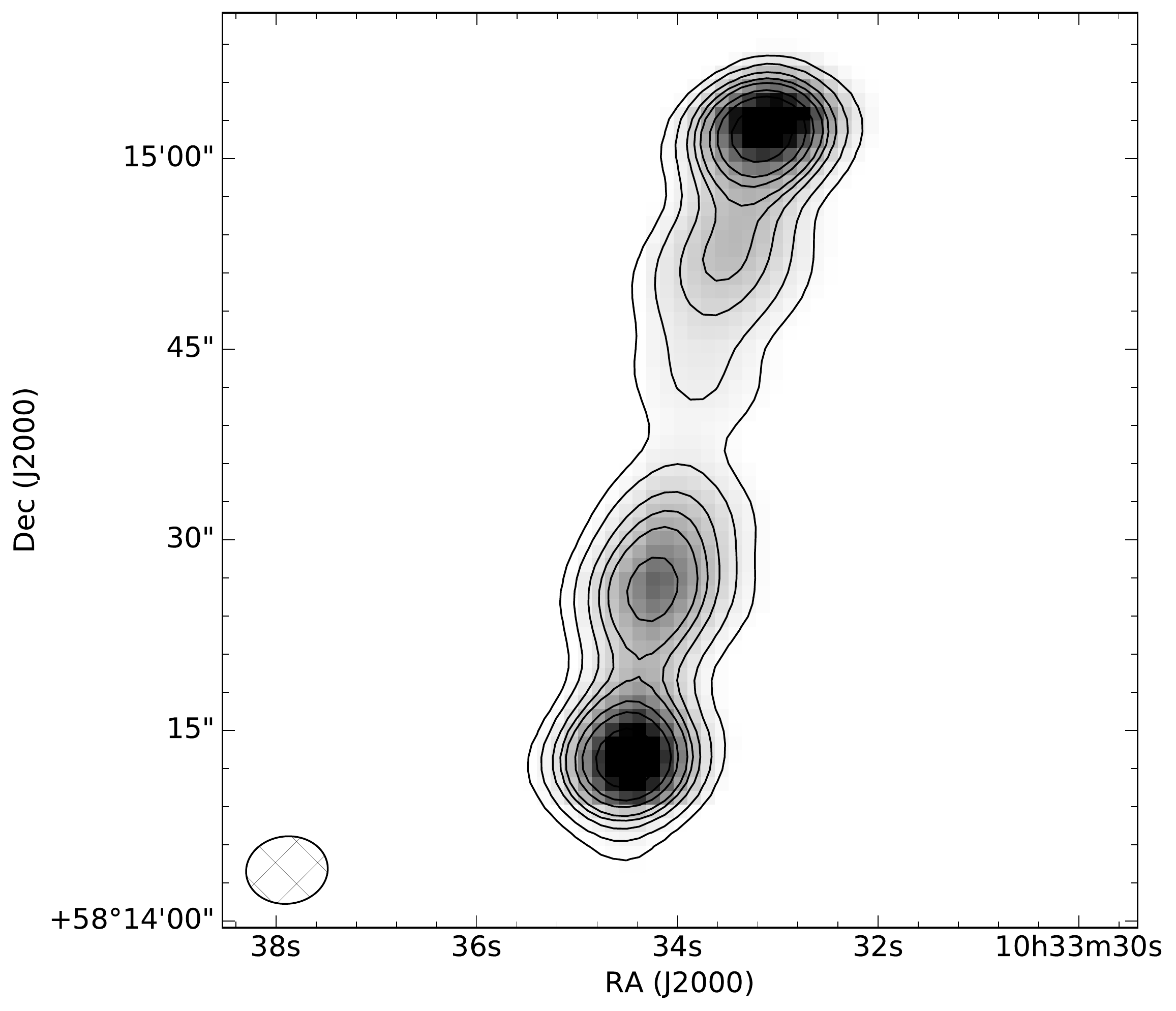} }
\caption{The bright source 3C244.1 that falls in the LOFAR f.o.v. This source has a flux density of $\sim35$\,Jy\,beam$^{-1}$ and has been removed from the LOFAR images due to the artefacts it produced. The contour levels shown are 0.1, 0.25, 0.5, 0.75, 1.0, 1.5, 2.0, 3.0, 4.0 Jy. This image was made from 8\,hrs observation with LOFAR using a bandwidth of 195\,kHz at 150.7\,MHz. The beam size is $5.3 \times 6.4$\,arcsec and the rms reached in this image is 2.6\,mJy\,beam$^{-1}$.\label{3c244}}
\end{figure}

During the imaging step it became clear that a single source (3C244.1, 1.78\,deg from the pointing centre) was dominating the visibilities at these frequencies and producing artefacts across the full field of view. To remove these artefacts 3C244.1 was `peeled' by first subtracting all other sources in the field, calibrating only 3C244.1 using a model derived from separate LOFAR observations of this source, and subtracting these visibilities. In order to obtain an accurate model we observed 3C244.1 for 8\,hrs at 150\,MHz with LOFAR. A single subband at 150.7\,MHz was reduced following the procedure described in Sec.~\ref{calib} and then imaged in CASA using multi-scale, multi-frequency synthesis (with nterms=2) CLEAN. The best image obtained of 3C244.1 is shown in Fig.~\ref{3c244}. 
After 3C244.1 had been successfully peeled, all other sources in the Lockman Hole field were added back and another round of phase calibration performed (this time excluding 3C244.1\footnote{3C244.1 has been excluded from all following images and analysis}). A new skymodel for the target field was extracted from this data using {\sc PyBDSM} \citep{pybdsm} and the same process repeated on the remaining groups of 10 subbands. 

\subsubsection{Imaging the data}

Once calibrated, and 3C244.1 successfully peeled, each 10 subband block was imaged with the {\sc AWImager} using Briggs weighting and a robust parameter of $-0.5$ in order to inspect the image quality across the full bandwidth. Images with obvious artefacts were excluded from further analysis (66 subbands were excluded, most of which were at the edges of the LOFAR band where the sensitivity decreases). The remaining 300 subbands were then averaged by a factor of 2 in both frequency and time and re-imaged in groups of 50 subbands (10\,MHz bandwidth) in order to detect fainter sources. At the time of reducing this data the software available did not allow us to image large bandwidths taking into account the spectral index of the radio sources (i.e. multi-frequency synthesis with nterm higher than 1). As such, we chose to image the data in chunks of 10\,MHz to optimise the depth to which we could {\sc CLEAN} the image without introducing too many errors. 

This imaging was carried out in 3 steps following the procedure presented in \citeauthor{Shimwell2016}, 2016, submitted. First an initial image was created using a pixel size of 3.4\,arcsec, robust=$-0.5$ and longest baseline 12\,k$\lambda$. This image was deconvolved down to a relatively high threshold of 20\,mJy to create a CLEAN mask which was made from the restored image. Due to artefacts around brighter sources the deconvolution was then done in two stages to better enable CLEANing of the fainter sources, but avoid CLEANing bright artefacts. The data was first reimaged using the `bright-source' CLEAN mask down to a threshold set by the largest rms measurement in the noise map. This output image (with the bright sources already deconvolved) was then reprocessed using the `faint-source' CLEAN mask to CLEAN down to a lower threshold approximately equal to the median value of the rms map. Since the synthesised beam changes as a function of frequency, each 50 SB image was then convolved to the same beam size and all images combined in the image plane, weighted by the variance. This results in a central frequency of 148.7\,MHz, but we refer to this image as the 150-MHz LOFAR image hereafter for simplicity.

\begin{figure*}
\centering{\includegraphics[width=\linewidth]{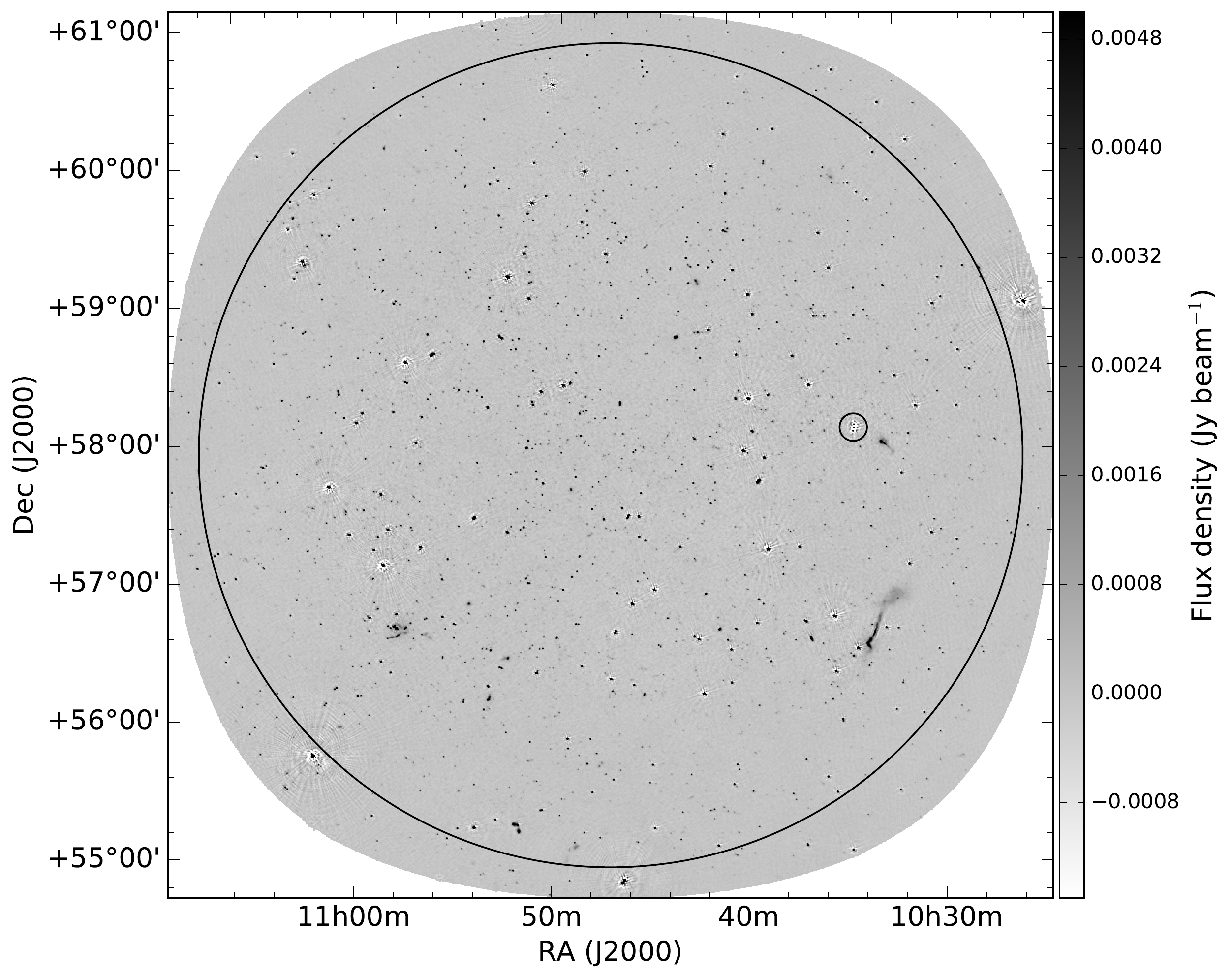}} 
\caption{LOFAR image of the Lockman Hole field at 150\,MHz. This image has a resolution of 18.6$\times$14.7\,arcsec and reaches a noise level of 0.16\,mJy\,beam$^{-1}$ in the centre of the field. Almost 5000 sources are detected in this image above the 5$\sigma$ peak flux density level. The smaller circle marks the position of 3C244.1 which has been subtracted from this image and the larger circle denotes the 3 degree radius from which we extract sources. \label{lofarimages}} 
\end{figure*}

\begin{figure*}
\centering{\includegraphics[width=\linewidth]{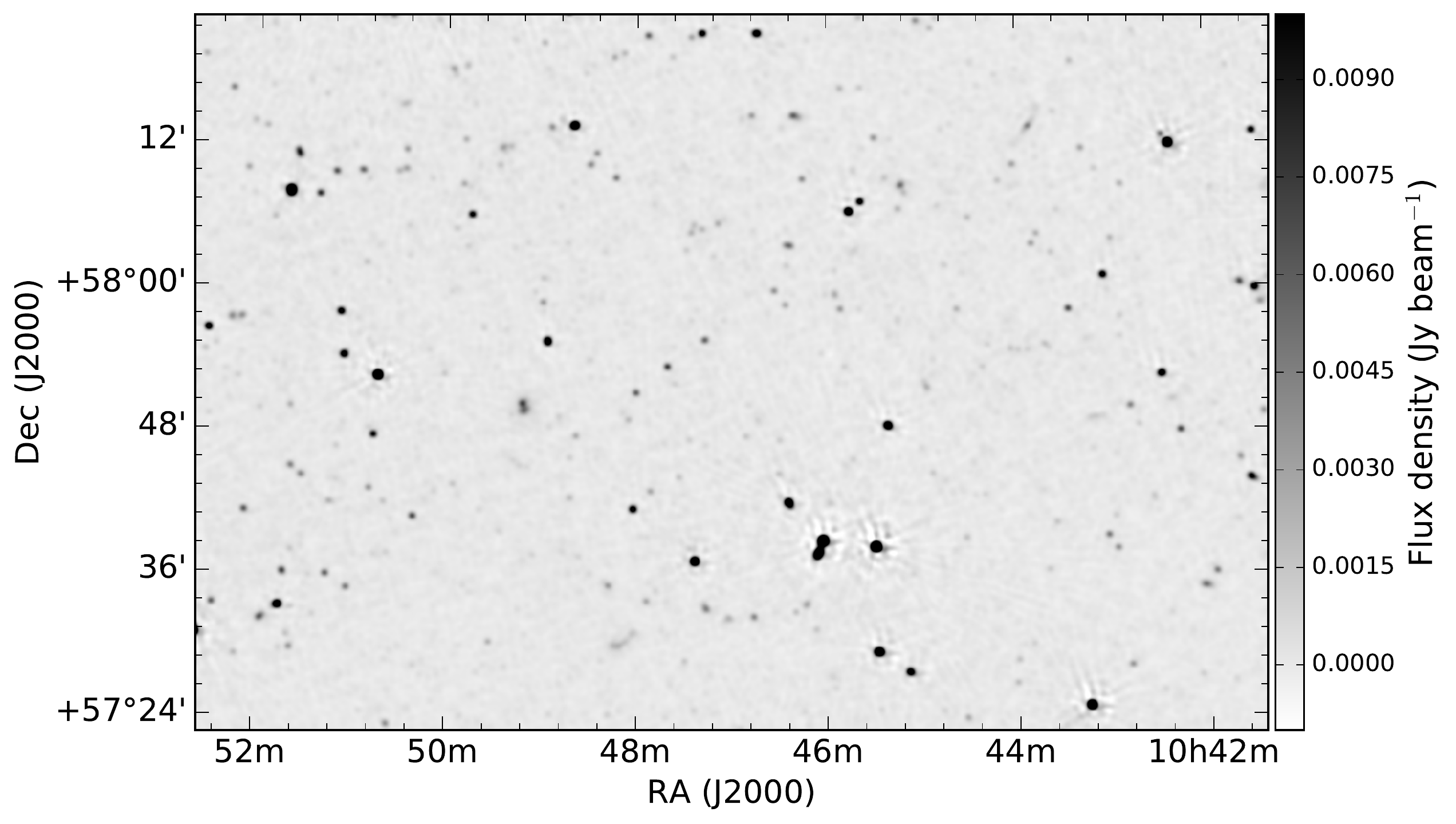}} 
\caption{Zoomed-in region of the Lockman Hole field covering an area of 1.5 $\times$ 1 degree. The image details are the same as given in Fig.~\ref{lofarimages}.\label{lofarimageszoom}} 
\end{figure*}

The resulting image has a final beam size of 18.6$\times$14.7\,arcsec with PA$=85.7$ and an rms of $\sim 160 \mu$Jy\,beam$^{-1}$ in the centre of the beam. Figs.~\ref{lofarimages} and \ref{lofarimageszoom} show the 150-MHz LOFAR image of the Lockman Hole field. The full $\sim30$ square degree field is shown in Fig.~\ref{lofarimages} and a zoomed-in region (approximately $1.5\times1$ degree) is shown in Fig.~\ref{lofarimageszoom}.

\subsection{Ionospheric distortions}

One of the biggest challenges with wide-field imaging, particularly at low frequencies, is correcting for the changing ionosphere across the field of view. As a result of applying the same phase solutions to the full field of view, phase errors are still evident around bright sources, particularly in regions furthest away from the pointing centre. In addition, errors in the beam model (in both amplitude and phase) mean that artefacts become worse farther from the pointing centre. In order to image the full field of view at high resolution (i.e. $\sim$5 arcsec using all of the Dutch stations) direction-dependent effects need to be corrected for, requiring a more thorough calibration strategy such as the `facet-calibration' technique presented by \citet{Vanweeren2016} and \citet{Williams2016}. However, standard calibration techniques are still adequate down to a resolution of $\sim$20\,arcsec as presented here (see also Shimwell, et al., submitted). 

Due to the significant computational time required to apply the facet-calibration method, and the fact that for the analysis presented here we are focused on a statistical cross-matching of sources with multi-band radio data (which are typically at $\sim$15\,arcsec resolution or lower), we have elected to limit the resolution to $\sim$20\,arcsec. Calibrating and imaging this data at higher resolution using the facet-calibration technique will be the subject of a future paper.

\subsection{Noise analysis and source extraction}

\begin{figure}
\centering
\includegraphics[width=\linewidth]{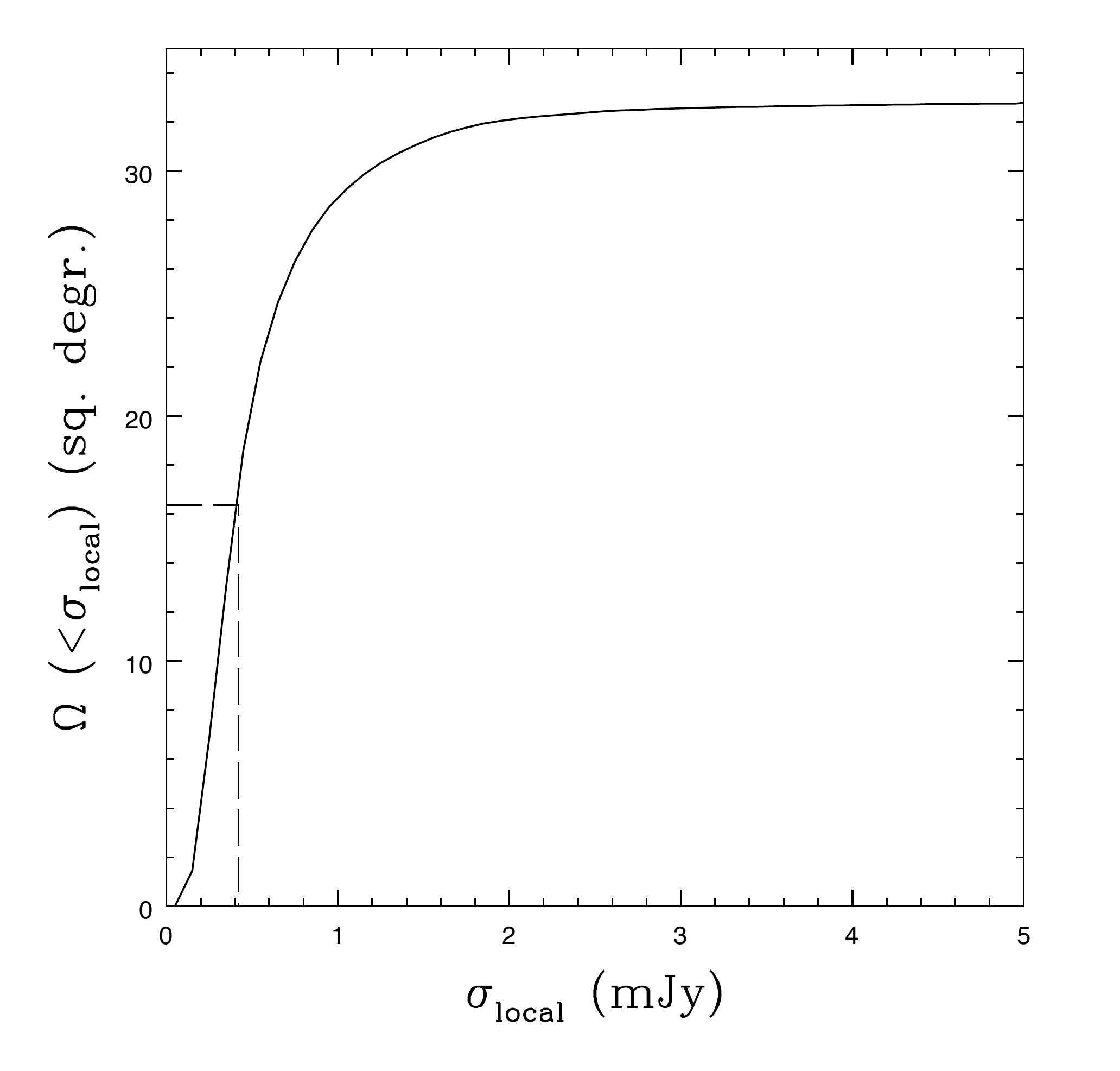} 
\caption{Visibility area of the LOFAR HBA image within 3 degrees of the phase centre. Cumulative fraction of the total area of the noise map characterized by a measured noise lower than a given value. Dotted lines indicate the noise value measured over 50\,per\,cent of the total area.}
\label{fig-noise}
\end{figure}

A source catalogue was extracted using the LOFAR source extraction package {\sc PyBDSM} (Mohan \& Rafferty 2015). To limit any effects of bandwidth or time smearing, we only extracted sources within 3 degrees of the phase centre\footnote{Following the equations given in \citet{Bridle99, MSSS}, the combined time and bandwidth smearing at 18.6\,arcsec resolution, 3 degrees from the pointing centre is $I/I_0 = 0.93$, where $I/I_0$ refers to the reduction in peak response of a source in the image.}. {\sc PyBDSM} initially builds a noise map from the pixel data using variable mesh boxes. The noise ($\sigma_{\rm local}$) increases from $\sim 160$ $\mu$Jy\,beam$^{-1}$ at the phase center to $\sim 700$ $\mu$Jy\,beam$^{-1}$ at the maximum radial distance of 3 degrees. However phase errors result in regions of much higher noise (up to $\sim 5$\,mJy\,beam$^{-1}$) around bright sources. The cumulative distribution of $\sigma_{\rm local}$ over the region of the map considered for source extraction is shown in Fig.~\ref{fig-noise}. Fifty per cent of the total area has $\sigma_{\rm local}< 400$ $\mu$Jy (see dotted lines in Fig.~\ref{fig-noise}). This value can therefore be considered as representative of our HBA image.

{\sc PyBDSM} extracts sources by first identifying islands of contiguous emission above a given threshold, then decomposing this into Gaussian components. A peak threshold of 5$\sigma_{\rm local}$ was used to define sources and an island threshold of 4$\sigma_{\rm local}$ was used to define the island boundary. The `wavelet-decomposition' option was used during the source extraction, meaning that the Gaussians fitted were then decomposed into wavelet images of various scales. This method is useful for extracting information on extended objects. Sources were flagged according to the Gaussian components fitted; `S' means the source was fit by a single Gaussian component, `M' denotes that multiple Gaussian components were needed to fit the source and `C' refers to a single Gaussian component that is in a shared island with another source.

The final source catalogue consists of 4882 sources above a 5$\sigma$ flux limit of 0.8\,mJy. Of these, 3879 are flagged as `S' (i.e. well described by a single Gaussian), 391 are flagged as `C' (i.e. probable close-spaced double-lobed radio galaxies) and 612 are marked as `M', indicating a more complicated source structure.

\subsection{Flux scale and positional accuracy} \label{fluxcheck}

As discussed in Section \ref{calib}, flux densities have been calibrated according to the \citet{scaife+heald} flux scale. While this in theory should mean that the LOFAR flux densities are consistent with other radio surveys, in practice a number of instrumental and observational effects can combine leading to uncertainties in the absolute flux calibration. These can include errors associated with uncertainties in the LOFAR beam model (which also change with time and frequency), the transfer of gain solutions and ionospheric smearing effects. To confirm that the LOFAR flux densities are consistent with previous low-frequency observations we have compared the LOFAR flux densities with sources detected in the 151-MHz Seventh Cambridge (7C) survey \citep{Hales2007} and the alternate data release of the TIFR GMRT Sky Survey (TGSS; \citealt{tgss}\footnote{http://tgss.ncra.tifr.res.in/ and tgssadr.strw.leidenuniv.nl/}). We first crossmatched the LOFAR 150-MHz catalogue with the 151-MHz 7C catalogue. Due to the difference in resolution ($\sim$20\,arcsec compared to $\sim$70\,arcsec), we have only selected point sources in the LOFAR catalogue (since this has been derived at the higher resolution) for this comparison. Although we are only comparing point sources, here, and in any following analysis, we use the total flux densities for the flux comparison as the peak flux densities might be affected by ionospheric smearing (see discussion in Sec.~\ref{sec-sizes}). Using a matching radius of 20\,arcsec we find a total of 60 LOFAR sources that have a counterpart in the 7C survey with the median ratio of the LOFAR flux density to 7C flux density being 1.07 with a standard deviation of 0.25. Due to this systematic offset, the LOFAR 150-MHz flux densities are corrected by 7\,per\,cent. 

\begin{figure}
\centering{\epsfig{file=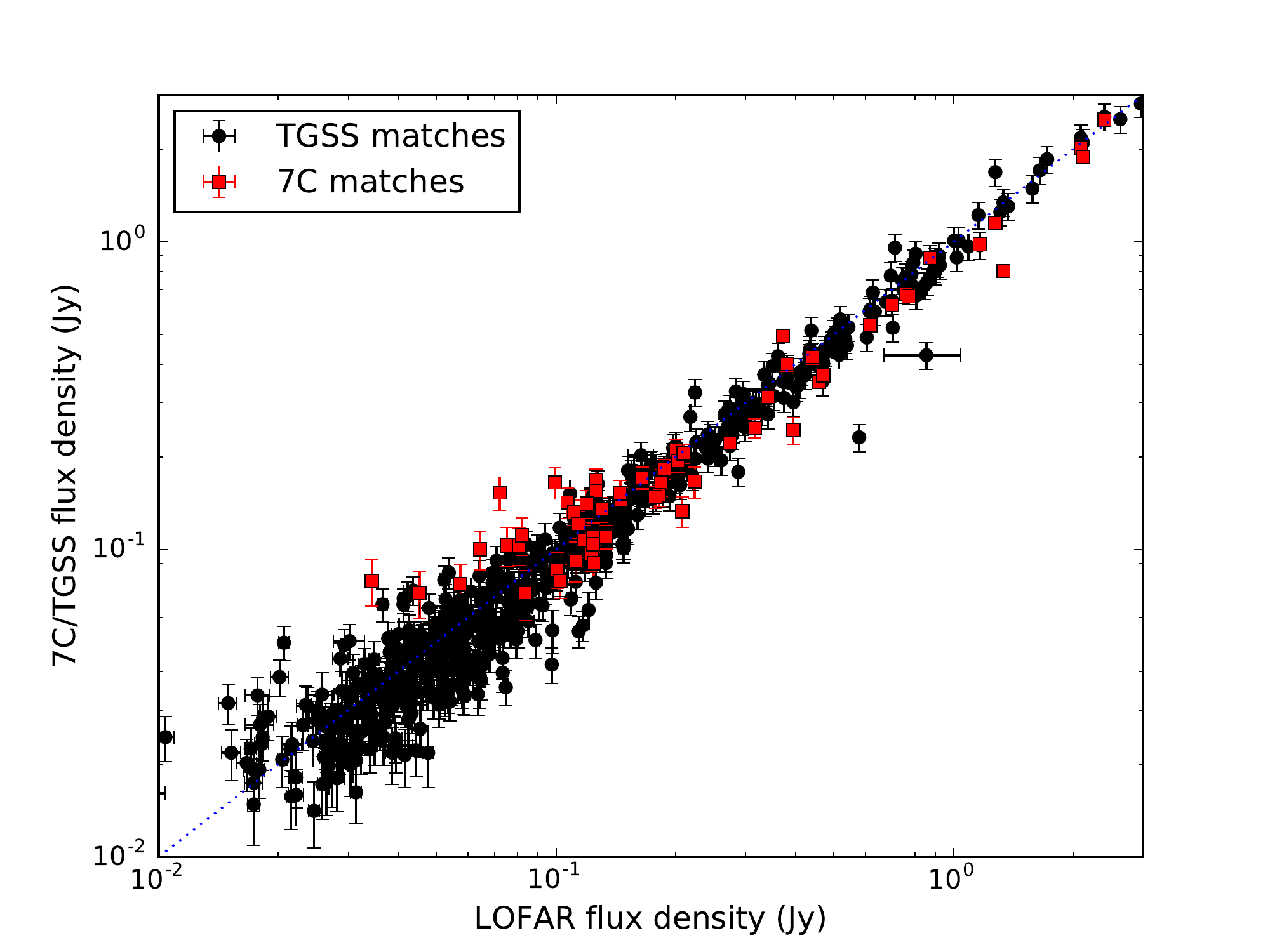, width=\linewidth}}
\caption{Flux comparison between the LOFAR and the 7C (red squares) and TGSS (black circles) surveys. Only point sources are included when crossmatching with the 7C survey due to the different resolutions of the two surveys, but we use integrated flux densities as the peak flux densities are more heavily affected by ionospheric smearing. \label{fluxcomp} }
\end{figure}

To verify this correction we compared the corrected LOFAR flux densities with the TGSS survey. Due to the similar resolutions of the LOFAR data presented here and the TGSS survey we can also include resolved sources in the crossmatching of these two catalogues, resulting in 631 matches (using a matching radius of 10\,arcsec). Using the corrected LOFAR fluxes we obtain a median flux ratio of LOFAR/TGSS=1.00 with standard deviation 0.27. Using the uncorrected flux densities we obtain a median flux ratio of 1.07, in agreement with the comparison of the 7C survey. Fig. \ref{fluxcomp} shows the flux density comparisons with both the 7C (red squares) and TGSS surveys (black circles). For a direct comparison we plot the uncorrected LOFAR flux densities for both the 7C and TGSS matches.

Given the uncertainties associated with imperfect calibration at these frequencies, and the fact that comparisons with other 150\,MHz datasets reveal flux offsets of 7\,per\,cent, a global 7\,per\,cent flux error is added in quadrature to the flux density errors associated with the source extraction (typically of order 10\,per\,cent, but this can vary significantly depending on the flux density of the source). 

\begin{figure}
\centering{\epsfig{file=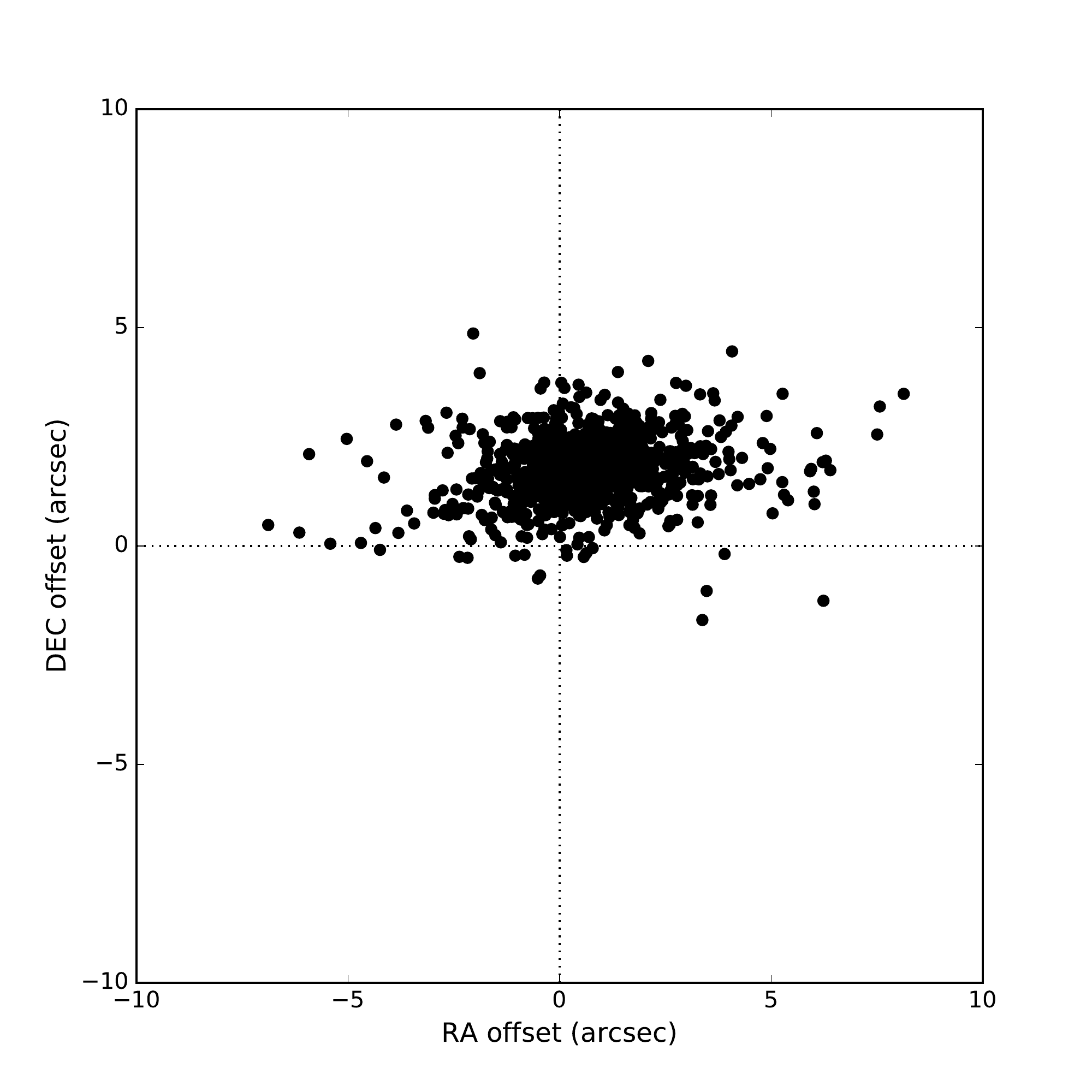, width=\linewidth}}
\caption{Offset in positions of LOFAR sources compared to FIRST catalogued positions. \label{poscomp} }
\end{figure}

We also checked the positional accuracy by crossmatching with the FIRST catalogue \citep{first}, again only including point sources. Fig.~\ref{poscomp} shows the offset in right ascension and declination between the LOFAR positions and FIRST positions. This shows a clear systematic offset, primarily in declination, for all sources, not uncommon when doing phase-only self-calibration which can result in positions being shifted by up to a pixel. The offsets are 0.6\,arcsec in right ascension and 1.7\,arcsec in declination, well below the adopted pixel size. As such, we do not correct for these positional offsets in this work, but care should be taken when using these positions to crossmatch with higher resolution observations, in particular when searching for optical or InfraRed (IR) counterparts. Any optical/IR counterparts presented in this paper were crossmatched based on more accurate positions provided by deeper observations at 1.4\,GHz (see Sec.~\ref{wsrtcrossmatch}).

\subsection{LOFAR LBA observations} \label{lba}

The Lockman Hole field was also observed at lower frequencies using the Low-Band Antenna (LBA) array. The LBA observations were carried out at 22--70~MHz on 15 May 2013, using the LBA\_OUTER station configuration\footnote{The LBA stations can only use 48 of the 96 elements. The choice is provided between the inner 48 and the outer 48 (a ring-like configuration)}. The integration time was 1 sec. and each subband had 64 channels. Using the multi-beaming capabilities of the LBA, the flux calibrator 3C196 was observed simultaneously using the same frequency settings (248 subbands of 195.3~kHz each). Flagging and averaging of the data were performed in the same way as for the HBA observations. Due to the larger field of view for the LBA observations, demixing was also carried out on these observations using the observatory's pre-processing pipeline \citep{Heald2010}. For a preliminary analysis, a set of 10 subbands around 60~MHz were selected for further processing, using the same packages as for the HBA data reduction (see Section \ref{data_reduction}).

The amplitude calibration was carried out in a similar fashion to the HBA data, but in this case using the model of 3C196 provided by V.N. Pandey. The amplitude gains were smoothed in time to remove the noise and applied to the corresponding target subband. To correct for clock offsets found in the observations, the phase solutions from a single timestamp were also applied. Subsequently, the ten subbands were merged into a single 2 MHz dataset, while maintaining the 40 channel spectral resolution. The merged dataset was then phase calibrated using the sky model derived from the 150-MHz HBA observations. Again, 3C244.1 was peeled to remove strong artefacts from this source and another round of phase calibration performed with the same 150-MHz sky model (this time without 3C244.1). No direction-dependent ionospheric phase solutions have been derived. 

The resolution and noise of the LBA images were optimised using tapering and weighting in order to minimise the effect of moderate to severe ionospheric disturbances, allowing us to trace the spectral properties of the bright sources detected in the HBA image (Fig.~\ref{lofarimages}) down to lower frequencies. The images were produced using {\sc AWimager}, with a cell size of 5\,arcsec and an image size of 6000 pixels. The primary beam (FWHM) at this frequency is $\sim$4.5$^{\circ}$. The maximum $uv$ range was set at 5 k$\lambda$, and the Briggs robust parameter was set to 0. The noise level in the 60-MHz map is 20\,mJy\,beam$^{-1}$ at a resolution of 45\,arcsec. Sources were extracted using PyBDSM with the same parameters as used for the HBA source extraction, resulting in a catalogue of 146 sources.

\subsubsection{Verification of LBA flux densities} \label{lbaflux}

\begin{figure}
\centering{\epsfig{file=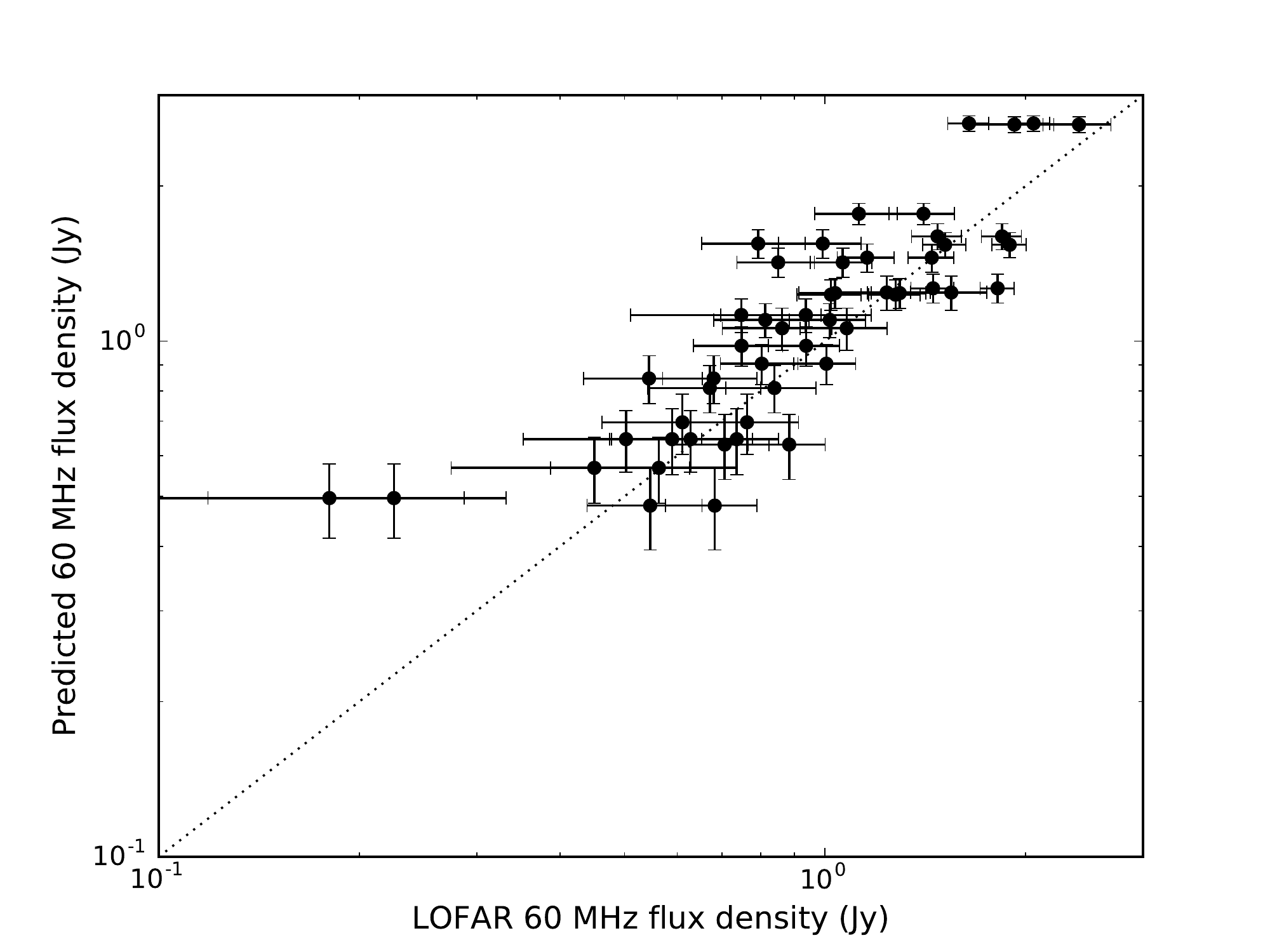, width=\linewidth}}
\caption{Flux comparison between LOFAR 60-MHz data and predicted 60-MHz fluxes from the VLSSr survey (assuming a spectral index of $\alpha_{60}^{74} = -0.8$. Only point sources are included due to the different resolution of the two surveys, but integrated flux densities are used to avoid the ionospheric smearing effects associated with the peak flux densities. \label{fluxcomplba}} 
\end{figure}

To check the reliability of the flux densities extracted from the LBA image we crossmatched the 60-MHz LOFAR catalogue with the 74-MHz VLA Low-Frequency Sky Survey Redux (VLSSr) catalogue \citep{vlss}. To account for the difference in frequency we predict 60-MHz flux densities from the VLSSr survey assuming a spectral index of $\alpha_{60}^{74} = -0.8$. Based on these predicted flux densities we calculate a median flux ratio of LOFAR/VLSSr=0.80 with a standard deviation of 0.23. Fig. \ref{fluxcomplba} shows the flux density comparison for LOFAR 60-MHz sources against predicted 60-MHz flux densities from the VLSSr survey. Only point sources are included in this analysis due to the difference in resolution (VLSSr has a resolution of 80\,arcsec), but we use the integrated flux densities as the peak flux densities are more heavily affected by ionospheric smearing.

The underestimation of the LOFAR LBA flux densities is not unexpected due to the ionospheric conditions during the observations and the fact that these have not been corrected for during the data reduction. Based on the comparison with the VLSSr survey we scale the LBA fluxes by a factor of 1.25. We also increase the flux densities errors by 20\,per\,cent (added in quadrature to the flux errors reported by {\sc PyBDSM}) to account for the uncertainties associated with the absolute flux calibration at these frequencies.

\section{Source Counts at 150\,MHz} \label{counts}

In this section we present the source counts derived from our 150-MHz catalogue. In order to derive the source counts we first need to correct for resolution bias and incompleteness at low flux densities. We do this following the procedures outlined by \citet{Prandoni2001, Prandoni2006} and by \citet{Williams2016} in deriving the HBA counts in the Bo\"otes field. An early analysis of the source counts in the Lockman Hole region is also presented by \citet{Guglielminothesis}.

\subsection{Visibility area}

To derive the source counts we weight each source by the reciprocal of its visibility area ($A(<\sigma_{\rm local})/A_{\rm tot}$) as derived from Fig.~\ref{fig-noise}. This takes into account the varying noise in the image by correcting for the fraction of the total area in which the source can be detected. However, due to the Gaussian noise distribution there is still some incompleteness in the lowest flux density bins (i.e. if a source happens to fall on a noise dip the flux will either be underestimated or the source will potentially go undetected). As demonstrated through Monte Carlo simulations by \citet{Prandoni2000}, incompleteness can be as high as 50\,per\,cent at the 5$\sigma$ threshold, reducing down to 15\,per\,cent at 6$\sigma$, and to 2\,per\,cent at 7$\sigma$. However, such incompleteness effects can be (at least partially) counterbalanced by the fact that sources below the detection threshold can be pushed above it when they sit on a noise peak. Williams et al. (2016) have shown through Monte Carlo simulations undertaken in a LOFAR HBA field (Bo\"otes) with a similar noise level (rms $\ge$ $120-150$ $\mu$Jy) that such incompleteness effects become negligible above 2\,mJy\footnote{While the noise levels reached are similar, the Bo\"otes field was reduced using the facet-calibration technique meaning that the contribution of artefacts will be less in this image. The impact of artefacts on the source counts is mentioned at the end of Sec.~\ref{sec-counts}.}. As such, we only derive the source counts down to this flux density limit.

\subsection{Source size distribution and resolution bias} 
\label{sec-sizes}

\begin{figure}
   \centering
   \resizebox{9cm}{!}{\includegraphics[]{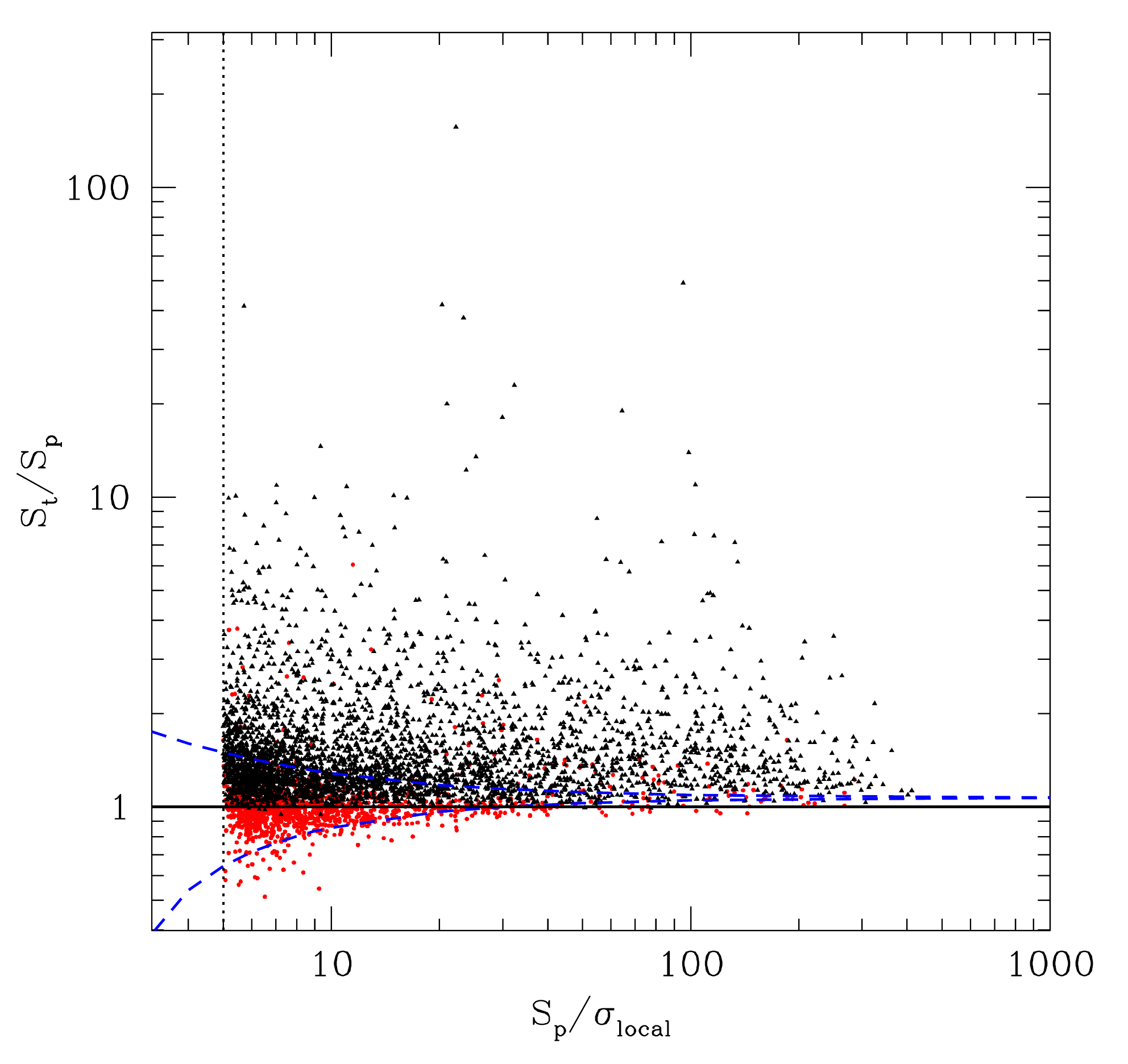}}
      \caption{Flux density ratio ($S_t/S_p$) as a function of signal-to-noise of the source. The envelope function defined by Eq.~(\ref{eq-resolv}) is shown by the upper dashed line while the dotted line refers to the $S/N=5$ cut-off used in the source extraction. Red dots refer to sources that were detected as unresolved by {\sc PyBDSM}; black triangles refer to resolved or partially resolved sources (according to the model extracted using {\sc PyBDSM}). Only sources that lie above the upper dashed line are marked as resolved in the final catalogue. The lower dashed line is obtained by mirroring the upper dashed line.}
         \label{fig-sratio}
   \end{figure}

To measure the extension of a radio source we can use the following relation:
    \begin{equation}
      S_{t}/S_{p} = \theta_{\rm maj}\theta_{\rm min}/b_{\rm maj}b_{\rm min} 
\label{eq-sratio}
   \end{equation}
   where $S_{t}/S_{p}$ is the ratio of the integrated to peak flux density, $\theta_{\rm maj}$ and $\theta_{\rm min}$ are the source sizes and $b_{\rm maj}$ and $b_{\rm min}$ refer to the synthesised beam axes (assuming a Gaussian-shaped source). Plotting this flux density ratio ($S_{t}/S_{p}$) against signal-to-noise we can establish a criterion for determining if a source is extended. This is shown in Fig.~\ref{fig-sratio}. Since the integrated flux density must always be equal to or larger than the peak flux density, sources with $S_{t}/S_{p}<1$ provide a good measure of the error fluctuations. This allows us to determine the 90\,per\,cent envelope function which is characterised by the following equation:
 
   \begin{equation}
      S_{t}/S_{p} = 1.07 + 2.13 \left( \frac{S_{p}}{\sigma_{\rm local}}\right)^{-1}
  \label{eq-resolv}
   \end{equation}
   
    where $\sigma_{\rm local}$ refers to the local rms. Here we have assumed two-dimensional elliptical Gaussian fits of point sources in the presence of Gaussian noise following the equations of error propagation given by \citet{Condon1997}. We have also incorporated the correction for time and bandwidth smearing, which causes a maximum underestimation of the peak flux of 0.93.

This envelope function is shown by the upper dashed line in Fig.~\ref{fig-sratio}. Sources that lie above this line are classified as extended or resolved and sources below the line are considered to be point sources. Note that this is different to the criterion used by {\sc PyBDSM} during the source extraction (red points are classified as unresolved by {\sc PyBDSM} and black points show resolved, or partially resolved, sources). 

\begin{figure}
	\centering
	\resizebox{9cm}{!}{\includegraphics[]{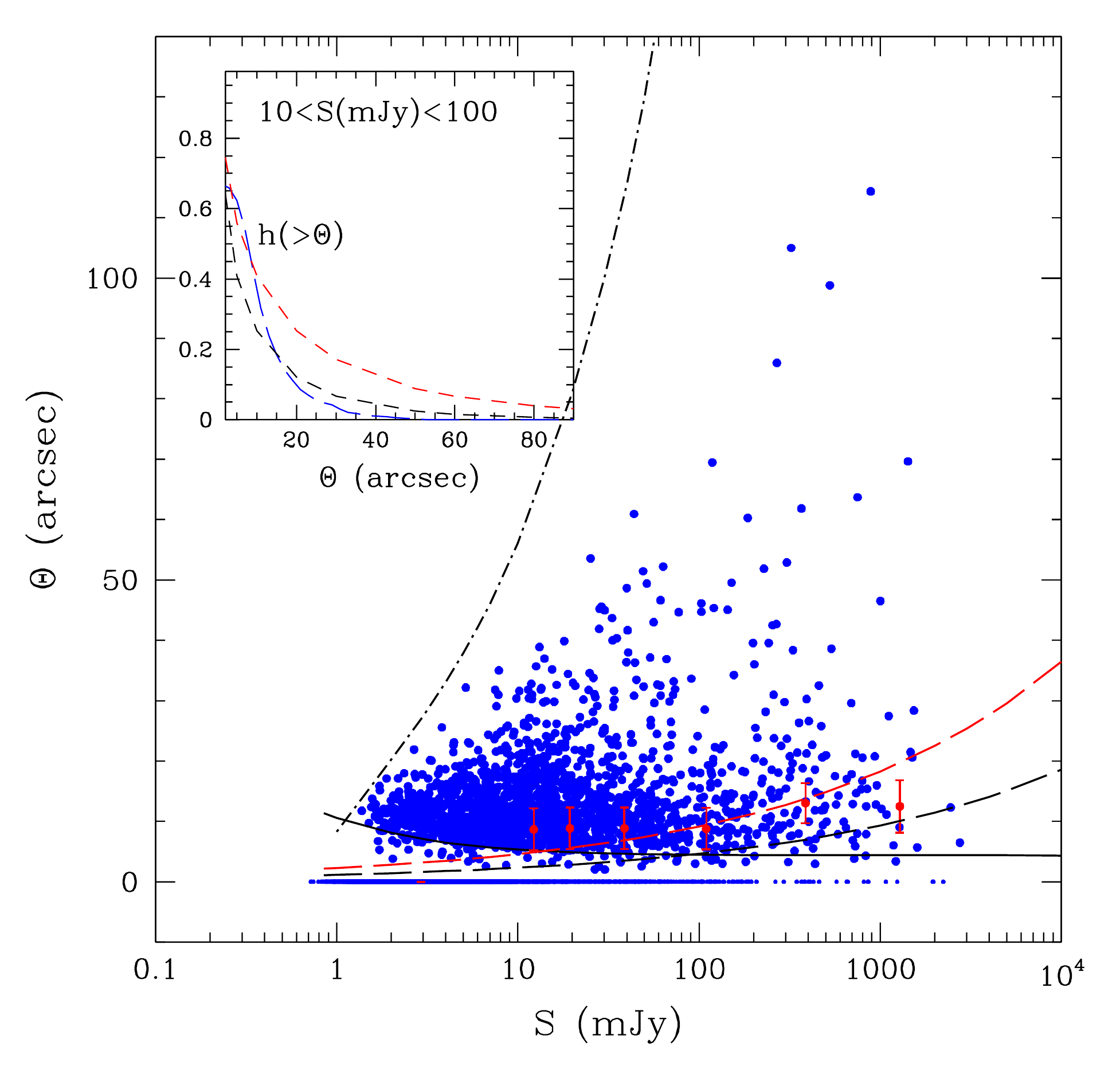}}
		\caption{Deconvolved angular size ($\Theta=\sqrt{\theta_{\rm maj}\cdot\theta_{\rm min}}$) as a function of integrated flux density. The red points with inter-quartile bars denote the median source sizes of the sample. The black dot-dashed line represents the maximum size a source can have ($\Theta_{\rm max}$) and still be detected in the source catalogue (i.e. have $S_{p} =5\sigma_{\rm local}$). Above this line the sample is incomplete due to resolution bias. The black sold line shows the smallest size detectable in these observations ($\Theta_{\rm min}$). The two dashed lines show the median source size relations presented by \citet{Windhorst1990}; the black shows the median source size rescaled to 150 MHz using $\alpha=-0.8$ which better describes the larger source sizes, while the red line assumes a factor 2 larger normalization factor which better describes the smalled sources in our sample. These are also shown in the inner panel which plots the angular size distribution. The distribution obtained from this sample is shown in blue and the Windhorst relations shown in red and black. }
	\label{fig-sizeflux}
\end{figure}

Using Eqs.~\ref{eq-sratio} and~\ref{eq-resolv} we can derive the minimum angular size, $\Theta_{\rm min}$ detectable in these observations. This is shown by the solid line in Fig.~\ref{fig-sizeflux} where we plot the deconvolved angular sizes against flux density for sources detected in the Lockman Hole. Resolved sources account for $\sim$40\,per\,cent of the full sample, increasing to $\sim$70\,per\,cent for sources with $S>10$\,mJy and $\sim$80\,per\,cent for sources with $S>60$\,mJy. We compare the median angular sizes (for sources with $S>10$\,mJy) and the angular size integral distribution (for sources with $10 \leq S_{\rm mJy} \leq 100$, inner panel) with the relations presented by \citet{Windhorst1990} for deep 1.4-GHz samples: $\Theta_{\rm med} = 2\times (S_{\rm{1.4 GHz}})^{0.30}$ ($S$ in mJy and $\Theta_{\rm med}$ in arcsec) and $h(>\Theta)= {\rm exp}[-\ln 2 (\Theta/\Theta_{\rm med})^{0.62}$], where $\Theta_{\rm med}$ is rescaled to 150~MHz by assuming a spectral index of $\alpha=-0.8$\footnote{$\Theta$ and $\Theta_{\rm med}$ represent the geometric mean of the source major and minor axes.}. Our sources tend to have larger median sizes with respect to the ones expected from the \citet{Windhorst1990} relation (see black dashed line in Fig~\ref{fig-sizeflux}). A better description of the median size distribution of our sources is obtained by assuming a $2\times$ larger normalization factor (see red dashed line).

While somewhat larger sizes can be expected going to lower frequency, such a discrepancy can be explained by the presence of residual phase errors affecting our LOFAR image 
in the absence of direction-dependent calibration. Phase errors can broaden or smear out sources by a larger factor than bandwidth and time averaging smearing alone. In addition, at these frequencies the Point Spread Function (PSF) is a combination of the synthesised beam and ionospheric smearing effects which varies across the field of view and is not taken into account by the above equations. This hypothesis seems to be supported by the fact that for well resolved sources ($>10-15$ arcsec), where size measurements are less affected by phase errors, the integral size distribution of our sample is in good agreement with the one proposed by \citet{Windhorst1990}. 

It is also worth noting that larger source sizes were also noticed by \citet{Williams2016} in their analysis of the Bo\"otes field. In that case facet calibration was performed, but their higher resolution HBA image was affected by larger combined bandwidth and time averaging smearing, resulting respectively in radial and tangential size stretching. 

The black dot-dashed line in Fig.~\ref{fig-sizeflux} represents the resolution bias limit. While we are using total flux densities for the source counts, the extraction of the source catalogue is based on the peak flux densities (i.e. $S_{p} =5\sigma_{\rm local}$ to be detected). The resolution bias takes into account the fact that an extended source with flux $S_t$ will fall below the detection limit of the survey before a point source of the same $S_t$.

Following \citet{Prandoni2001, Prandoni2006}, a correction {\sl c} has been defined to account for incompleteness due to resolution bias:

\begin{equation}
      c=1/[1-h(>\Theta_{\rm lim})]
   \end{equation}
where  $h(>\Theta_{\rm lim})$ is the assumed integral angular size distribution and $\Theta_{\rm lim}$ is the angular size upper limit. This limit is defined as a function of the integrated source flux density: 

\begin{equation}
      \Theta_{\rm lim}=max[\Theta_{\rm min},\Theta_{\rm max}]
   \end{equation}
where $\Theta_{\rm min}$ and $\Theta_{\rm max}$ are the parameters defined in Section~\ref{sec-sizes}. Above this limit ($\Theta_{\rm lim}$) we expect to be incomplete. We introduce $\Theta_{\rm min}$ in the equation as this accounts for the effect of having a finite synthesized beam size. This becomes important at low flux densities where $\Theta_{\rm max}$ approaches 0.

\subsection{Source Counts} \label{sec-counts}

The normalized 150-MHz differential source counts are listed in Table~\ref{tab-counts}. Here we list the flux interval used ($\Delta S$), the geometric mean of that interval ($\langle S\rangle$), number of sources in that bin ($N_S$), the differential counts normalised to a non-evolving Euclidean model (${\rm d}N/{\rm d}S$ $S^{2.5}$) and the Poissonian errors (calculated following \citealt{Regener1951}). Two determinations are provided for the normalised counts: the one obtained by correcting for resolution bias using the \citet{Windhorst1990} relation and the one obtained by modifying the Windhorst et al. $\Theta_{\rm med} - S$ relation as discussed in the text. For the counts derivation we used all sources brighter than $\sim$2\,mJy, to minimize the incompleteness effects at the source detection threshold discussed above. Artefacts around bright sources can still contaminate our source counts above this threshold as discussed later.

\begin{figure*}
   \centering
   \resizebox{12cm}{!}{\includegraphics[]{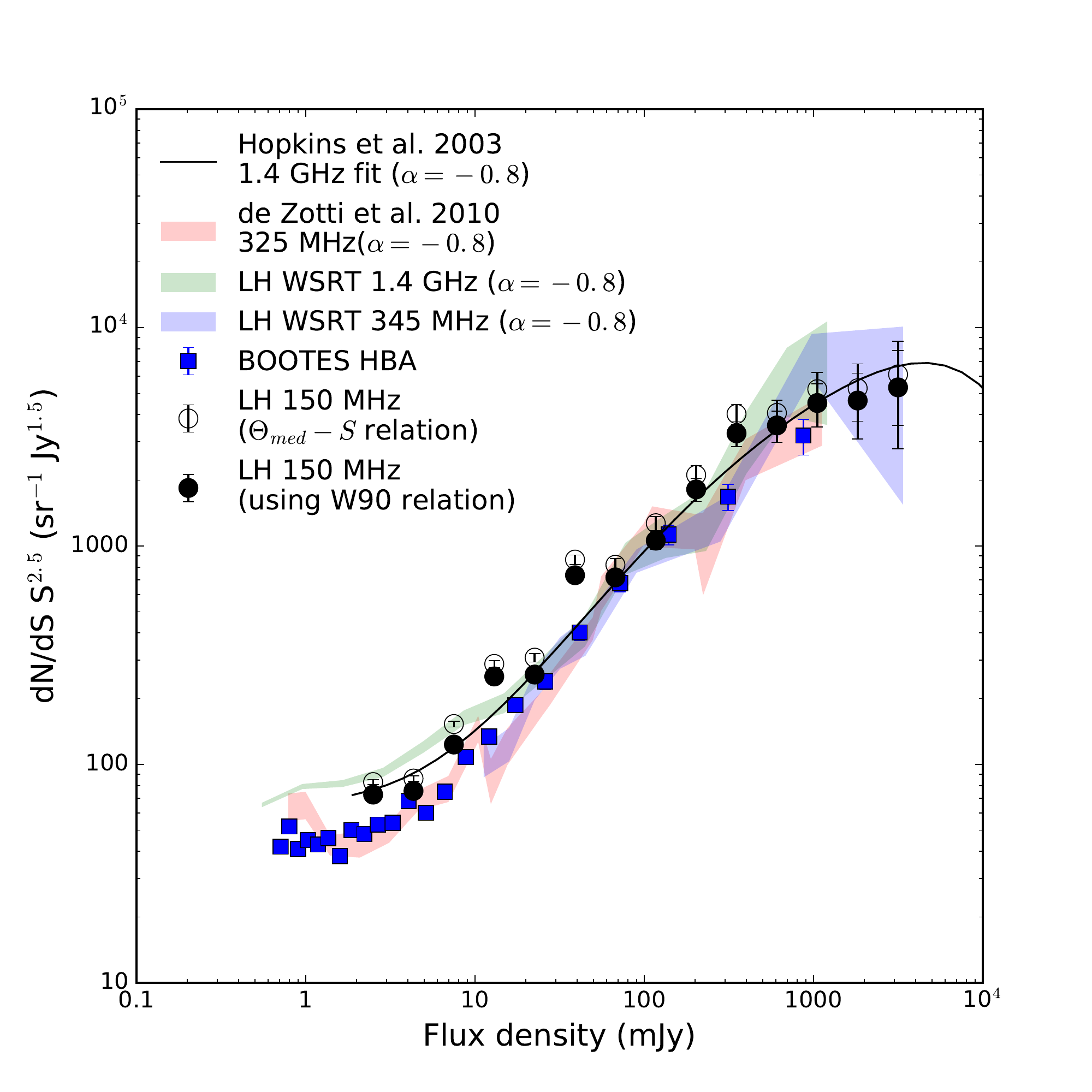}}   
      \caption{Normalized 150-MHz differential source counts as derived from the catalogue  discussed in this work (black points): the two different symbols represent the two recipes adopted for resolution bias correction: \citet{Windhorst1990} relation (filled circles) and modified $\Theta_{\rm med} - S$ relation (empty circles, see text for detail). Vertical bars represent Poissonian errors on the normalized counts. Our counts are compared to the ones derived in the LOFAR HBA Bo\"otes field (\citealt{Williams2016}, blue squares) and from extrapolations from higher frequencies. All extrapolations assume a spectral index of $\alpha=-0.8$. The green and blue shaded regions respectively refer to extrapolations from 1.4 GHz and 345\,MHz of the recent source counts derived in the same region as part of the Lockman Hole Project (Prandoni et al. 2016a,b, in preparation). The red shaded region refers to the 325-MHz data collection discussed by \citet{DeZotti2010}, while the black solid line shows the polynomial best fit to 1.4-GHz counts derived by \citet{Hopkins2003}. }
         \label{fig:diffcounts}
\end{figure*}

\begin{table*}
       \caption{150 MHz Source Counts derived from our LOFAR HBA survey. The normalised counts are derived using both the \citet{Windhorst1990} relation (W90 relation) and the modified $\Theta_{\rm med} - S$ relation (mod. $\Theta_{\rm med}$) for resolution bias as discussed in Sect.~\ref{sec-sizes}.}
         \label{tab-counts} 
     \begin{center}
      \begin{tabular}{crrrrc}
      \hline
      $\Delta S$ & $\langle S\rangle$ & $N_{S}$ & \multicolumn{2}{c}{${\rm d}N$/${\rm d}S$ $S^{2.5}$}  &  $\pm \sigma$ \\
      (mJy) & (mJy) &  & (${\rm sr^{-1} Jy^{1.5}}$) & (${\rm sr^{-1} Jy^{1.5}}$) \\
      &&& W90 relation & mod. $\Theta_{\rm med}$ & \\
      \hline 
 
1.9 - 3.3 & 2.5 & 981 & 72.7 & 83.0 & +2.3, -2.4 \\
3.3 - 5.7 & 4.3 & 872 & 75.6 & 86.1 & +2.6, -2.7 \\
5.7 - 9.9 & 7.5 & 717 & 123.3 & 153.0 & +4.6, -4.8 \\
9.9 - 17 & 13 & 561 & 252.7 & 288.4 & +10.7, -11.1 \\
17 - 30 & 23 & 349 & 257.9 & 308.1 & +13.8, -14.5 \\
30 - 50 & 39 & 283 & 734.8 & 866.3 & +43.7, -46.3 \\
50 - 90 & 68 & 165 & 718.3 & 820.4 & +55.9, -60.3 \\
90 - 150 & 117 & 123 & 1058 & 1271 & +95, -104 \\
150 - 270 & 203 & 69 & 1816 & 2116 & +219, -245 \\
270 - 460 & 351 & 59 & 3280 & 4025 & +427, -483 \\
460 - 800 & 608 & 37 & 3565 & 4068 & +586, -682 \\
800 - 1400 & 1052 & 20 & 4515 & 5224 & +1010, -1235 \\
1400 - 2400 & 1823 & 9 & 4636 & 5274 & +1545, -2060 \\
2400 - 4200 & 3157 & 4 & 5324 & 6105 & +2542, -4206 \\
	\hline
    \end{tabular}
      \end{center}

      \end{table*}

Fig.~\ref{fig:diffcounts} shows our 150 MHz source counts in comparison with other determinations from the literature. We notice that at high flux densities the counts obtained assuming the $\Theta_{\rm med} - S$ relation of \citet{Windhorst1990} are more reliable (black filled circles), while at low flux densities, where most sources are characterized by intrinsic angular sizes $<10-15$ arcsec, the counts derived assuming $2\times$ larger median sizes (black empty circles) should provide a better representation. 

Our 150-MHz counts broadly agree with the counts obtained in the Bo\"otes HBA field \citep{Williams2016}, as well as with extrapolations from higher frequencies, with the exception of a few points that are higher than expected. We note that contamination by artefacts (approximately 3.6\,per\,cent, see Sec.~\ref{crossmatchfull}) can have some impact on our source counts at low/intermediate flux densities, where this effect is of comparable size to the counts associated errors. In addition, extrapolations from 1.4\,GHz counts in the Lockman Hole region also show somewhat higher counts at these flux densities, suggesting that cosmic variance effects could play a role in explaining the differences in the source counts between the Lockman Hole and Bo\"otes fields. 

\section{Spectral index properties of low-frequency radio sources} \label{lockman-wide}

Studying the spectral index properties of low-frequency radio sources allows us to gain insight into the source populations detected at these frequencies. In order to carry out an unbiased analysis we have defined different sub-samples, each with different flux density limits, to best match the corresponding multi-frequency radio information available and represent a complete sample. 

We first crossmatched the full LOFAR catalogue with all-sky surveys to form the `Lockman--wide' subsample. Whilst this sample covers the entire field of view, the depth is limited due to the flux density limits of these surveys. In order to investigate the properties of low-frequency radio sources at fainter flux densities we also formed the `Lockman--WSRT' sample by crossmatching the LOFAR 150-MHz catalogue with deeper 1.4-GHz observations carried out with WSRT. The `Lockman--deep' sample was then formed by crossmatching this deeper Lockman--WSRT subsample with deeper surveys at other frequencies. Forming these subsamples is discussed in more detail in the following sections. An overview of the number of sources falling into each subsample is given in Table \ref{surveys}. 

We note that the observations used in this analysis are not contemporaneous so in some cases variability could lead to incorrect spectral indices, particularly for sources with peaked or rising spectra (where emission from the AGN core may be dominating the emission). However, since we select sources at 150\,MHz, where the radio emission is dominated by the steep spectrum lobes built up over timescales of Myrs--Gyrs, we do not expect variability to significantly affect the majority of sources.

\subsection{Crossmatching with wide-area sky surveys} \label{crossmatchfull}

To investigate the spectral index properties of low-frequency selected sources, we crossmatched the 150-MHz LOFAR catalogue with the 1.4-GHz NRAO VLA Sky Survey (NVSS; \citealt{nvss}), 325-MHz Westerbork Northern Sky Survey (WENSS; \citealt{wenss}) and 74-MHz VLA Low-Frequency Sky Survey Redux (VLSSr; \citealt{vlss}). Since these surveys are all at lower resolution we only include LOFAR sources that have deconvolved source sizes less than 40\,arcsec, approximately matching the resolution of NVSS. This excludes 25 sources (6.1\,per\,cent) from the following analysis, potentially biasing against some of the larger radio galaxies. However, given the additional complexities in obtaining accurate spectral indices for these sources, excluding these objects leads to a cleaner sample where the spectral indices are calculated in the same manner.

\begin{figure}
\centering{\epsfig{file=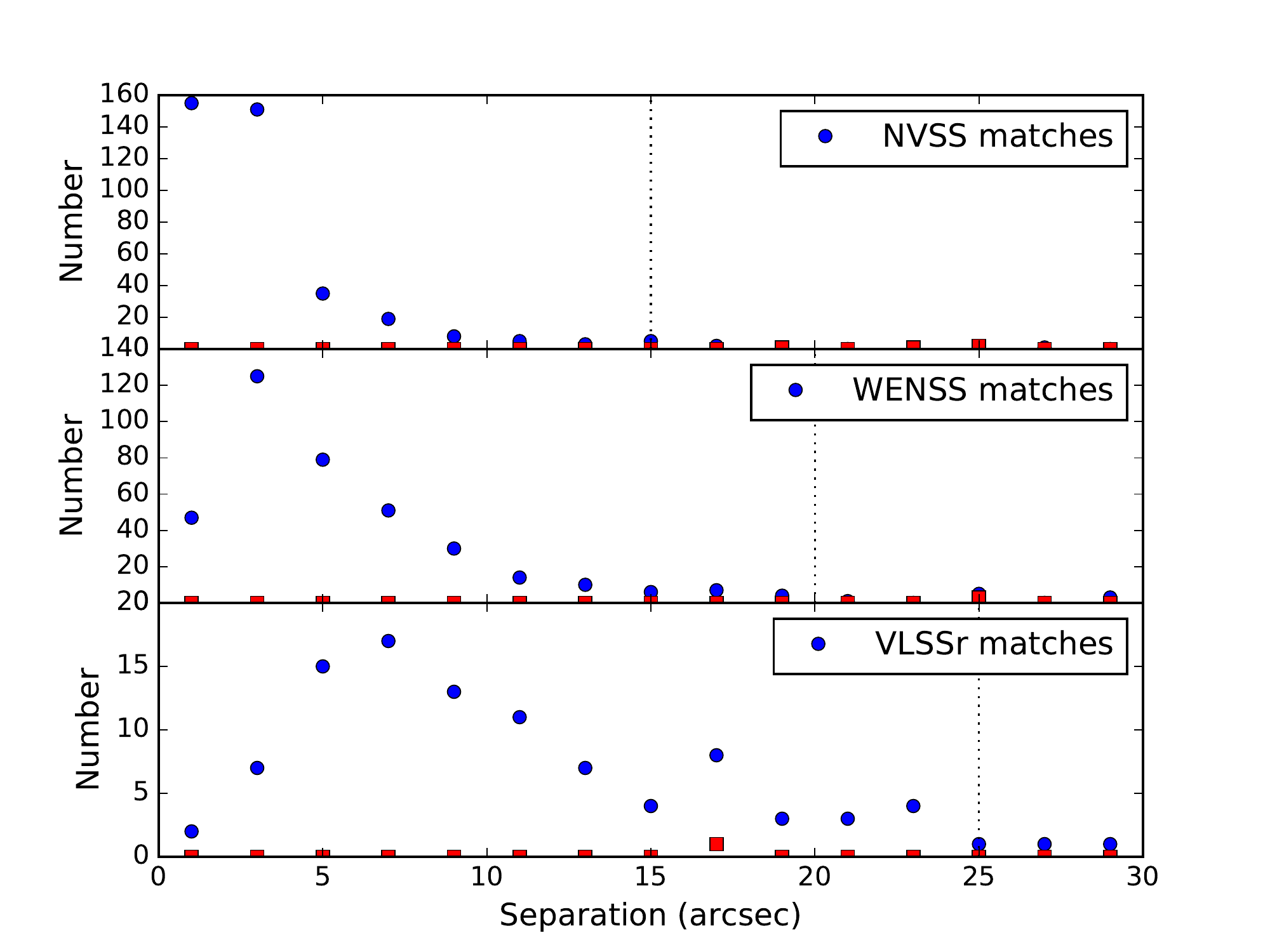, width=\linewidth}}
\caption{Monte Carlo tests to determine the optimal matching radius for automatically accepting associations in the NVSS, WENSS and VLSSr catalogues. The blue circles show the number of matches as a function of search radius using the LOFAR positions while the red squares show the number of matches for a random catalogue. The dashed lines mark the search radii chosen for each survey. \label{montecarlo} }
\end{figure}

For each survey we conducted a Monte Carlo test to compare how many spurious matches are included as a function of search radius. Fig. \ref{montecarlo} shows the results of these Monte Carlo tests for each of the NVSS, WENSS and VLSSr catalogues. From this analysis it was determined that the optimal search radius (to include the majority of real identifications, but limit the number of false IDs) was 15\,arcsec for the NVSS catalogue, 20\,arcsec for the WENSS catalogue and 25\,arcsec for the VLSSr catalogue. For the vast majority of sources there was only a single match within the search radius so we simply accepted the closest match when crossmatching these catalogues. Most sources are unresolved at the resolution of these surveys, but sources with extreme spectral indices were checked visually to confirm resolution effects were not affecting the spectral index calculation. 

All LOFAR sources that were not detected in NVSS were checked by eye to exclude any artefacts that were catalogued during the LOFAR source extraction. This process revealed 14 sources that were identified as artefacts in the LOFAR image, corresponding to 3.6\,per\,cent of the crossmatched sample. This doesn't include any artefacts that happened to be associated with an NVSS source, but from Fig.~\ref{montecarlo} this number is expected to be minimal (i.e. only 1 random match within a 20\,arcsec search radius).

Due to the different flux limits of each of these surveys, we have only included sources with $S_{150}>40$\,mJy such that the sample is not dominated by too many unrestrictive limits on the spectral indices. This flux limit was chosen such that any LOFAR sources not detected in NVSS (above a flux limit of 2.5\,mJy) have spectra steeper than $\alpha_{150}^{1400}=-1.2$, typically defined as Ultra-Steep Spectrum (USS) sources (see Section \ref{uss}). Similarly, sources not detected in WENSS (flux limited at 18\,mJy) have an upper limit of the spectral index of $\alpha_{150}^{345}<-1.0$. Due to the higher flux limit of the VLSSr survey only a small number of sources are detected at 74~MHz. For a complete comparison, we have identified a subset of the Lockman--wide sample with $S_{150}>300$\,mJy which, if undetected in the VLSSr survey, gives us a lower limit on the spectral index of $\alpha_{74}^{150}>-0.7$. All limits were confirmed by visual inspection to ensure that they were true non-detections.

This leaves us with 385 sources that form the Lockman--wide sample. Of this sample, 377 have counterparts at 1.4\,GHz in NVSS, 367 have counterparts in WENSS and 93 have VLSSr matches. A summary of the number of matches found in each catalogue is shown in Table \ref{surveys}. \citet{scaife+heald} note that the WENSS flux densities need to be scaled by a factor of 0.9 for agreement with the flux scale of \citet*{rcb}. However, by comparing with the spectral indices from 150\,MHz to 1400\,MHz we found that this correction resulted in underestimated flux densities at 325\,MHz for sources in the Lockman Hole. As such, we do not apply this correction in the following analysis. 

\subsection{Spectral analysis of the Lockman--wide sample.} \label{spec_wide}

\begin{figure}
\centering{\epsfig{file=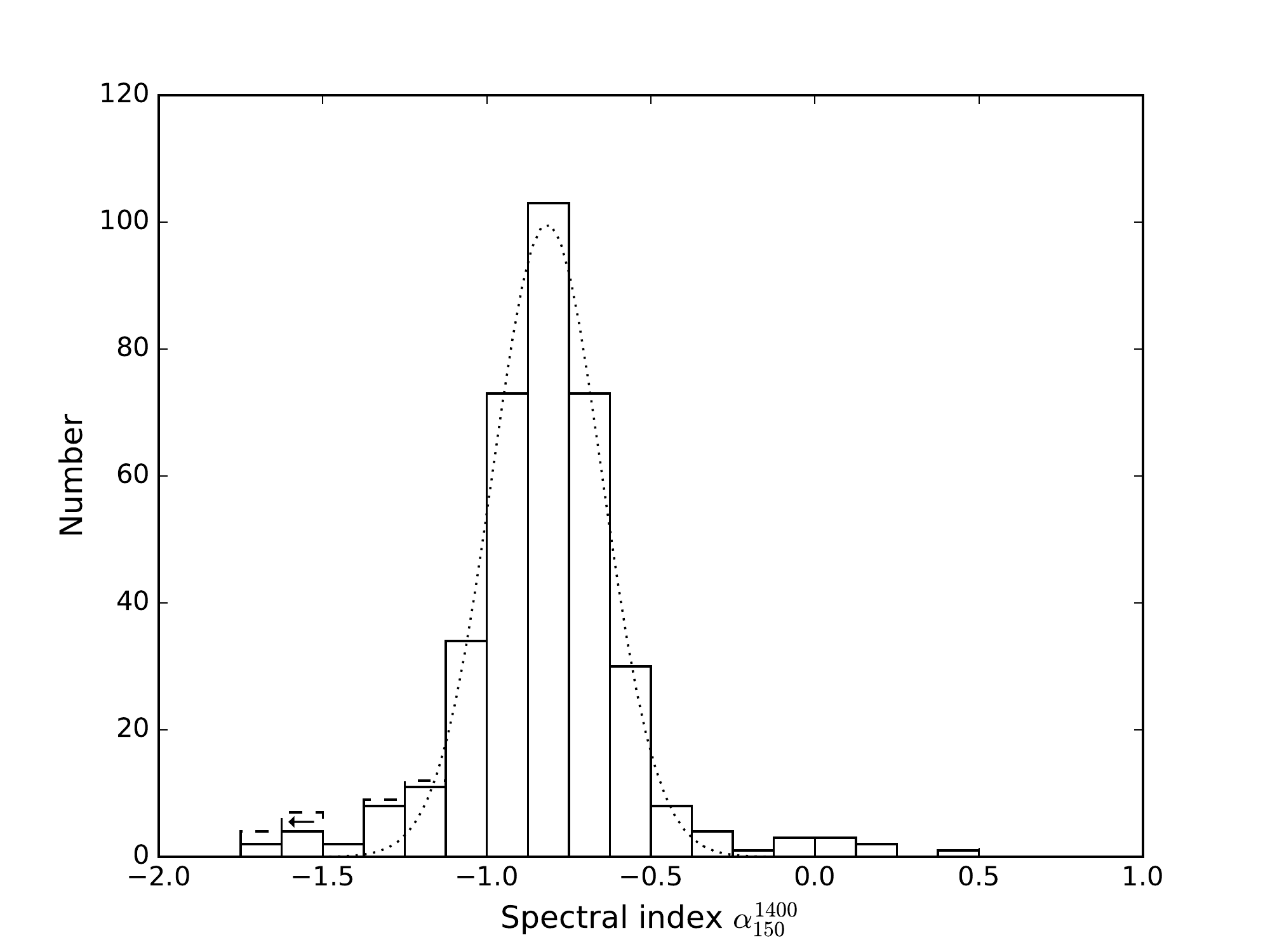, width=\linewidth}}
\caption{Histogram of spectral indices from 150\,MHz -- 1.4\,GHz for sources in the Lockman--wide sample (385 sources). The dashed line indicates upper limits on the spectral index for sources that were not detected at 1.4\,GHz. The dotted line marks a Gaussian fit to the distribution showing that there are more ultra-steep and flat-spectrum sources than would be expected in the tail of the distribution (if the spectral indices followed a normal distribution). This indicates that the source population probed by these surveys cover a wide range of source properties which can not be explained by a single population.\label{alphahistnvss} }
\end{figure}

We first study the 2-point spectral index from 150~MHz to 1.4~GHz for sources in the Lockman--wide sample. Fig. \ref{alphahistnvss} shows the spectral index distribution which has a median spectral index of $\alpha_{150}^{1400}=-0.82\pm0.018$ (errors from bootstrap) and an interquartile range of [$-0.94$, $-0.70$]. This is slightly steeper than found in previous studies which find the spectral index over these frequencies to be typically around $-0.78$ \citep{Williams2016, Williams2013, Intema2011, Ishwara-Chandra2010} but can range from $-0.76$ \citep{Hardcastle2016} to $-0.85$ \citep{Ishwara-Chandra2007}.

Based on the spectral indices between 150\,MHz and 1.4\,GHz we can classify sources into three different categories; flat-spectrum sources, which we define as $\alpha_{150}^{1400}>-0.5$, make up 5.7\,per\,cent of the sample, steep-spectrum sources ($-1.2<\alpha_{150}^{1400}<-0.5$) make up 89.4\,per\,cent and ultra-steep spectrum sources ($\alpha_{150}^{1400}<-1.2$) account for 4.9\,per\,cent of the sample. As expected for a low-frequency survey, the sample is predominately comprised of steep-spectrum radio sources. To determine the fraction of sources that are genuinely ultra-steep, we fit a Gaussian to the distribution shown by the dotted line in Fig. \ref{alphahistnvss}. From this distribution we would expect only 2\,per\,cent of sources to have $\alpha<-1.2$ if we were probing a single population with finite S/N. This confirms that the source populations revealed in these observations follow a more complicated distribution in source properties and suggests that the majority of outlying spectral indices are real and an intrinsic property of the radio source. 

\begin{figure*}
\begin{minipage}{0.45\linewidth}
\centering{\includegraphics[width=\linewidth]{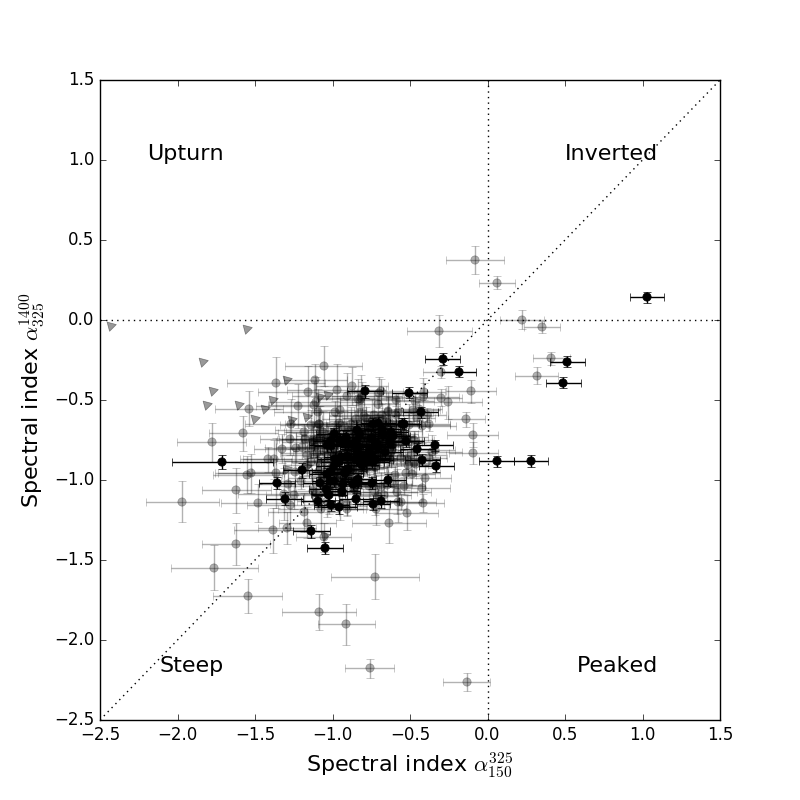}}  \end{minipage}
\begin{minipage}{0.45\linewidth}
\centering{\includegraphics[width=\linewidth]{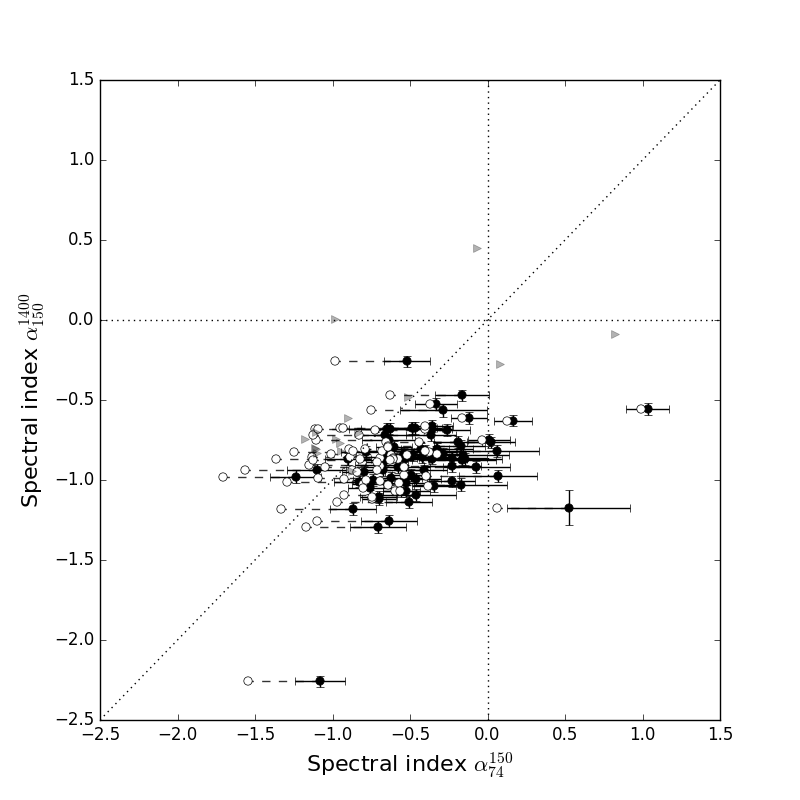}}
\end{minipage}
\caption{Radio colour-colour plots for sources in the Lockman--wide sample. Left: The spectral indices of the sample between 150 and 325\,MHz against the spectral indices from 325\,MHz--1.4\,GHz. The black points identify the subset of sources which are also shown on the right (i.e. S$_{150}>300$\,mJy). The grey points show the remaining sources in the Lockman--wide sample and the grey triangles denote upper limits for sources not detected at 325\,MHz. Right: Spectral indices between 74 and 150\,MHz against 150\,MHz--1.4\,GHz spectral indices. The filled symbols show the spectral indices using the catalogued VLSSr flux densities while the open circles show the spectral indices if the flux density corrections reported by \citet{vlss} are applied. The grey triangles denote upper limits for sources not detected at 74\,MHz. The diagonal dashed lines mark a power law across the full frequency range plotted. \label{nvssalpha-alpha}} 
\end{figure*}

Whilst the 2-point spectral index can provide information on the dominant source of the radio emission, assuming a power-law over such a large frequency range does not probe any curvature that may be present. To investigate this, Fig.~\ref{nvssalpha-alpha} shows the radio colour-colour plots, which compare the spectral indices between different frequency ranges, for sources in the Lockman--wide sample. The errors on the spectral index were calculated using the following formula:

\begin{equation}
\alpha_{\rm err} = \frac{1}{\rm ln\frac{\nu1}{\nu2}}\sqrt{\Big( \frac{S_{1,\rm err}}{S_1} \Big)^2 + \Big( \frac{S_{2,\rm err}}{S_2} \Big)^2 }
\end{equation}

where $\nu_{1,2}$ refers to the frequencies and $S_{1,2}$ the corresponding flux densities at those frequencies. This takes into account the larger errors associated with spectral indices calculated over a smaller frequency range. 

The left-hand plot in Fig.~\ref{nvssalpha-alpha} shows the spectral indices between 150 and 345~MHz compared to the spectral indices from 345--1400~MHz. Nearly all of the sources lie on the diagonal line indicating that they exhibit power-law spectra across the entire frequency range (i.e. 150~MHz to 1.4~GHz). The figure on the right probes the spectral indices at the lowest end of the frequency range studied: 74--150~MHz compared to 150--1400~MHz for sources that fall into the VLSSr-subset of the Lockman--wide sample. Sources that fall into this subset are marked in black in the left-hand plot, while the full sample is shown by the grey points. Sources that are not detected are shown as limits indicated by the triangles. 

Although there is no indication of any spectral curvature between 150\,MHz and 1.4\,GHz, when going to lower frequencies there is a tendency for objects to lie slightly to the right of the diagonal line, suggesting that the radio spectra of these objects begin to flatten between 74 and 150\,MHz. However, to verify if this trend is an intrinsic property of the source population, we first need confidence in our absolute flux calibration. 

While the VLSSr flux densities have already been corrected to bring them onto the \citet{scaife+heald} flux scale, \citet{vlss} noted that the VLSSr fluxes were slightly underestimated compared to the predicted flux density extrapolated from the 6C and 8C catalogues. Predicting the 74\,MHz flux densities from the 6C and 8C surveys \citet{vlss} reported flux density ratios of VLSSr/predicted$=0.97$ for sources with $S_{74}>1$\,Jy and  VLSSr/predicted$=0.72$ for sources fainter than 1\,Jy. Applying these corrections to the 74\,MHz flux densities results in steeper spectral indices as shown by the open circles in Fig.~\ref{nvssalpha-alpha}. Due to the uncertainities associated with the absolute flux scale at these frequencies we show both the corrected (open symbols) and uncorrected (filled symbols) spectral indices. These large uncertainties make it difficult to ascertain if there is any intrinsic spectral flattening at these frequencies. This is discussed further in Section \ref{spec_deep}.

\subsection{Crossmatching with deeper radio surveys in the Lockman Hole field}

\begin{figure}
\includegraphics[width=\linewidth]{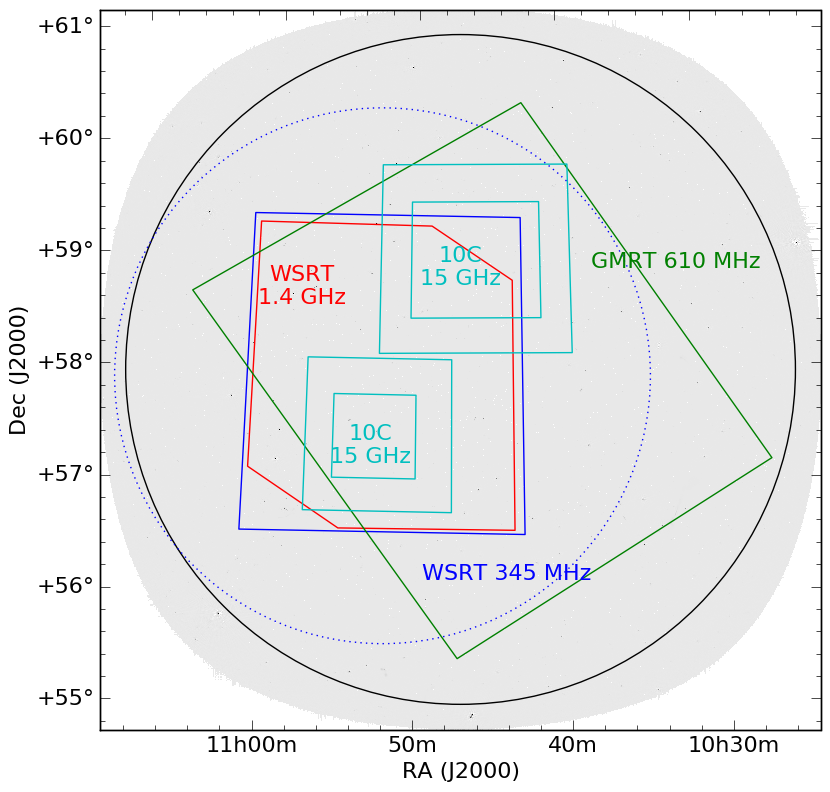} 
\caption{The footprints of each survey used in the spectral index analysis. The greyscale shows the LOFAR HBA primary beam and the black circle shows the area that was included in the source catalogue (i.e. up to 3 degrees from the pointing centre). Footprints of other survey areas are labelled accordingly. The primary beam of the LOFAR LBA observations is not shown here, but it has the same pointing centre as the HBA data and, since it is at lower frequency, covers a larger area. \label{footprints}} 
\end{figure}

One of the limitations in carrying out this analysis using all-sky surveys such as VLSSr and NVSS is that these surveys do not reach sufficient depths for all the LOFAR sources to be detected at these frequencies. However, the advantage of observing well-studied fields such as the Lockman Hole is that there is extensive, deeper multi-band data available in these regions as discussed in Section \ref{intro}. The survey area covered by each of these observations is shown in Fig.~\ref{footprints} and a summary of the different surveys used in this analysis is given in Table \ref{surveys}. Crossmatching with deeper high-frequency data allows us to exploit the depth of the LOFAR data and push this study of the low-frequency spectral properties to much fainter radio source populations.

To obtain a complete subset of sources that probe fainter flux densities we first crossmatch the full LOFAR 150-MHz catalogue with WSRT observations at 1.4~GHz. We then use this catalogue to crossmatch with other deep radio surveys of the Lockman Hole field. In this section we briefly describe each of these surveys and how they were matched to their LOFAR counterparts. 

\begin{table*}
\caption{Radio surveys that were used to form the multi-frequency samples. The numbers listed for each survey in the Lockman--deep sample refer to sources that have $S_{150}>9$\,mJy. The crossmatching of these surveys was carried out for the full Lockman--WSRT sample so the total number of detections for each survey is slightly higher. See Sections \ref{w90}--\ref{lbacrossmatch} for the total number of matches in each catalogue.  \label{surveys}}
\begin{tabular}{llcccccc}
  \hline                       
& Survey & Frequency & Resolution & Area covered & rms & No. matches \\
&  &  & (arcsec) & (sq. degrees) & (mJy) &  \\
  \hline
& LOFAR & 150 MHz & 18.6 $\times$ 14.7 & 34.7 & 0.16 & 4882 \\
  \hline	
{\bf Lockman--wide:} & LOFAR & 150 MHz & 18.6 $\times$ 14.7 & 34.7 & 0.16 & 385 \\
& \multicolumn{3}{l}{(Point sources with $S_{150}>40$~mJy)} & & & \\
& NVSS & 1.4 GHz & 45 & 34.7 & 0.5 & 377 \\
& WENSS & 345 MHz & 54 $\times$ 64 & 34.7 & 3.6  & 367 \\
& VLSS & 74 MHz & 80 & 34.7 & 100  & 93	 \\
  \hline
{\bf Lockman--WSRT:} & LOFAR & 150 MHz & 18.6 $\times$ 14.7 & 6.6 & 0.16 & 1302 \\
& \multicolumn{3}{l}{(all sources in same area as WSRT field)} & & &\\
& WSRT & 1.4 GHz & 11 $\times$ 9 & 6.6 & 0.011 & 1289 \\
\hline
{\bf Lockman--deep:} & LOFAR & 150 MHz & 18.6 $\times$ 14.7 & 6.6 & 0.16 & 326 \\
& \multicolumn{3}{l}{(Sources in WSRT field with $S_{150}>9$~mJy)} & & &\\
& WSRT & 1.4 GHz & 11 $\times$ 9 & 6.6 & 0.011 & 322 \\ 
& 10C & 15 GHz & 30 & 4.6 & 0.05--0.1  & 81 \\
& GMRT\footnotemark & 610 MHz & 6 $\times$ 5 & 13 & 0.06  & 103\\
& WSRT 90cm (W90) & 345 MHz & $70 \times 44$ & 7 & 0.8 & 211 \\
& LBA & 60 MHz & 45 & 6.6 & 20 & 42 \\ 
  \hline 
\end{tabular}
\end{table*}

\footnotetext{For the GMRT catalogue only point sources with flux densities above 5 mJy were used in the crossmatching (see Section \ref{gmrt}).} 

\subsubsection{Westerbork 1.4\,GHz mosaic} \label{wsrtcrossmatch}

A 16-pointing mosaic of the Lockman Hole field was observed at 1.4\,GHz with the Westerbork Synthesis Radio Telescope (WSRT). This covers a 6.6 square degree field down to a a rms noise level of 11$\mu$Jy at the centre of the field with a resolution of 11$\times$9\,arcsec. For more details of the Westerbork observations we refer the reader to the accompanying paper (Prandoni et al., 2016a, in preparation). 

These WSRT observations resulted in a catalogue of over 6000 sources which was then crossmatched with the 150\,MHz LOFAR catalogue. To find the optimal search radius a Monte Carlo test was again carried out. Comparing the number of real matches to random matches revealed that a search radius of 10\,arcsec was the ideal cutoff to limit the number of spurious matches whilst also including the maximum number of real matches. 

Although the majority of sources are catalogued as point sources in both the WSRT and LOFAR catalogues, simply finding the closest match is not as reliable for sources with complex morphology or sources which have been separated into multiple components. To ensure that objects were not being excluded due to the 10\,arcsec search radius, and to check the reliability of the matches, we applied the following criteria to determine the best matched catalogue: 

\begin{itemize}
\item If there was a single WSRT match within 10\,arcsec of the LOFAR source, this match was automatically accepted.
\item If there were multiple WSRT matches within 10\,arcsec of the LOFAR source, both matches were visually inspected to choose the best match by eye. This was particularly useful for cases where the LOFAR sources had been catalogued as separate components, but were listed as a single source in the WSRT catalogue (note that these are flagged in the catalogue as discussed at the end of this section).
\item For LOFAR sources which did not have a WSRT counterpart, the search-radius was expanded to 15\,arcsec. This was primarily needed for extended sources where the fitted peak position differed between the LOFAR and WSRT catalogues. All sources matched in this way were visually inspected to confirm the match was likely to be real. 
\item Sources that were identified as extended in either the LOFAR or WSRT source catalogues were checked by eye to confirm the correct counterparts were identified. 
\end{itemize}

This resulted in a final sample of 1302 sources, which we refer to hereafter as the Lockman--WSRT sample. An additional flag was added to the catalogue referring to how the match was identified. The numbers in brackets refer to the number of sources which fall into each category:

\begin{itemize}
\item V - This match was identified and/or confirmed by visual inspection (299 sources)
\item A - This match was automatically accepted (974 sources)
\item C - A complex match (16 sources). Likely to be a genuine match, but a mismatch between the two source catalogues. For example, a double-lobed radio galaxy may have been classified as a single source in the WSRT catalogue, but catalogued as two separate components in the LOFAR catalogue. These sources are included in the catalogue\footnote{Sources were catalogued based on the LOFAR source extraction. For example, if a source was catalogued as two components in the LOFAR catalogue, then both these are listed and matched to the same WSRT source.}, but have been excluded from further analysis.
\item L - A LOFAR source with no WSRT counterpart (6 sources). These are discussed further in Section \ref{uss_deep}. 
\item LC - A LOFAR component with no WSRT counterpart (7 sources). Also discussed further in Section \ref{uss_deep}. 
\end{itemize}

\subsubsection{Lockman--WSRT catalogue format} \label{cat}

The full Lockman--WSRT catalogue is included as supplementary material in the electronic version of the journal. A sample of the catalogue is shown in Table \ref{lofarwsrtcat} and below we list the columns included in the published catalogue. Note that this catalogue only includes LOFAR sources in the area covered by the WSRT 1.4-GHz mosaic since virtually all of the LOFAR sources have a match in the WSRT catalogue. This reduces the risk of artefacts around bright sources in the larger LOFAR field being included. The catalogue of the full LOFAR field at higher resolution will be published in a future paper. \\

\noindent \emph{Column 1:} LOFAR name \\
\emph{Columns 2, 3:} LOFAR 150\,MHz source position (RA, Dec) \\
\emph{Columns 4, 5:} Integrated 150\,MHz flux density and error \\
\emph{Column 6:} local rms \\
\emph{Column 7:} Source flag (from PyBDSM - S/C/M) \\
\emph{Column 8:} WSRT name \\
\emph{Columns 9, 10:} WSRT position (RA, Dec) \\
\emph{Columns 11--12:} Integrated 1.4\,GHz flux density and error \\
\emph{Columns 13--14} Peak 1.4\,GHz flux density and error\\
\emph{Columns 15--17:} Deconvolved 1.4~GHz source size and orientation\\ 
\emph{Column 18:} Spectral index between 150\,MHz and 1.4\,GHz ($\alpha_{150}^{1400}$) \\

\begin{landscape}
\begin{table}
\caption{First 10 lines of the Lockman--WSRT catalogue. The full catalogue is available in the online version of this paper. \label{lofarwsrtcat}}
\centering
\scriptsize
\begin{tabular}{lllrrlllllrrrrrrrll}
\hline
\textbf{LOFAR name} & \textbf{RA} & \textbf{DEC} & \textbf{St$_{150}$} & \textbf{err} & \textbf{rms} & \textbf{Flag} & \textbf{WSRT name} & \textbf{RA} & \textbf{DEC} & \textbf{Sint$_{1.4}$} & \textbf{err} & \textbf{Spk$_{1.4}$} & \textbf{err} & \textbf{maj}{\textdagger} & \textbf{min} & \textbf{pa} & \textbf{$\alpha_{150}^{1400}$} & \textbf{ID flag} \\
& & & (mJy) & (mJy) & (mJy) & & & & & (mJy) & (mJy) & (mJy) & (mJy) & $"$ & $"$ & $^{\circ}$ & & \\
\hline
J104316+572452 & 10:43:16.3 & +57:24:52.8 & 277.8 & 19.54 & 0.73 & M & LHWJ104316+572453 & 10:43:16.11 & +57:24:53.7 & 42.4 & 0.39 & 39.2 & 0.36 & 4.05 & 0.00 & 39.1 & -0.8 & V \\
J104321+583440 & 10:43:21.6 & +58:34:40.3 & 19.8 & 1.44 & 0.22 & S & LHWJ104321+583438 & 10:43:21.65 & +58:34:38.3 & 5.7 & 0.21 & 5.5 & 0.20 & 0.00 & 0.00 & 0.0 & -0.6 & A \\
J104325+581854 & 10:43:25.6 & +58:18:54.5 & 6.3 & 0.55 & 0.19 & S & LHWJ104325+581852 & 10:43:25.49 & +58:18:52.0 & 2.3 & 0.25 & 2.0 & 0.22 & 4.62 & 3.04 & 46.3 & -0.4 & A \\
J104328+575809 & 10:43:28.1 & +57:58:09.1 & 23.1 & 1.66 & 0.21 & S & LHWJ104328+575807 & 10:43:28.16 & +57:58:07.3 & 13.4 & 0.23 & 12.8 & 0.22 & 2.71 & 1.51 & -22.0 & -0.2 & A \\
J104336+574017 & 10:43:36.4 & +57:40:17.9 & 5.2 & 0.51 & 0.21 & S & LHWJ104336+574015 & 10:43:36.35 & +57:40:15.4 & 1.3 & 0.19 & 1.1 & 0.16 & 0.00 & 0.00 & 0.0 & -0.6 & A \\
J104343+575513 & 10:43:43.0 & +57:55:13.6 & 5.1 & 0.50 & 0.21 & C & LHWJ104342+575509 & 10:43:42.93 & +57:55:09.4 & 0.9 & 0.16 & 0.8 & 0.14 & 0.00 & 0.00 & 0.0 & -0.8 & V \\
J104344+574258 & 10:43:44.3 & +57:42:58.5 & 1.8 & 0.37 & 0.20 & S & LHWJ104344+574252 & 10:43:44.89 & +57:42:52.5 & 0.8 & 0.13 & 0.7 & 0.13 & 0.00 & 0.00 & 0.0 & -0.4 & A \\
J104348+580427 & 10:43:48.1 & +58:04:28.0 & 7.3 & 0.62 & 0.20 & S & LHWJ104348+580412A & 10:43:48.10 & +58:04:25.8 & 9.5 & 0.13 & 8.8 & 0.12 & 3.41 & 1.76 & 12.8 & 0.1 & V \\
J104351+565550 & 10:43:51.0 & +56:55:50.3 & 6.5 & 0.77 & 0.36 & S & LHWJ104350+565552 & 10:43:50.93 & +56:55:52.8 & 2.1 & 0.37 & 2.2 & 0.39 & 0.00 & 0.00 & 0.0 & -0.5 & A \\
J104351+580338 & 10:43:51.2 & +58:03:38.1 & 9.8 & 0.76 & 0.20 & S & LHWJ104348+580412B & 10:43:51.32 & +58:03:35.7 & 3.4 & 0.12 & 3.3 & 0.12 & 0.00 & 0.00 & 0.0 & -0.5 & V \\

\hline
\end{tabular}

\begin{tablenotes}
\item{\textdagger}{Objects listed with source sizes of 99.9 have been detected as two separate components in the WSRT mosaic, but combined together and catalogued as one object in post-processing. See Prandoni et al., (2016a, in preparation) for further details. }
\end{tablenotes}
\end{table}
\end{landscape}

\emph{Column 19:} Identification flag (see Section \ref{wsrtcrossmatch}) \\

We use this Lockman--WSRT catalogue for crossmatching with other frequency datasets discussed in Sections \ref{w90}--\ref{lbacrossmatch}. 

\subsubsection{WSRT 90-cm observations} \label{w90}

Deep 345-MHz observations of the same region imaged at 1.4~GHz were obtained with the WSRT in maxi-short configuration during winter 2012. The expected thermal noise in our 345-MHz image (uniform weighting) was $\sim 0.11$ mJy/beam. However, given the poor resolution of Westerbork at 90-cm band ($\sim 70$\,arcsec $\times 44$\,arcsec in our case), confusion limited the effective noise to $\sim 0.5$ mJy/beam. These data are presented in further detail in an accompanying paper (Prandoni et al., 2016b, in preparation).

A catalogue of 234 sources was extracted from the inner $\sim$7 square degrees of the image (extending to $\sim 90$ arcmin distance from the image centre), where the noise is lowest and flattest. About half of this inner region is characterized by noise values $\leq 0.8$ mJy, assumed as a reference noise value for the source extraction. This catalogue was then crossmatched with the WENSS catalogue to verify the extracted flux densities at this frequency. A correction factor of 1.1 was applied to the WSRT 345-MHz sources to align with the WENSS flux density scale.

The region covered by the extracted catalogue at 345\,MHz covers the same region as the deep WSRT 1.4-GHz catalogue. This is marked by the rectangular area in Fig.~\ref{footprints} while the primary beam of the 345-MHz observations is shown by the dotted circle. An association radius of 25 arcsec was used resulting in 225 matches to the LOFAR-WSRT catalogue. Due to the large difference in resolution between the WSRT 345-MHz data and the LOFAR and WSRT 1.4-GHz data, sources with extreme spectral indices were visually inspected and flagged in cases where this was a problem (typically associated with multiple LOFAR sources being matched to the same WSRT 345-MHz source). This process excluded 16 sources. Due to the varying rms of the image (in particular around bright sources), any sources not detected at 345\,MHz were given upper limits of 5$\times$ the local rms. For the majority of sources this was approximately 4\,mJy.

\subsubsection{GMRT 610\,MHz mosaic} \label{gmrt}

A large mosaic of the Lockman Hole field was made from observations at 610\,MHz with the Giant Metre Wavelength Telescope (GMRT) from 2004--2006 covering a total area of 13 deg$^2$ \citep{Garn2008, Garn2010}. This mosaic covers the central region of the LOFAR primary beam and overlaps with the majority of the deep 1.4-GHz WSRT mosaic (excluding a small area in the south-eastern corner of the WSRT coverage) and reaches an rms of 60\,$\mu$Jy in the central regions at a resolution of $5\times6$\,arcsec.

When crossmatching this catalogue with the LOFAR--WSRT catalogue, only point sources at 610~MHz were included due to the difference in resolution. On inspection of the spectral indices between 150\,MHz, 610\,MHz and 1.4\,GHz it was also discovered that GMRT sources fainter than 5\,mJy tend to have underestimated flux density measurements so a flux density limit of $S_{610, pk}=$5\,mJy was applied for the crossmatching. The underestimation of fluxes for the fainter sources in this 610-MHz mosaic was also noted by \citet{Whittam2013}. One possible explanation for this is that the mosaic was not {\sc CLEAN}ed deeply enough therefore the fainter sources were not deconvolved sufficiently. We used a search radius of 10\,arcsec which resulted in 121 matches at 610\,MHz.

\subsubsection{10C survey}

The 10C survey was observed with the Arcminute Microkelvin Imager (AMI) at 15.7\,GHz. This survey covers 27 deg$^2$ at 30\,arcsec resolution across 10 different fields, two of which are in the Lockman Hole region. These two fields cover 4.64 deg$^2$ down to an rms noise level of 0.05\,mJy in the central regions of each field and 0.1\,mJy in the outskirts. For more details on the observations and data reducion for the 10C survey we refer the reader to \citet{10c}. When crossmatching with the LOFAR--WSRT catalogue we used a search radius of 15\,arcsec to match the analysis carried out by \citet{Whittam2013}. There are 119 LOFAR sources detected in the 10C survey.

\subsubsection{LBA LOFAR observations} \label{lbacrossmatch}

The LBA observations, data reduction and source extraction is discussed in Section \ref{lba}. Although the 60-MHz LOFAR data do not reach the fainter flux density limits of the surveys listed in previous sections, the advantage of crossmatching with the Lockman--WSRT catalogue is that it limits the chance of including artefacts in the spectral analysis. For sources not detected in the LBA image we place a 5$\sigma$ upper limit on the 60-MHz flux density of 100\,mJy. Using a matching radius of 20\,arcsec we find 46 matches in the LOFAR 60-MHz catalogue. 

\subsection{Multi-frequency spectral analysis of the LOFAR--WSRT and Lockman--deep samples}  \label{lockmandeep}

Using the multi-frequency information available in the Lockman Hole field we can study the spectral properties of low-frequency radio sources in a similar manner as was done for the Lockman--wide sample. As much of the complementary data goes to fainter flux density limits, albeit in a smaller area of sky, we can study a larger sample of radio sources to confirm if the trends found earlier are significant, and also investigate if the spectral behaviour of these radio sources changes at lower flux densities. 

Due to the similar, or lower resolution, of most of the surveys, we were able to include all the LOFAR sources in this analysis. Again, all upper limits were confirmed by visual inspection and in cases of a clear mismatch these were excluded from the analysis. 

\subsubsection{Spectral analysis from 150 MHz to 1.4 GHz.} \label{lofar-wsrt}

\begin{figure}
\centering{\epsfig{file=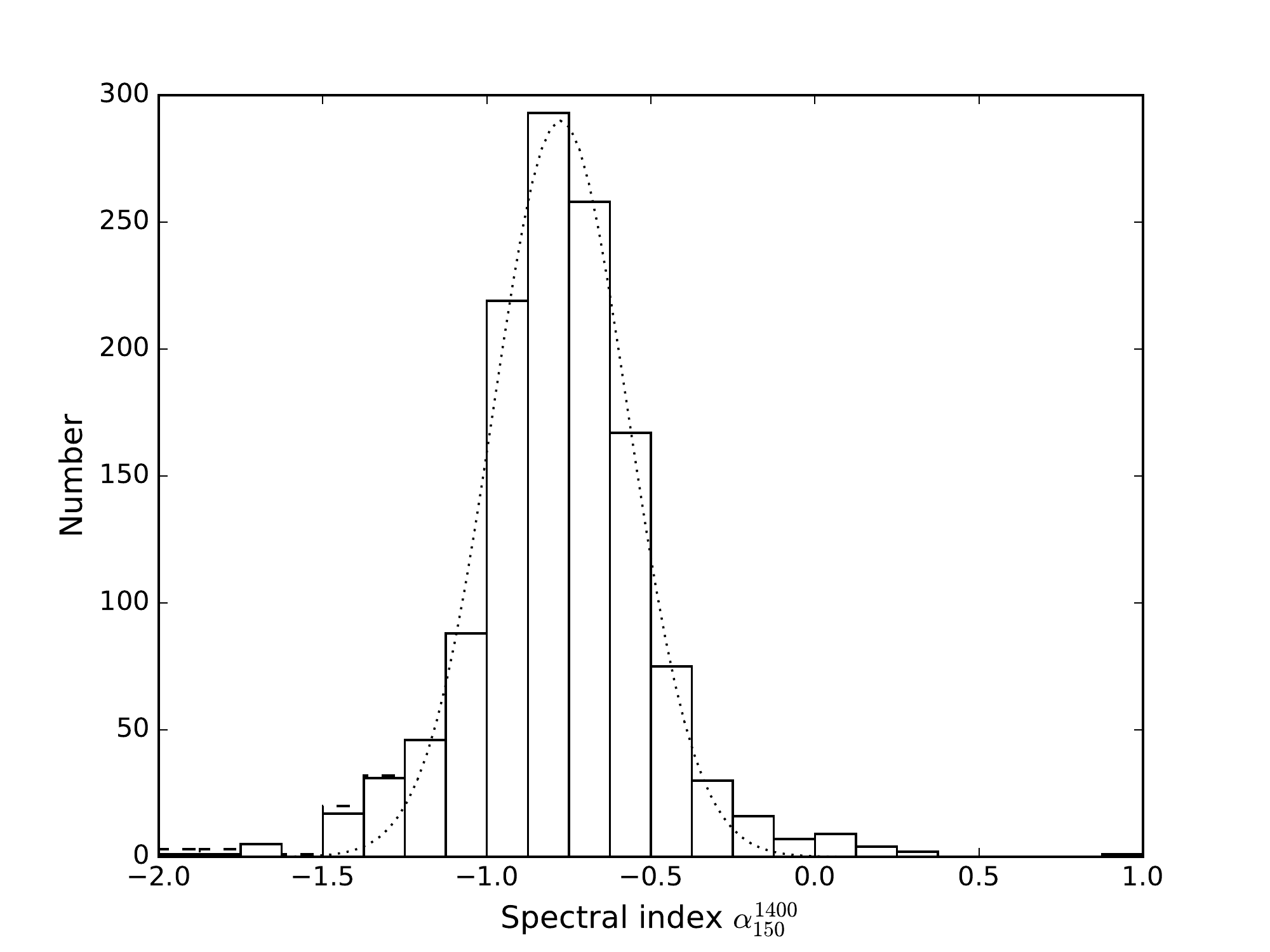, width=\linewidth}}
\caption{Histogram of spectral indices from 150\,MHz -- 1.4\,GHz for sources in the Lockman--WSRT sample (1302 sources). Upper limits of the spectral index are marked by the dashed line. The dotted line marks a Gaussian fit to the distribution, showing that a large number of the ultra-steep spectrum and flat-spectrum sources can not be attributed to the tail of the distribution.  \label{alphahist} }
\end{figure}

We first investigate the spectral indices between 150\,MHz and 1.4\,GHz in the Lockman--WSRT catalogue. Due to the flux limit reached in the WSRT mosaic, any LOFAR sources not detected at 1.4~GHz have upper limits of $\alpha_{150}^{1400}<-1.2$. The distribution of the 150\,MHz--1.4\,GHz spectral index is shown in Fig.~\ref{alphahist} with upper limits denoted by the dashed line. The median spectral index is $\alpha_{150}^{1400}=-0.78\pm0.015$ (errors from bootstrap) with an interquartile range of [$-0.95$, $-0.65$],  consistent with previous studies of 150~MHz-selected radio sources \citep{Intema2011, Williams2013, Williams2016, Hardcastle2016}. 

Separating the Lockman--WSRT sample into the three primary categories discussed earlier we find that flat-spectrum sources ($\alpha_{150}^{1400}>-0.5$) represent 11.3\,per\,cent of the sample, steep-spectrum sources ($\alpha_{150}^{1400}<-0.5$) make up 82.1\,per\,cent and ultra-steep spectrum sources ($\alpha_{150}^{1400}<-1.2$) account for 6.6\,per\,cent of the sample. The majority of these sources are genuinely ultra-steep, with only 2\,per\,cent of sources with $\alpha_{150}^{1400}<-1.2$ expected in the tail of a Gaussian distribution (shown by the dotted line in Fig.~\ref{alphahist}).

These ultra-steep spectrum sources are discussed further in Section \ref{uss_deep}. The division of the Lockman--WSRT sample into these three categories is roughly the same as for the Lockman--wide sample, suggesting that the source populations probed at these frequencies do not significantly change by going deeper at 150\,MHz. 

Other studies of spectral index properties of radio sources have reported a flattening of the spectral index as a function of flux density. For example, \citet{Prandoni2006} found that the spectral indices between 1.4\,GHz and 5\,GHz became flatter with decreasing flux density down to $\sim$0.5\,mJy. On the other hand, \citet{Randall2012} did not see any evidence for a flattening of the spectral index with decreasing flux density between 843\,MHz and 1.4\,GHz. However, it is important to note a key difference in the selection of these samples; \citet{Prandoni2006} selected sources at 5\,GHz which will be more dominated by flat spectrum sources than the 843-MHz selected sample studied by \citet{Randall2012}. 

At lower frequencies, \citet{Intema2011} found that the median spectral index between 153~MHz and 1.4~GHz became flatter with decreasing flux density between 5~mJy and 2~Jy. This is in agreement with the studies of \citet{Ishwara-Chandra2010} and \citet{Williams2013}, which both report a spectral flattening below approximately 20--50\,mJy using spectral indices from 150\,MHz to 610\,MHz and 150\,MHz to 1.4\,GHz respectively. However, the spectral index limits imposed by the flux density limits in these latter studies systematically biases against detecting steep-spectrum sources at the fainter end of the flux density distribution, as noted in \citet{Williams2013}. 

\begin{figure}
\includegraphics[width=\linewidth]{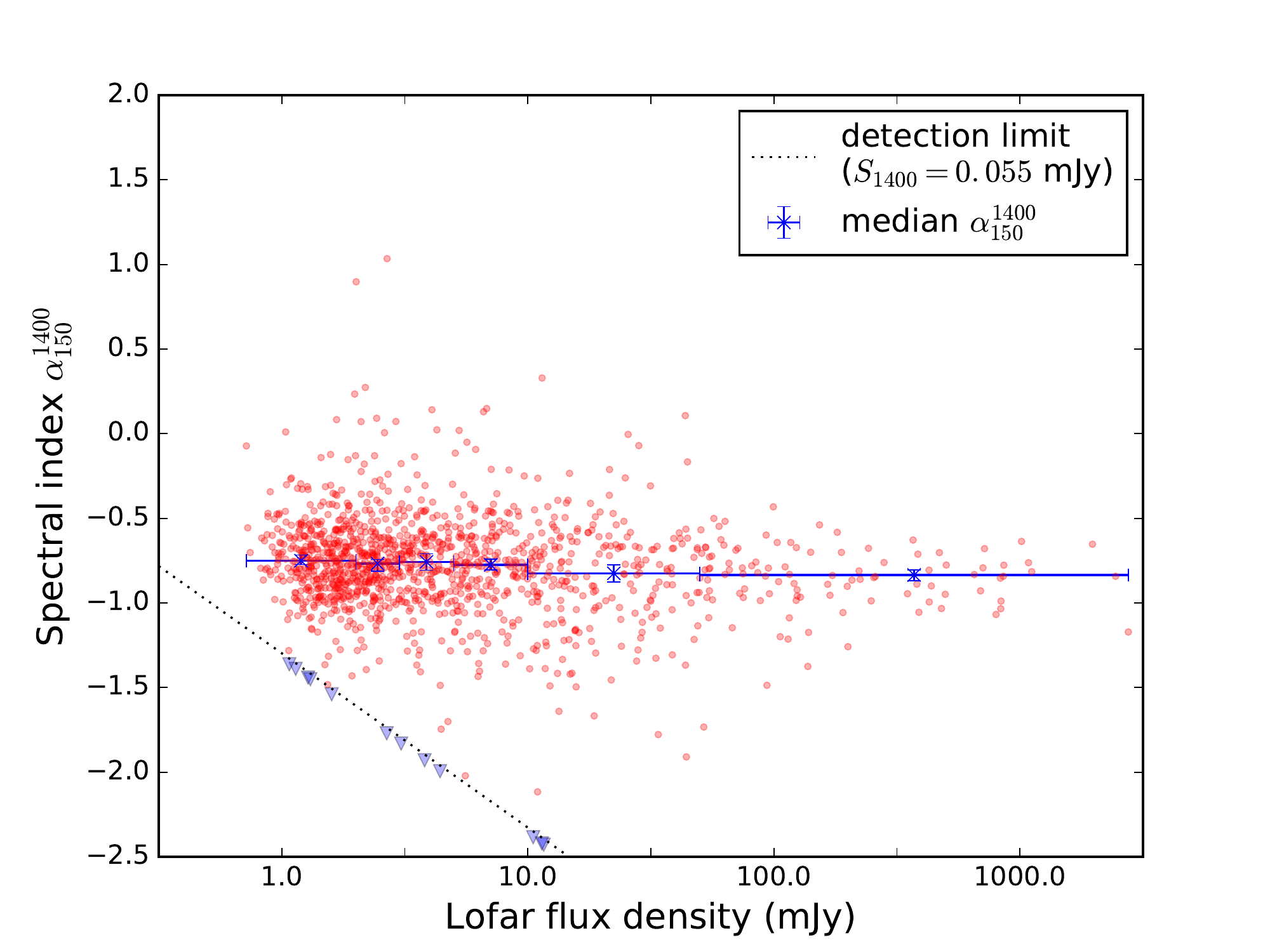}
\caption{ Median spectral index as a function of 150\,MHz flux density for sources in the Lockman--WSRT sample. The small red points shows the individual measurements for all sources in the sample while the blue squares show the median spectral index in each bin (these are shown in Table \ref{alphatable}). The horizontal error bars denote the bin widths and the vertical errorbars represent the 95\,per\,cent confidence interval of the median (from bootstrap). Blue triangles mark the upper limits of the spectral index for sources detected in the LOFAR catalogue that do not have a WSRT counterpart (see Section \ref{uss_deep}). \label{flattening}} 
\end{figure}

The Lockman--WSRT catalogue provides the ideal sample to conclusively determine whether there is any evidence for spectral flattening as a function of flux density, since virtually all LOFAR sources have a 1.4-GHz counterpart. We have divided the Lockman--WSRT sample into bins of flux density and plotted the median spectral index as shown in Fig.~\ref{flattening} and Table \ref{alphatable}. The errors shown are the 95\,per\,cent confidence interval determined by the bootstrap method. The median spectral indices become slightly flatter down to a flux density of 10\,mJy, in agreement with \citet{Intema2011}. Below these fluxes the median spectral index stays roughly constant. This change in median spectral index also explains the higher median $\alpha_{150}^{1400}$ noted in the Lockman--wide sample which is flux density limited at $S_{150}=40$\,mJy.

It is also interesting to note that the integrated spectral index of the brighter sources ($>50$\,mJy), where powerful AGN are likely to dominate, broadly agrees with that of bright radio galaxies in both large samples (e.g. \citealp{Laing83}, 178 to 750 MHz) and in detailed studies at very low frequencies (e.g. \citealp{Harwood16}, 10 to 1400 MHz) and so is likely to be robust.

\begin{table}
\begin{center}
\caption{Median spectral indices shown in Fig.~\ref{flattening}. \label{alphatable}}
\begin{tabular}{lcc}
\hline                       
{\bf Flux density bins} & {\bf Median $\alpha_{150}^{1400}$} & {\bf 95\,per\,cent CI} \\
\hline
$<$2\,mJy & -0.75 & -0.79, -0.73 \\
2--3 mJy & -0.77 & -0.79, -0.72 \\
3--5 mJy & -0.76 & -0.81, -0.71 \\
5--10 mJy & -0.77 & -0.81, -0.75\\
10--50 mJy & -0.83 & -0.87, -0.77 \\
$>$50 mJy & -0.84 & -0.86, -0.80 \\
\hline
\end{tabular}
\end{center}
\end{table}

\subsubsection{Spectral analysis of the Lockman--deep sample} \label{spec_deep}

To investigate any spectral curvature that may be occurring at lower flux densities we form a complete sample with multi-band radio information from 60\,MHz to 15\,GHz. Although the majority of these radio surveys go to fainter flux density limits, we still define a flux-limited sample at 150\,MHz in order to form a complete sample with which we can carry out a useful analysis of the low-frequency radio source population. 

We set a 150-MHz flux density limit of $S_{150}=9$\,mJy primarily driven by the $\sim4$\,mJy flux limit of the 345-MHz WSRT observations. This means that any sources not detected at 345\,MHz must have spectral indices steeper than $\alpha^{345}_{150}\sim-1.0$, analogous to the Lockman--wide sample. Using this flux threshold gives upper limits at other frequencies of $\alpha_{150}^{1400}<-1.25$ and $\alpha_{150}^{15000}<-0.6/-0.75$ (depending on where the object falls in the 10C field).

\begin{figure*}
\begin{minipage}{0.48\linewidth}
\includegraphics[width=\linewidth]{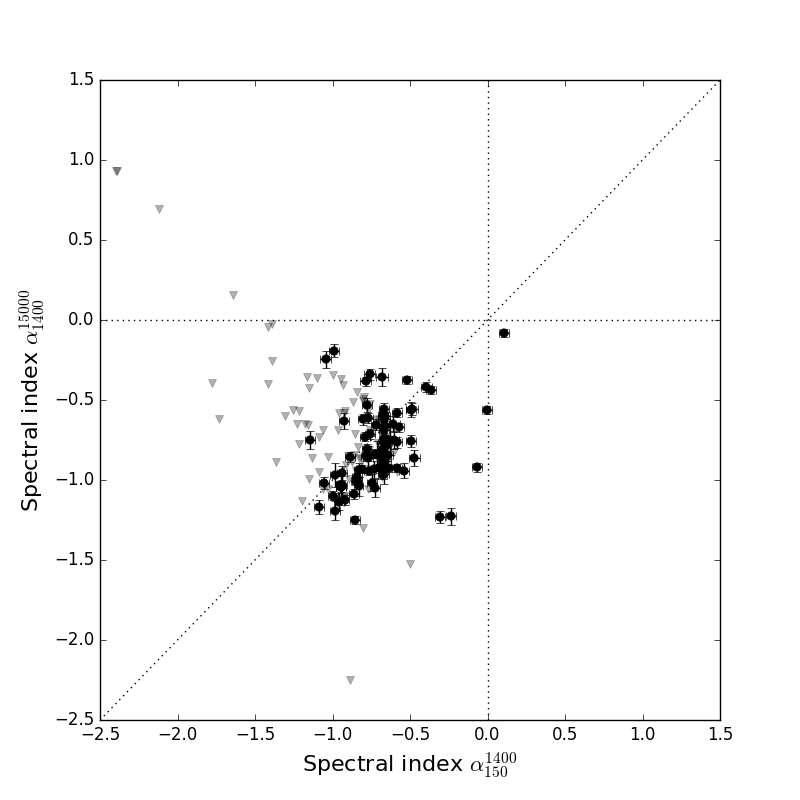}
\end{minipage}
\begin{minipage}{0.48\linewidth}
\includegraphics[width=\linewidth]{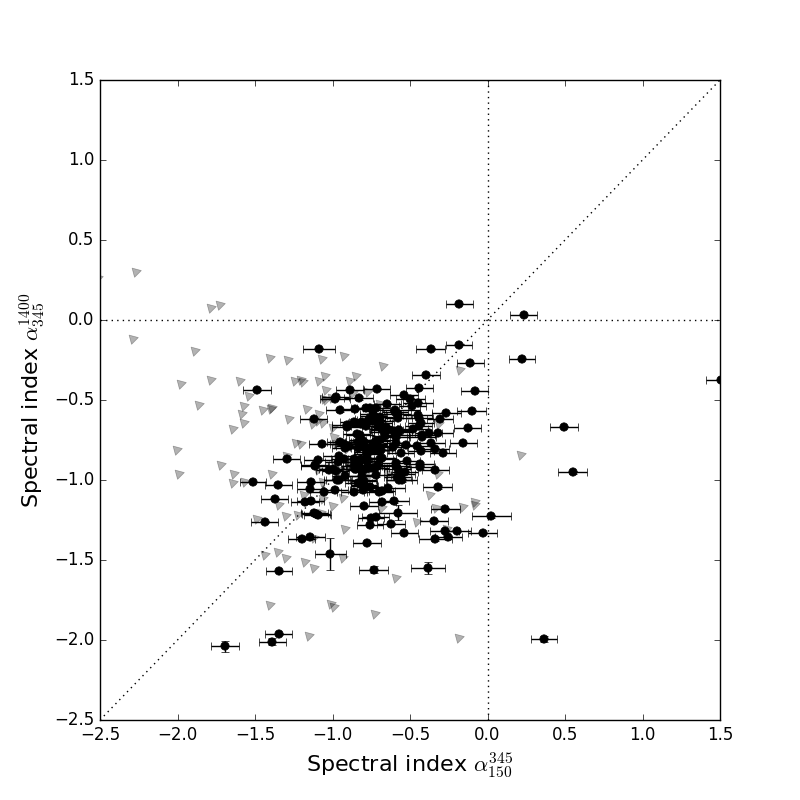}
\end{minipage}
\caption{Radio colour-colour plots of sources in the Lockman--deep sample. The left hand plot only shows sources that fall in the areas covered by the 10C survey. The grey triangles represent upper limits for sources undetected at 15\,GHz on the left and 345\,MHz in the right-hand plot.  \label{alpha-alpha}} 
\end{figure*}

The only frequency that does not have reliable limits is the GMRT 610-MHz data. Due to the higher resolution of this dataset, and underestimated flux densities for faint sources, we have only included point sources above $S_{610}=5$\,mJy in this crossmatching. As such, any source without a match at 610\,MHz could simply indicate that the source is extended at this frequency rather than below a certain flux density limit. We therefore include the 610-MHz information where available (i.e. when plotting spectra of individual sources), but cannot conclude anything significant about the source population at these frequencies.  Fig.~\ref{alpha-alpha} shows radio colour-colour plots for sources in the Lockman--deep sample. On the left we plot spectral indices between 150\,MHz and 1.4\,GHz against 1.4--15\,GHz spectral indices and on the right spectral indices between 150 and 345\,MHz against 345\,MHz--1.4\,GHz spectral indices. All sources that are undetected at either 15\,GHz or 345\,MHz are shown as limits indicated by the triangles. As with the Lockman--wide sample, the majority of sources exhibit a steep spectrum from the lowest to highest frequencies plotted, with a few sources showing evidence of curved spectra.  

We also investigate the trend for sources to become slightly flatter at lower frequencies  using the 60-MHz LOFAR LBA observations. Again, due to the high flux density cutoff at low-frequencies we define a subset of bright sources with the aim of forming a complete sample. Selecting sources with $S_{150}>80$\,mJy means that any sources not detected in the 60\,MHz observations are flatter than $\alpha_{60}^{150}=-0.7$.

\begin{figure}
\includegraphics[width=\linewidth]{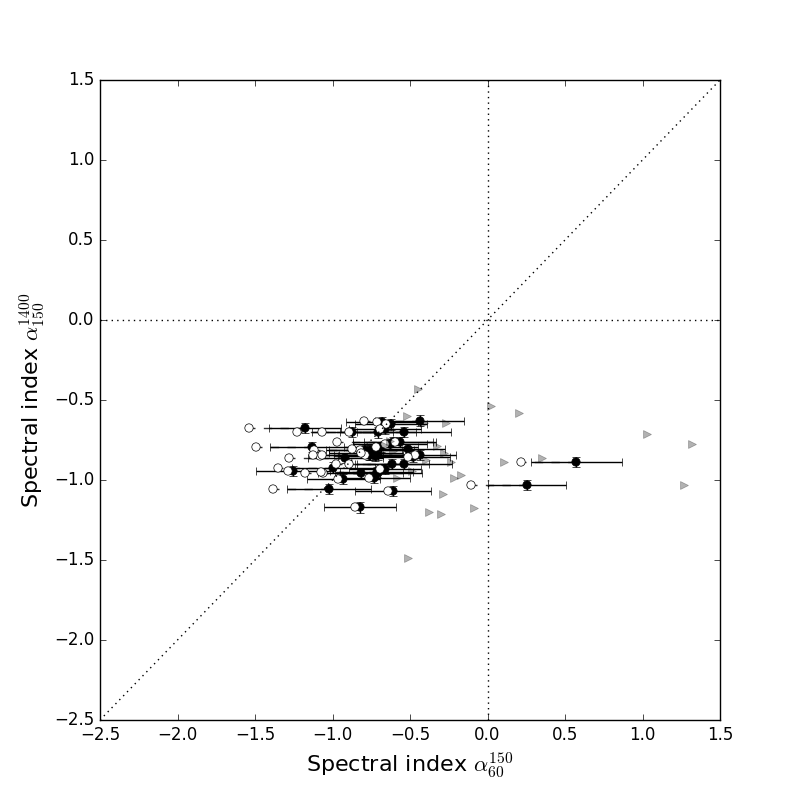}
\caption{Radio colour-colour plots from 60\,MHz to 1.4\,GHz using LOFAR LBA observations. Only sources with S$_{150}>80$\,mJy are shown here to form a complete sample. The filled sources show the original LBA flux densities and the open circles show the spectral indices if corrections are made to the LBA fluxes to bring them in line with predicted flux densities from the 6C and 8C surveys (see \autoref{spec_wide}). \label{lba_alpha}}
\end{figure}

Fig.~\ref{lba_alpha} shows the spectral index between 60--150\,MHz against spectral indices between 150--1400\,MHz. While many sources continue to exhibit a straight, power-law spectrum over this frequency range, an increasing number of sources fall to the right of this diagonal line. Since the 60-MHz flux densities have been scaled to the VLSSr flux densities we also applied the corrections noted in Section~\ref{spec_wide}. These are shown by the open symbols in Fig.~\ref{lba_alpha}. 

The broader distribution of spectral indices between 60 and 150\,MHz suggests that a number of sources begin to flatten at these frequencies, in agreement with previous low-frequency studies of the B\"{o}otes field \citep{Williams2013, vanWeeren2014}. However, it is difficult to draw a conclusive result from these plots due to the uncertainties associated with the absolute flux calibration at these frequencies. 

This is most clearly highlighted by the difference in spectral indices shown by the filled and open circles in Figs.~\ref{nvssalpha-alpha} and \ref{lba_alpha}. Even though they are tied to the same absolute flux scale \citep{scaife+heald}, the VLSSr and LOFAR LBA catalogues appear to have underestimated flux densities compared to that predicted from the 6C and 8C surveys (see Section~\ref{spec_wide} and \citealt{vlss}). This flux density offset of 20--30\,per\,cent can lead to differences in the spectral indices of up to $\Delta\alpha_{60}^{150}=0.36$. In addition, it is important to keep in mind that the LOFAR LBA fluxes have been scaled to the VLSSr flux scale assuming a spectral index of $\alpha_{60}^{74}=-0.8$ which also adds to the uncertainty of the intrinsic spectral index of the source. 

To confirm if this spectral flattening is real, a more thorough investigation is needed with careful consideration of the flux calibration issues associated with low-frequency radio observations. However, we note that spectral flattening at lower frequencies was also detected in the LOFAR Multifrequency Snapshot Sky Survey (MSSS) Verification Field \citep{MSSS}. Here it was found that the spectral indices were slightly flatter when including the full bandwidth (30--158\,MHz) compared to the spectral indices measured just using the High-Band Antennas (119--158\,MHz). 

\section{Interesting sources in the Lockman Hole field}

The radio colour-colour plots also serve as useful tools for identifying sources with more atypical spectra. For example, sources that are located in the lower-right quadrant of the diagram are peaking within the frequency range plotted, in this case anywhere between 60~MHz and 15~GHz. Alternatively, sources with very steep spectra lie towards the lower-left corner of the plot. In this section we discuss these two classes of sources in more detail and search for these objects in the Lockman Hole field. 

\subsection{Peaked-spectrum sources} \label{gps}

Radio sources with peaked spectra at around 1\,GHz are typically classified as Gigahertz-Peaked Spectrum (GPS) sources, thought to be the youngest radio galaxies \citep{Fanti1995, Odea1998}. The spectral peak marks the transition between the optically-thin and optically-thick emitting regimes and is generally associated with regions of dense nuclear material. As the radio source evolves, increasing in linear size up to scales of $\sim$10\,kpc, the spectral peak shifts to lower frequencies with the source then classified as a Compact Steep Spectrum (CSS) source \citep{Odea1998,Snellen2000,devries2009}. Studying these sources allows us to probe the intermediate stages of radio galaxy evolution, bridging the gap beween the young, compact sources detected at high-frequencies (generally GPS sources) and typical kpc-scale radio galaxies. Obtaining a complete census of CSS sources will provide insight into whether all GPS sources evolve into CSS sources and ultimately large radio galaxies, or whether this happens only under select conditions.

Sources with a spectral peak at low frequencies could also represent a population of high redshift GPS sources, where the intrinsic high-frequency spectral peak of the source has been redshifted into the LOFAR band \citep{Falcke2004, Coppejans2015}. These sources could represent the first generation of radio-loud AGN, providing insight into the formation of the earliest supermassive black holes. 

Observing GPS and CSS sources at low frequencies also allows us to investigate the cause of the spectral turnover. The generally favoured model is Synchrotron Self Absorption \citep{Fanti1990,devries2009}, but there is evidence for free-free absorption in some objects (see e.g. \citealt{Bicknell1997, Tingay2015, Callingham2015}). 

We identify candidate peaked-spectrum sources by selecting sources that lie in the lower-right hand quadrant of the plots shown in Figs.~\ref{nvssalpha-alpha}, \ref{alpha-alpha} and \ref{lba_alpha} - i.e. a flat or rising spectral index at the lower frequencies ($\alpha_{\rm low}>-0.2$) and a steep spectral index between the higher frequencies ($\alpha_{\rm high}<0.0$). The radio spectra of these sources were then inspected by eye to only include sources with a distinct spectral peak (primarily to exclude variable flat-spectrum sources that may appear peaked between certain frequencies due to the non-contemporaneous observations). In addition, detections at a minimum of four different frequencies (or at least a constraining upper limit) were required to classify a spectrum as peaked. 

This selection reveals 6 candidate GPS or CSS sources in the Lockman--wide sample and 7 in the Lockman--deep sample. These are shown in Fig.~\ref{gpsfig} and listed in Appendix A. While the low-frequency information is essential in identifying these objects as candidate GPS or CSS sources, ancillary data at other wavelengths is needed to determine the nature of these sources. In addition, repeat multi-frequency radio observations would confirm these sources as intrinsically peaked and limit the number of variable sources erroneously classified as peaked spectrum. A preliminary analysis of GPS and CSS sources in the Lockman Hole is also presented in \citet{Mahony2016}.

\begin{figure*}
\begin{minipage}{0.33\linewidth}
\centering{\epsfig{file=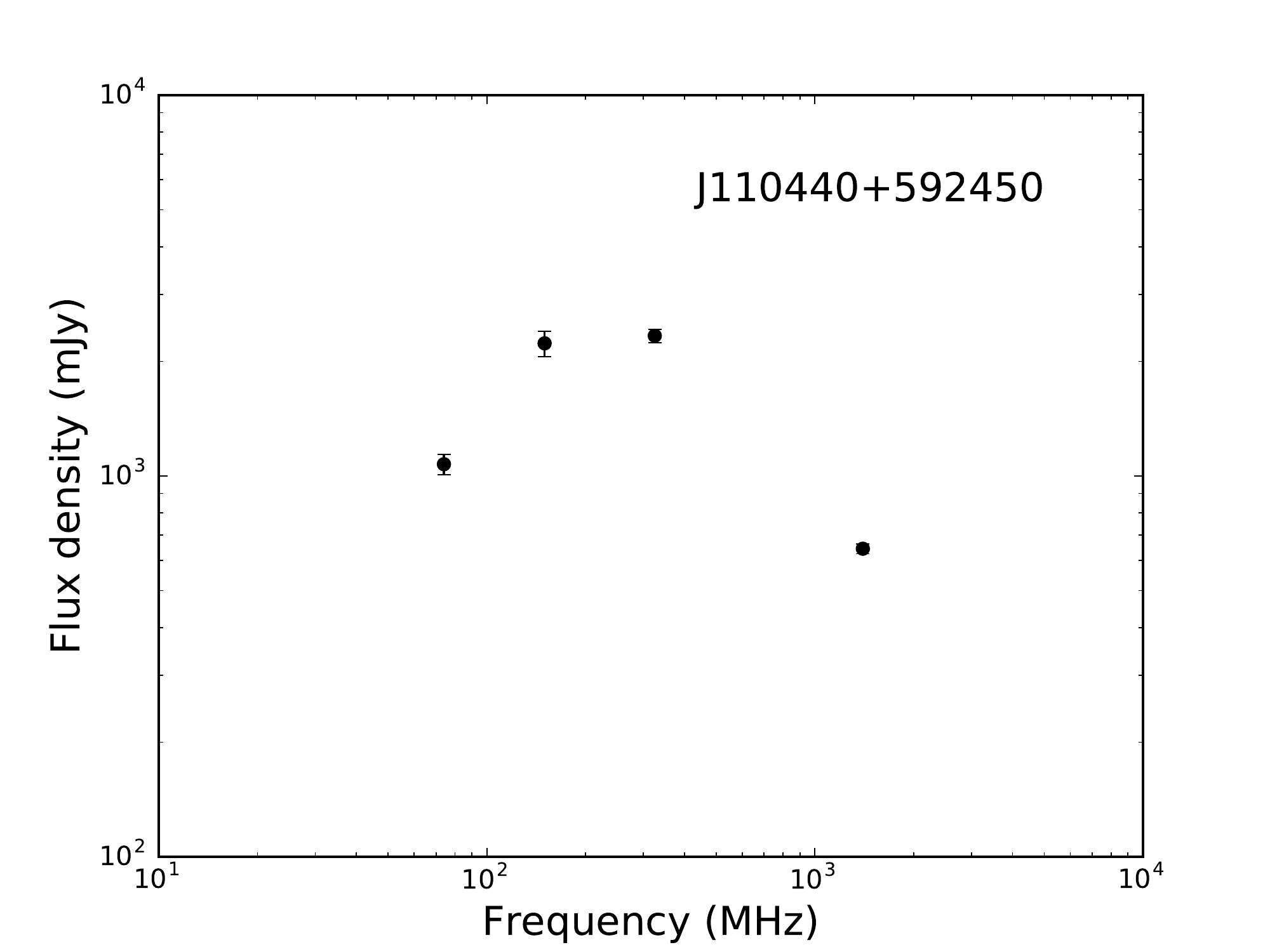, width=\linewidth}}
\end{minipage}
\hspace{-0.5cm}
\begin{minipage}{0.33\linewidth}
\centering{\epsfig{file=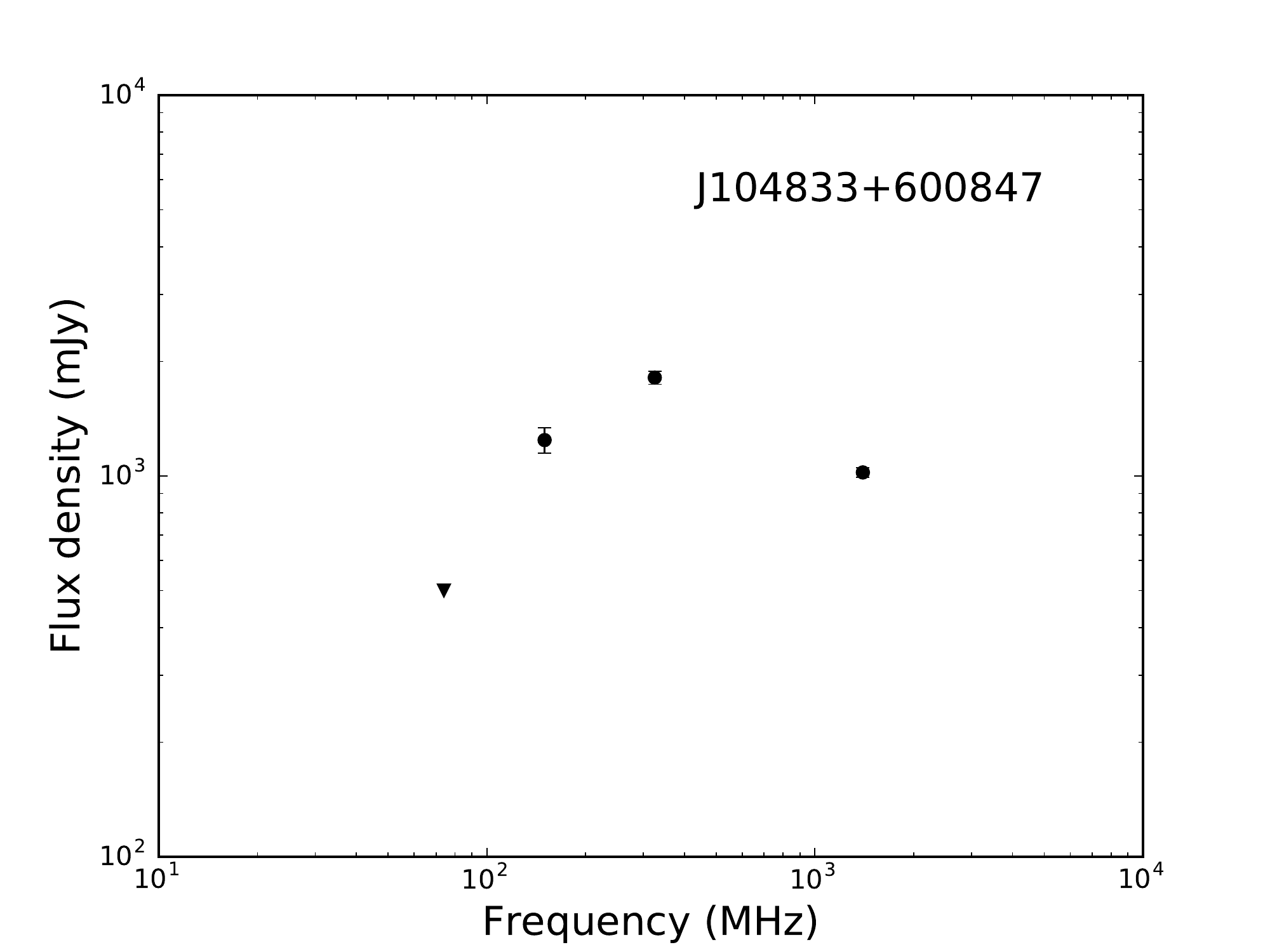, width=\linewidth}}
\end{minipage}
\hspace{-0.5cm}
\begin{minipage}{0.33\linewidth}
\centering{\epsfig{file=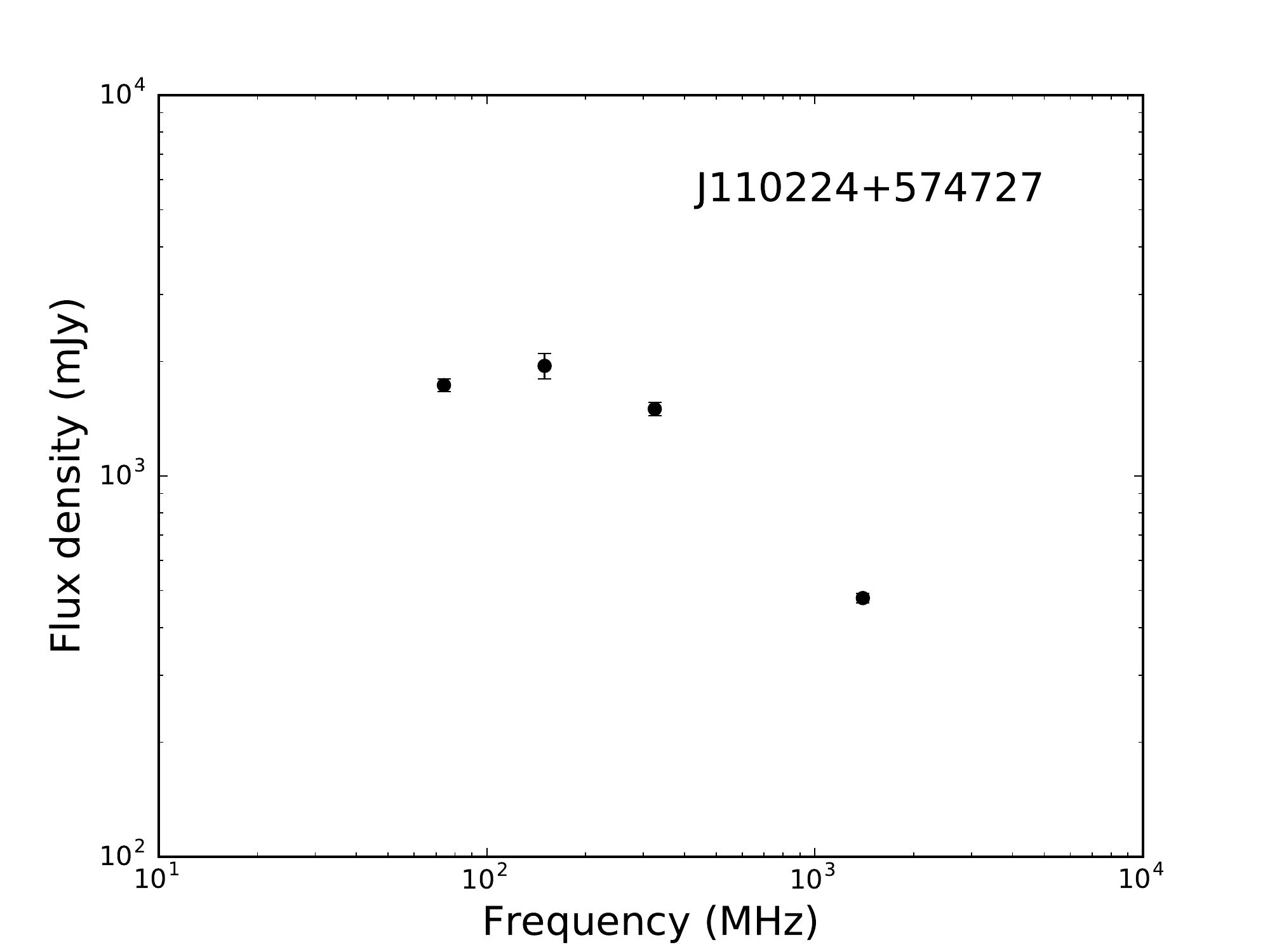, width=\linewidth}}
\end{minipage}
\hspace{-0.5cm}
\begin{minipage}{0.33\linewidth}
\centering{\epsfig{file=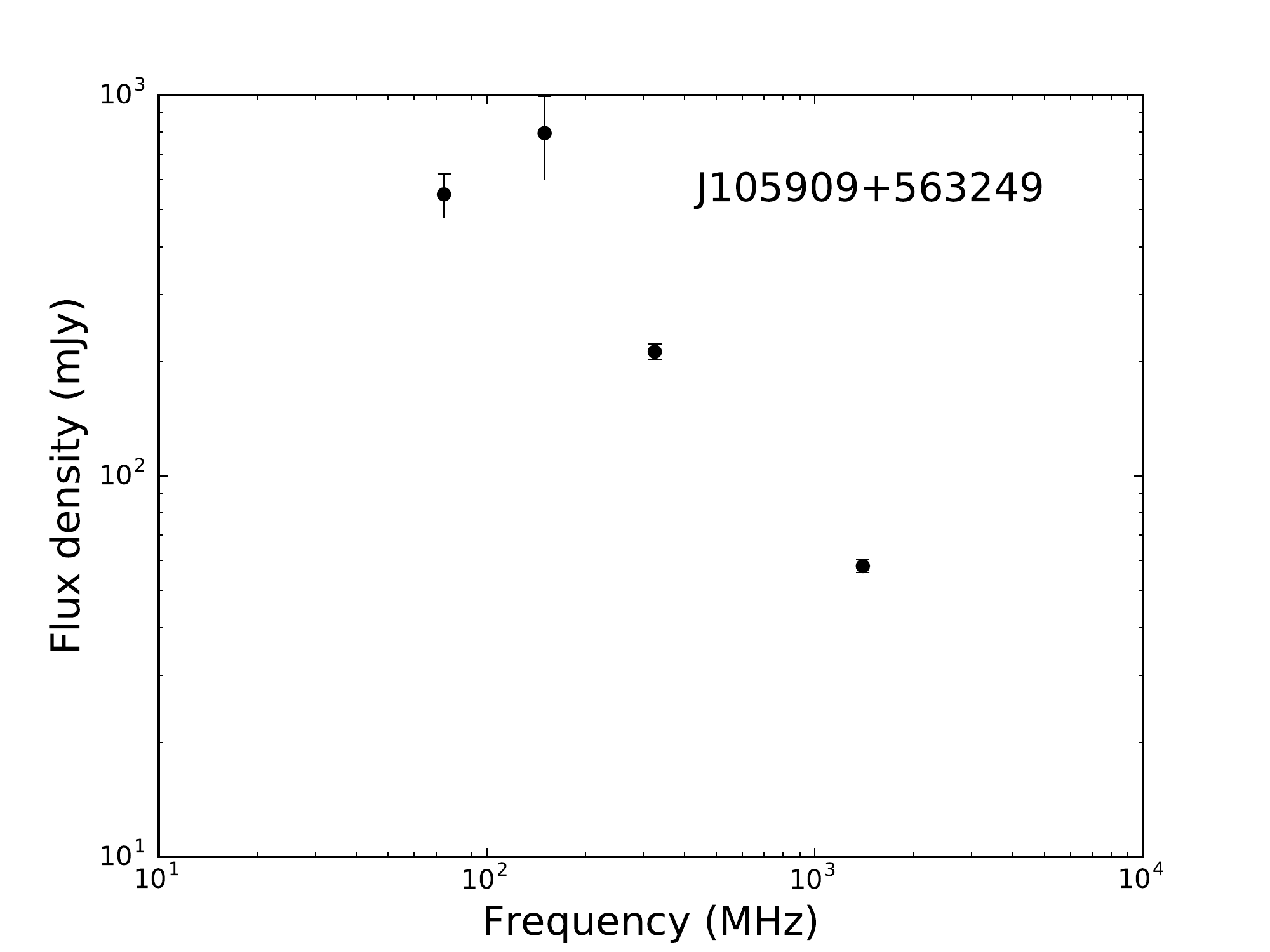, width=\linewidth}}
\end{minipage}
\hspace{-0.5cm}
\begin{minipage}{0.33\linewidth}
\centering{\epsfig{file=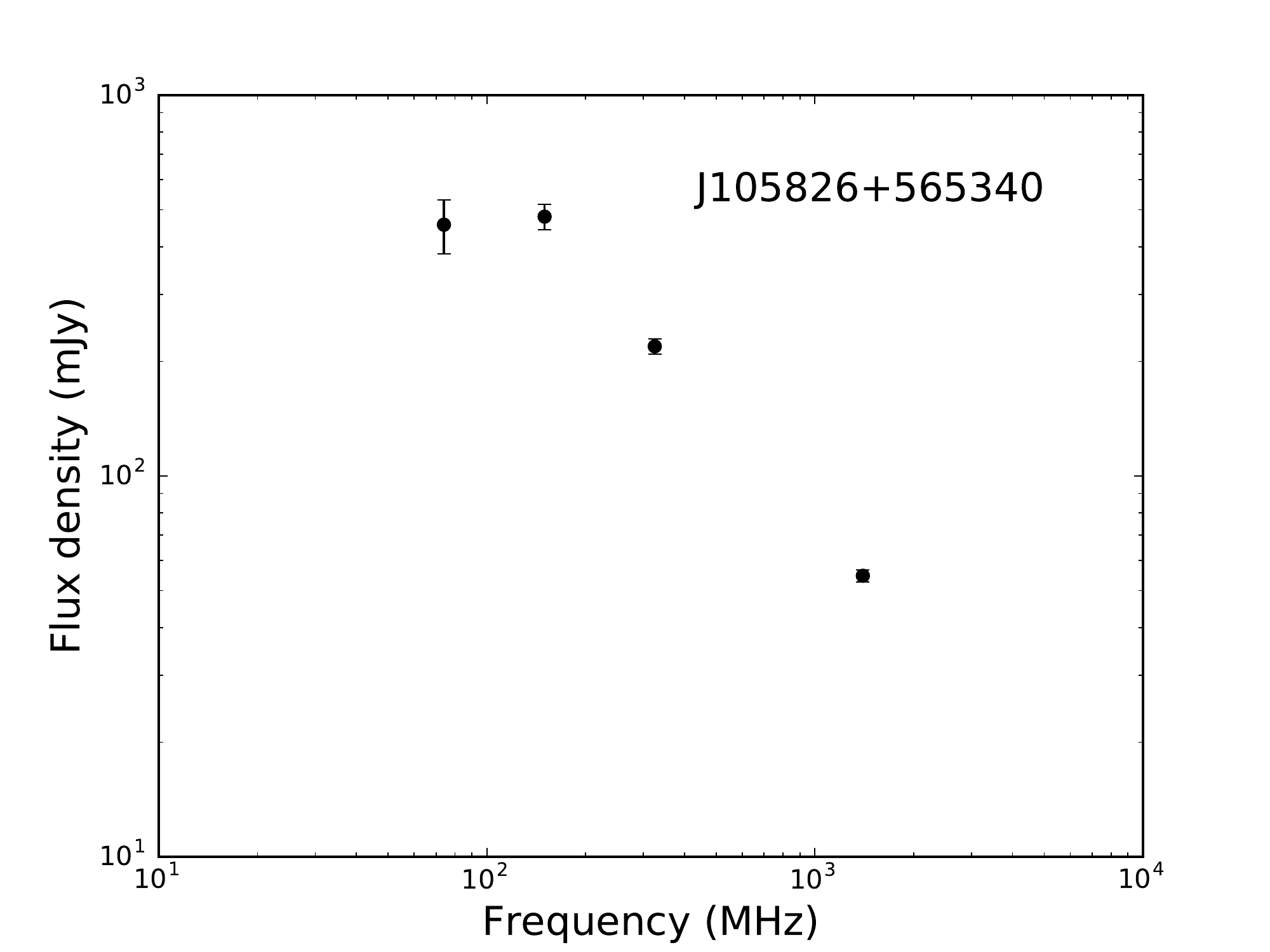, width=\linewidth}}
\end{minipage}
\hspace{-0.5cm}
\begin{minipage}{0.33\linewidth}
\centering{\epsfig{file=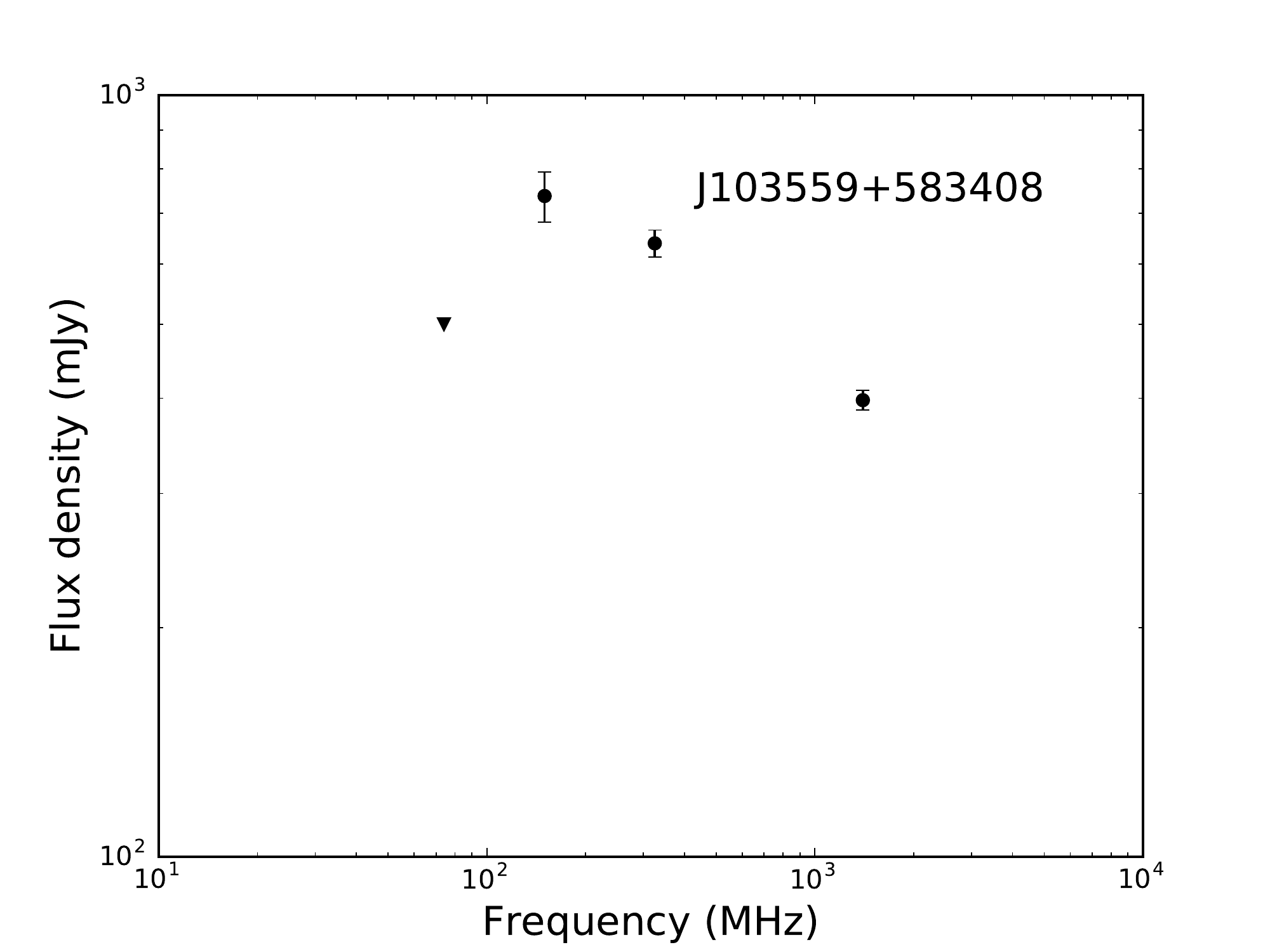, width=\linewidth}}
\end{minipage}
\hspace{-0.5cm}
\begin{minipage}{0.33\linewidth}
\centering{\epsfig{file=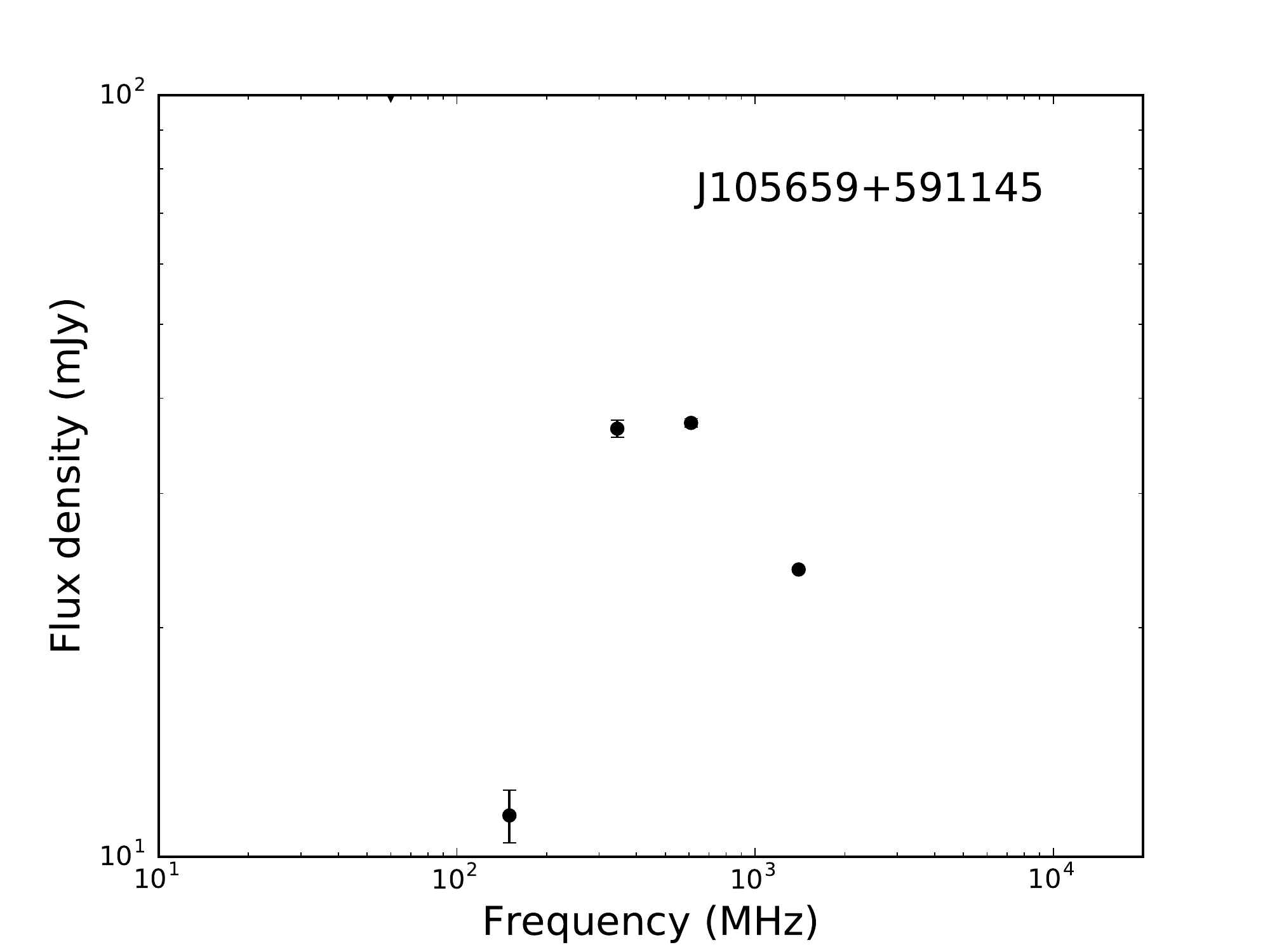, width=\linewidth}}
\end{minipage}
\hspace{-0.5cm}
\begin{minipage}{0.33\linewidth}
\centering{\epsfig{file=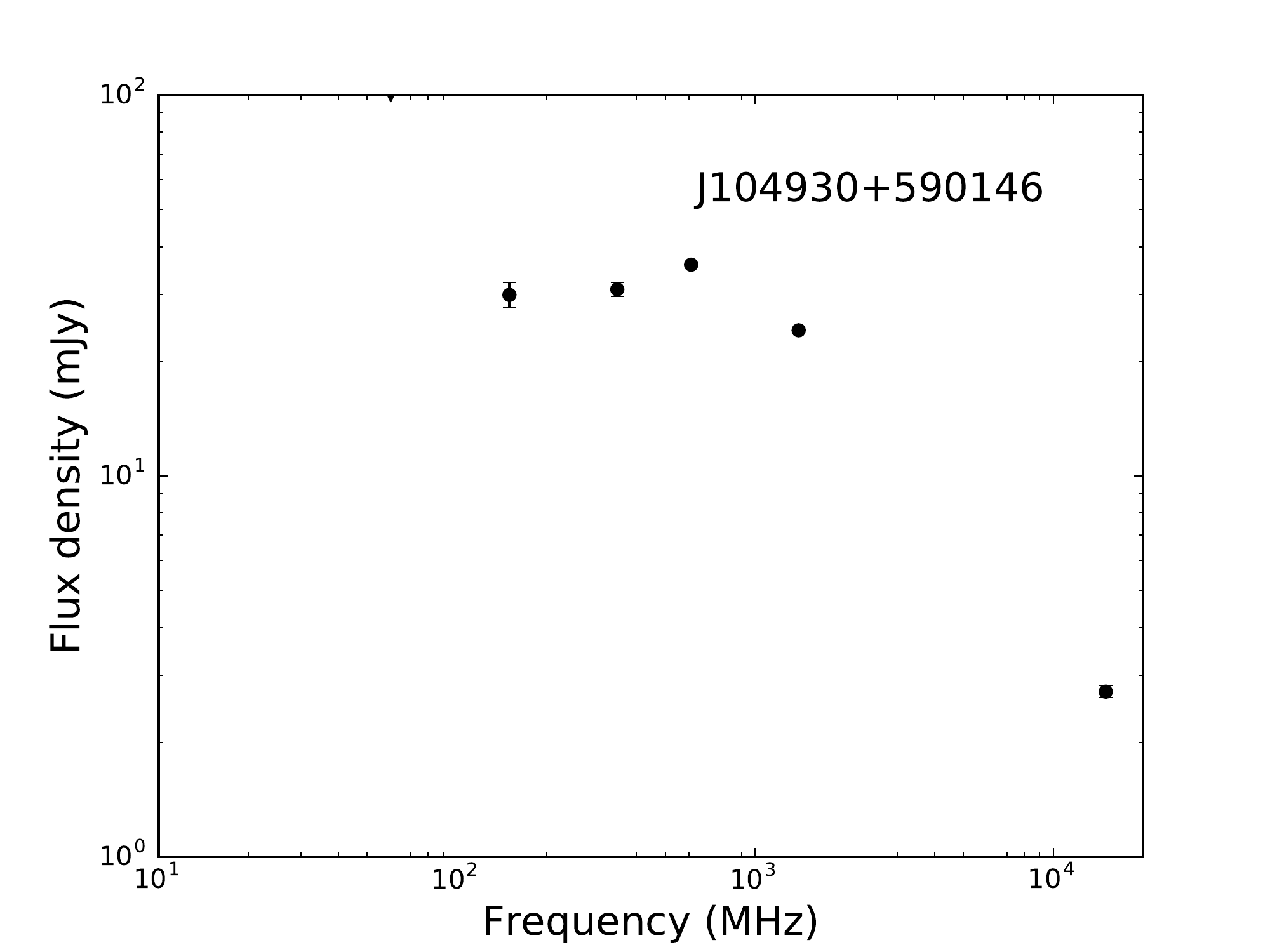, width=\linewidth}}
\end{minipage}
\hspace{-0.5cm}
\begin{minipage}{0.33\linewidth}
\centering{\epsfig{file=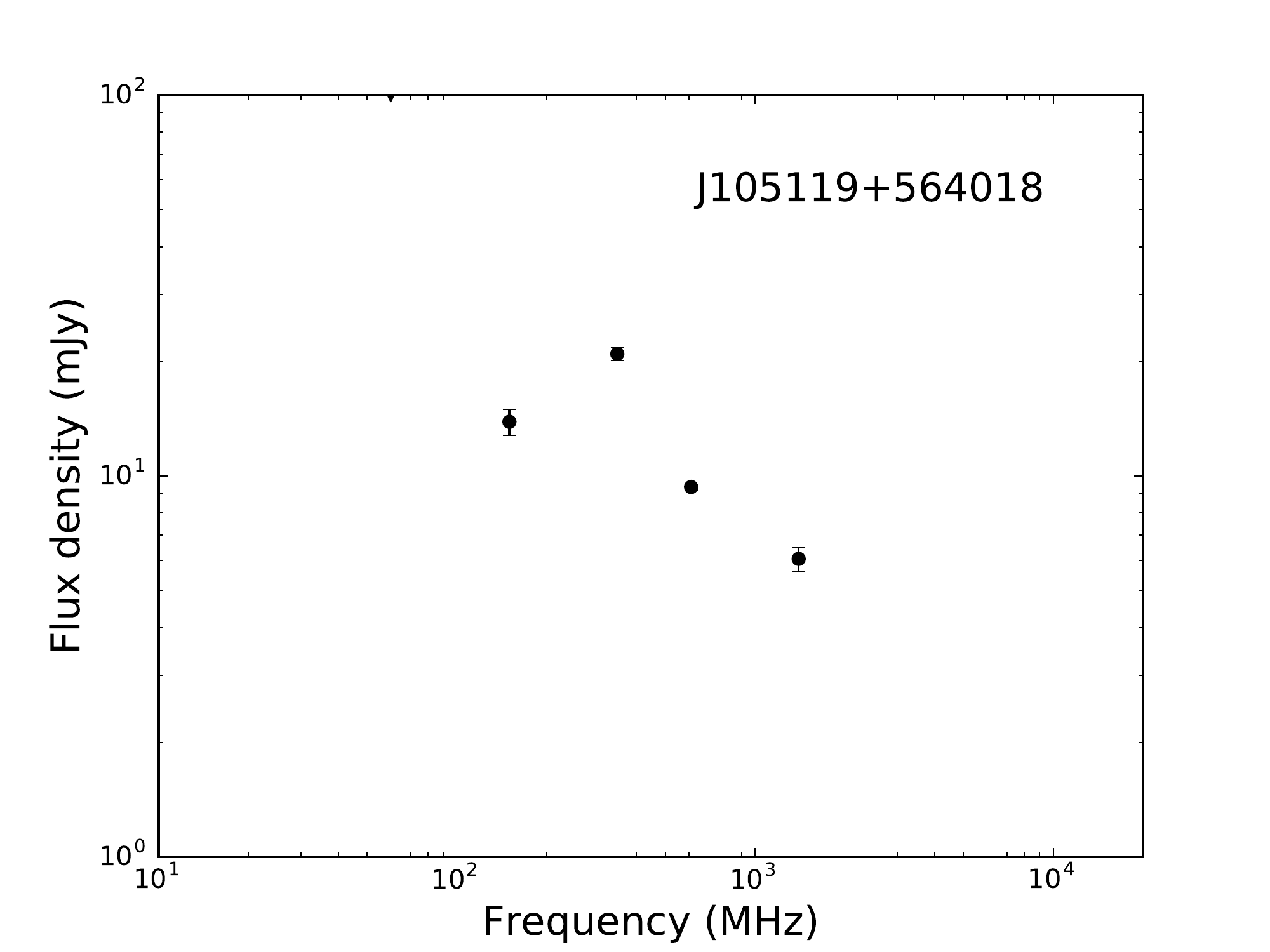, width=\linewidth}}
\end{minipage}
\hspace{-0.5cm}
\begin{minipage}{0.33\linewidth}
\centering{\epsfig{file=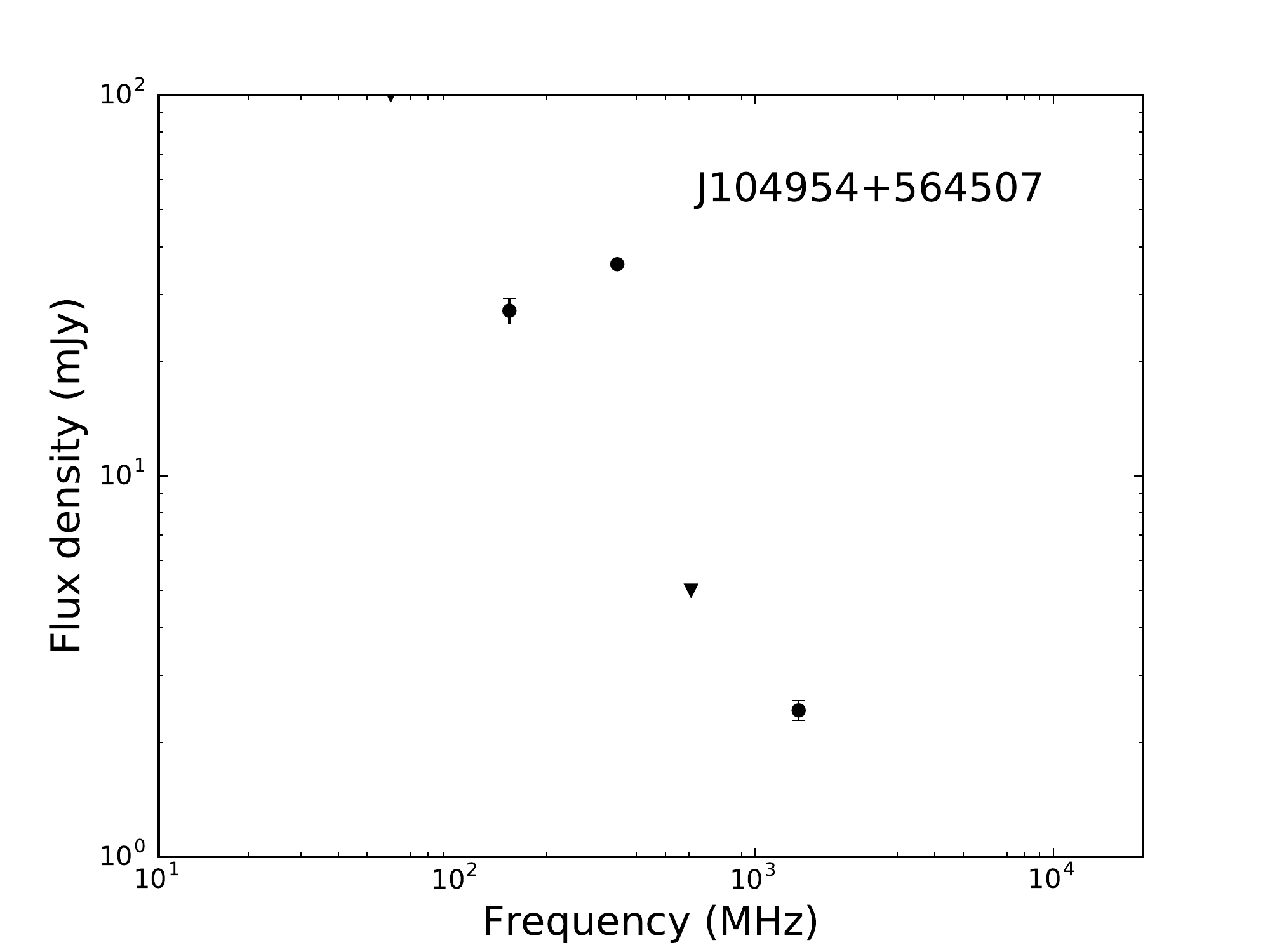, width=\linewidth}}
\end{minipage}
\hspace{-0.5cm}
\begin{minipage}{0.33\linewidth}
\centering{\epsfig{file=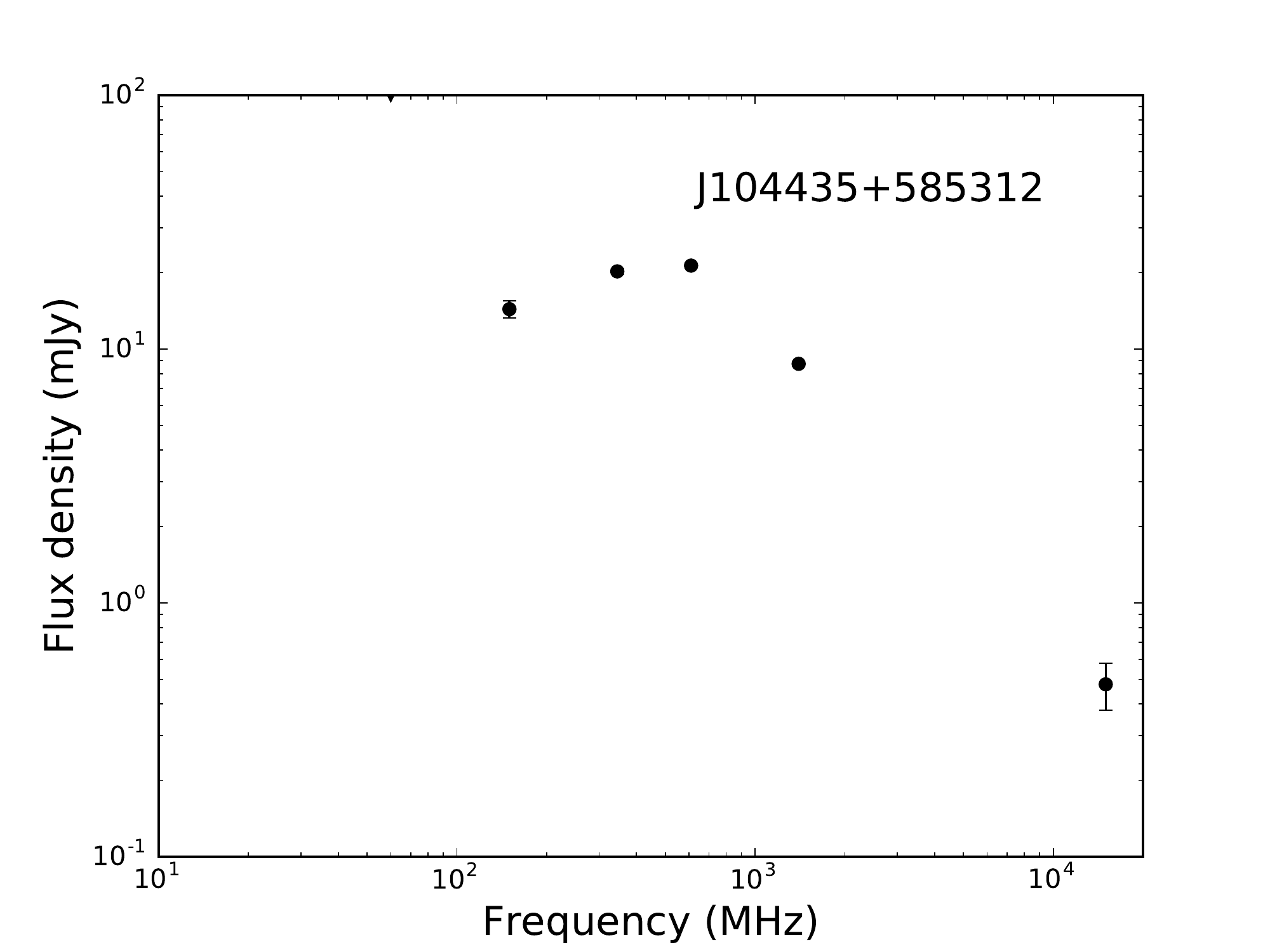, width=\linewidth}}
\end{minipage}
\hspace{-0.5cm}
\begin{minipage}{0.33\linewidth}
\centering{\epsfig{file=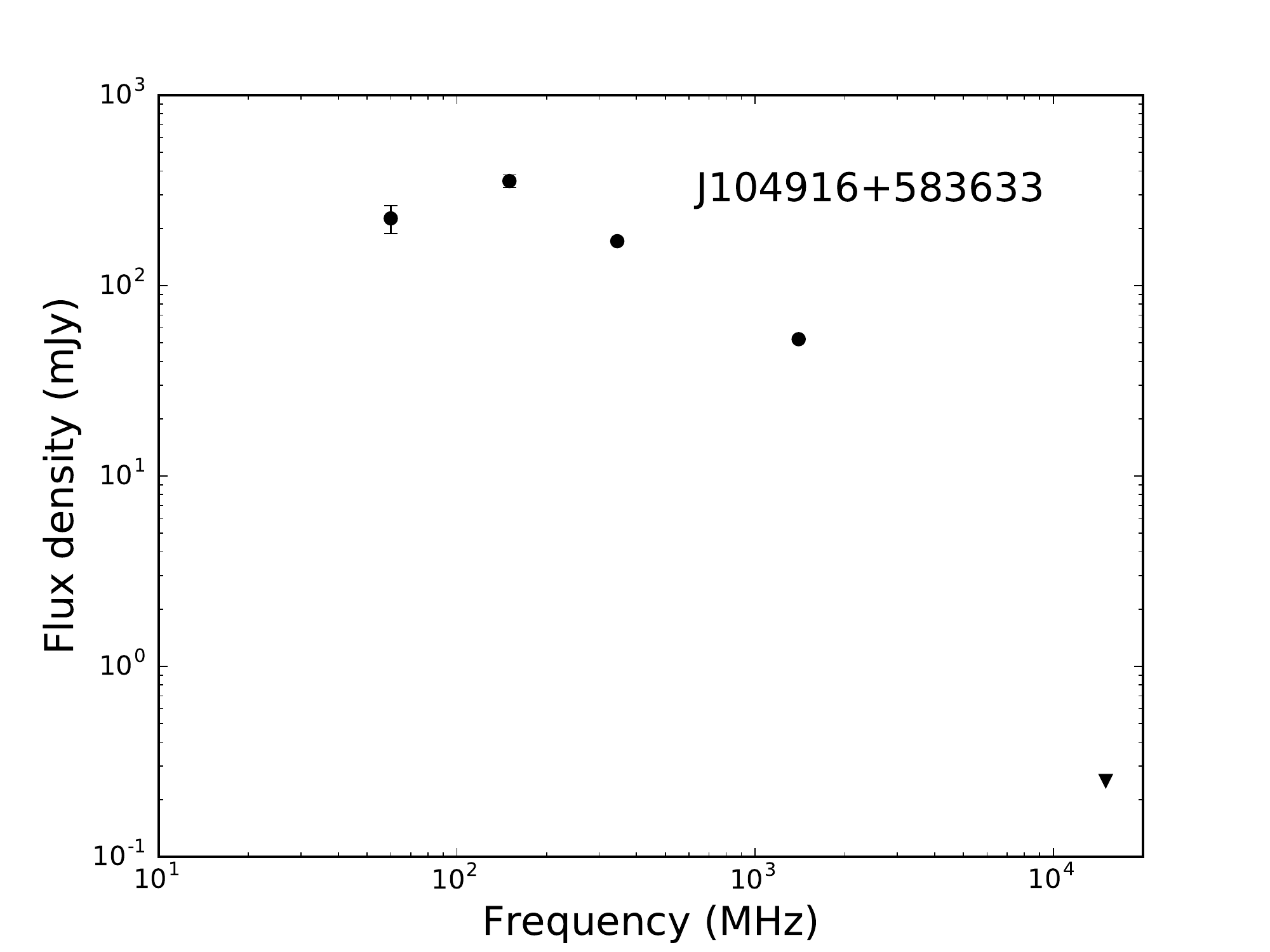, width=\linewidth}}
\end{minipage}
\hspace{-0.5cm}
\begin{minipage}{0.33\linewidth}
\centering{\epsfig{file=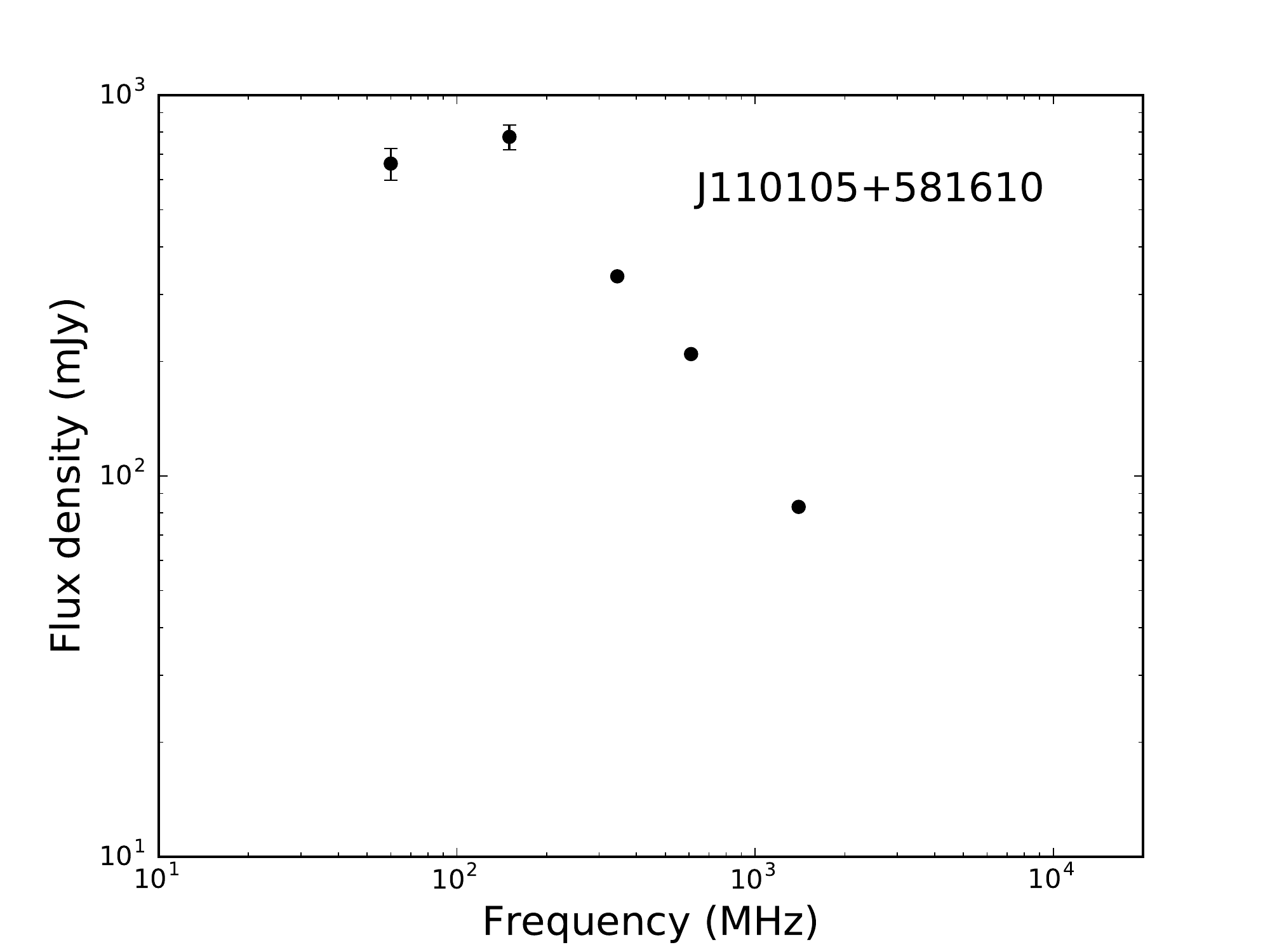, width=\linewidth}}
\end{minipage}
\caption{Peaked spectrum sources detected in the Lockman Hole. The first six sources are detected in the Lockman--wide sample and the remaining in the Lockman--deep sample. Errorbars are shown, but in many case are smaller than the datapoints. \label{gpsfig}}
 
\end{figure*} 

\subsubsection{Multi-wavelength properties of peaked-spectrum sources}

Searching the literature for existing observations of these targets revealed a number of optical/IR counterparts. Two out of the six candidates found in the Lockman--wide sample have known counterparts, both detected in the SLOAN Digital Sky Survey (SDSS; \citealt{sdss}): J104833+600845 is identified as a QSO at a redshift of $z=1.72$ and J103559+583408 is a candidate QSO with a photometric redshift of $z_{ph}=1.65$ \citep{Richards2009}\footnote{For sources in the Lockman--wide sample the optical counterparts were found using the NASA/IPAC Extragalactic Database where the radio-optical association was already made.}.

For sources in the Lockman--deep sample, a large fraction of the area has been observed by the Spitzer Extragalactic Representative Volume Survey (SERVS; \citealt{Mauduit2012a}) or the Spitzer Wide-Area Infrared Extragalactic Survey (SWIRE; \citealt{Lonsdale2003}) providing us with deep IR information in this area. This information has been collated in the SERVS data fusion catalogue \citep{Vaccari} which also includes IR information from the Two Micron All Sky Survey (2MASS; \citealt{2mass}) and the UKIRT Deep Sky Survey Deep eXtragalactic Survey (UKIDSS DXS; \citealt{ukidss}) as well as optical information from SDSS. This data fusion catalogue was crossmatched with the WSRT 1.4\,GHz mosaic to search for multiwavelength counterparts (for full details see Prandoni et al., 2016a, in preparation).

Of the candidate peaked-spectrum sources found in the Lockman--deep sample only one is detected in SDSS: J105119+564018 is identified as a $z=2.38$ QSO with $g=20.2$\,mag. Four sources are detected in the SERVS data fusion catalogue, three have $K$ (KRON) magnitudes ranging from 16.5 to 19.9 mag and the fourth has an upper limit of $K\sim21.0$. No additional redshift information is available, but were these galaxies to fall on the $K$-$z$ relation (see e.g. \citealt{Rocca-Volmerange2004}) we can infer that of these, two are potentially high-redshift GPS sources at $z>3$ (following the $K$-$z$ relation given in \citealt{Brookes2008}).

Sources that are not detected are outside the SERVS and SWIRE footprints, hence we cannot place useful limits on the IR properties of these sources. Further follow-up optical/IR observations are required to identify the host galaxies and determine the redshift. Another way to distinguish the high-$z$ GPS sources from lower redshift CSS sources would be through high-resolution imaging with Very Long Baseline Interferometry (VLBI). This would provide approximate ages of the radio source by placing limits on the linear size of these objects (see e.g. \citealt{Coppejans2016}). 

\subsection{Ultra-Steep Spectrum sources} \label{uss}

Another interesting class of sources detected in the Lockman Hole are the Ultra-Steep Spectrum (USS) sources. As the name suggests, these sources exhibit an extremely steep spectrum, generally attributed to radiation losses of relativistic electrons in the radio lobes, meaning they are most luminous at lower radio frequencies. LOFAR is therefore an ideal instrument to detect these sources. Finding sources with ultra-steep spectra is thought to be a good way of identifying a variety of different source populations including high-$z$ ($z>2$) radio galaxies (HzRGs; \citealt{BlumenthalMiley1979,Miley2008}), remnant emission from ageing radio galaxies \citep{Slee2001, Murgia2011, Brienza2016}, relic cluster emission (see e.g. \citealt{Ferrari2008, Feretti2012}) or radio halos in clusters \citep{Brunetti2008}.  

A number of different selection criteria are used to identify USS sources, ranging from $\alpha_{151}^{4850}<-0.981$ \citep{Blundell1998} to $\alpha_{843}^{1400}<-1.3$ \citep{DeBreuck2004}. To search for USS sources in the Lockman Hole we selected sources that had spectral indices between 150\,MHz and 1.4\,GHz steeper than $\alpha_{150}^{1400}=-1.2$, primarily driven by the upper limit of the spectral index for sources not detected in NVSS. Using this criterion we find 19 sources (4.9\,per\,cent) with ultra-steep spectra in the Lockman--wide sample, including eight upper limits for sources that are not detected in NVSS. Like the peaked-spectrum sources, follow up observations are needed to confirm these as either high-$z$ sources, remnant AGN, or cluster emission. There is also the possibility that some of these steep-spectrum sources are galactic sources (e.g. pulsars). However, given the high galactic latitude of the Lockman Hole (b$>50^{\circ}$) it is unlikely that these account for a significant fraction of the ultra-steep population found here.

\subsubsection{USS sources in the Lockman--WSRT sample} \label{uss_deep}

Since the WSRT 1.4\,GHz catalogue is more than an order of magnitude deeper than the NVSS catalogue, we can carry out a much more comprehensive search for USS sources using the Lockman--WSRT catalogue. Of the 1302 sources in the Lockman--WSRT catalogue, we find 85 candidate ultra-steep spectrum sources with $\alpha_{150}^{1400}<-1.2$, including 13 upper limits that were undetected in the WSRT 1.4\,GHz image\footnote{Four USS sources fall in both the Lockman--wide and Lockman--deep samples.}. These undetected sources can be separated into two different categories; faint LOFAR sources with no detectable emission at 1.4\,GHz - these are flagged as `L' in the Lockman--WSRT catalogue, or larger, more complex radio sources where excess emission has been detected in the LOFAR image, possibly indicative of an older phase of AGN activity (e.g. remnant lobes). These sources could be candidate restarted radio galaxies, where the timescales between the two sets of radio lobes can be used to probe the dutycycle of radio AGN activity, however, further follow-up observations are needed to confirm this. These have been flagged as `LC' in the catalogue. Some examples are shown in Appendix B.

Including the upper limits derived from the undetected sources at 1.4~GHz, this corresponds to 6.6\,per\,cent of the Lockman--deep sample being classified as ultra-steep. Previous searches for USS sources in the Lockman Hole field using deep 610~MHz and 1.4~MHz have revealed a similar percentage of 6.3\,per\,cent based on $\alpha_{610}^{1400}<-1.3$ \citep{Ibar2009,Afonso2011}. 

\subsubsection{Multiwavelength properties of USS sources}

To confirm if these LOFAR-selected USS candidates are high-$z$ radio galaxies or potential relic sources further follow-up observations at other wavelengths are required. A more complete analysis on the USS sources and the remnant AGN population in the Lockman Hole is discussed by \citet{Brienza2016} and will be further expanded in an upcoming paper  (Brienza et al., in preparation).

To investigate how many of these sources are potential high-$z$ radio galaxies we crossmatched the USS sources with the SERVS data fusion catalogue \citep{Vaccari} using a search radius of 1.75 \,arcsec. Of the 85 USS sources detected in the Lockman--deep sample, 82 of them overlap with the SERVS footprint. Of these, 26 (32\,per\,cent) are detected in SERVS and 19 sources are detected in the $K$-band as part of the UKIDSS DXS survey with $K$ (KRON) magnitudes ranging from 14.48 mag to 20.23 mag. While none of these have known redshift information, we can get a rough estimation of the redshift by placing them on the $K$-$z$ relation. Based on these $K$-band magnitudes, and following the $K$-$z$ relation given in \citet{Brookes2008}, we estimate that 8 (42\,per\,cent) objects have $z>3$ and 4 (21\,per\,cent) are at $z>4$. However, further spectroscopic follow-up would be required to confirm if these candidates are genuinely high-$z$ objects.

\begin{figure}
\centering{\includegraphics[width=0.9\linewidth]{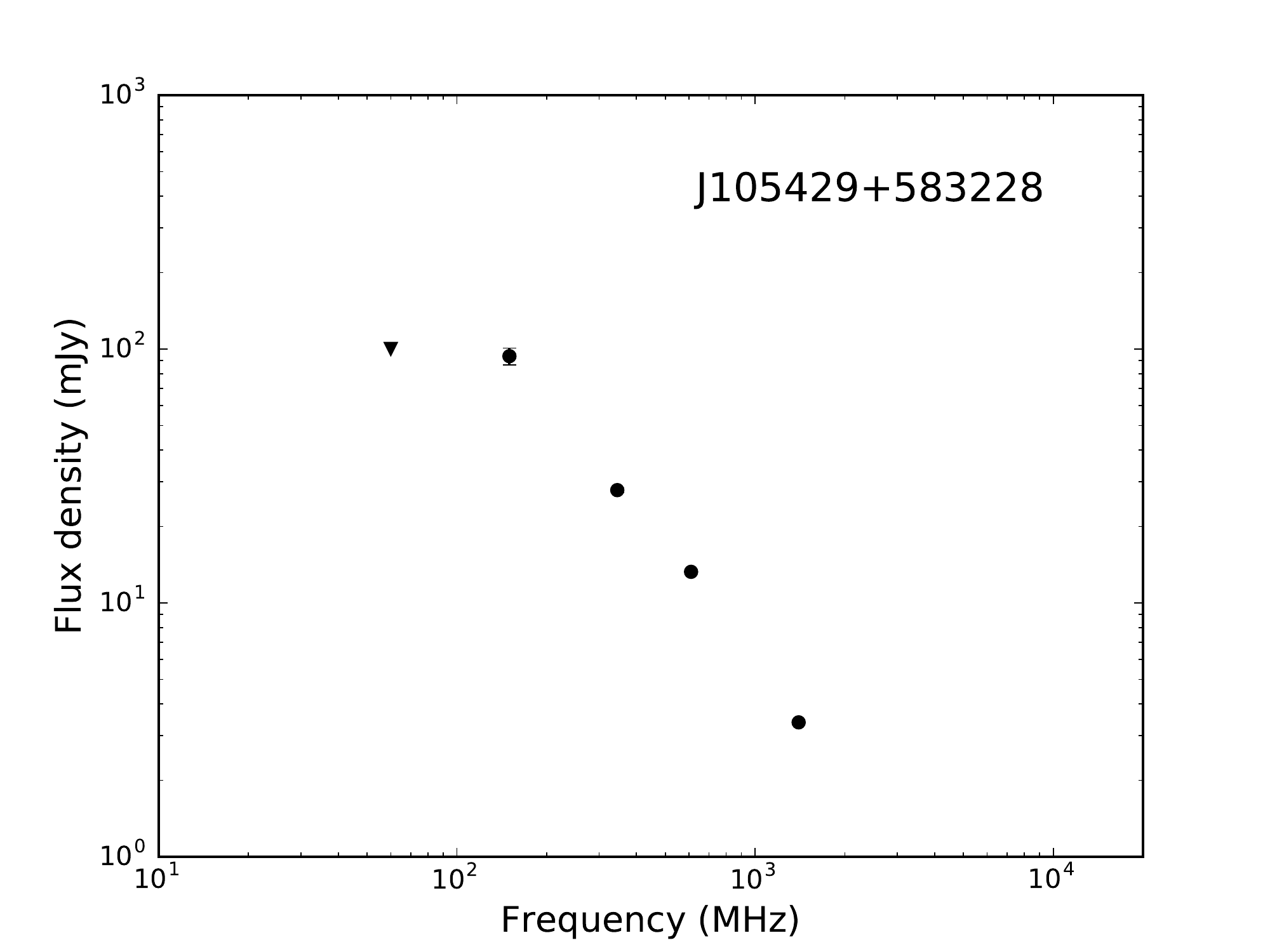}}
\caption{An example of one unusual USS source detected in the Lockman Hole. This source has an ultra steep spectrum of $\alpha_{150}^{1400}=-1.6$ between 150 and 1.4\,GHz, but is undetected at 60\,MHz and therefore has a much flatter spectral index from 60--150\,MHz of $\alpha_{60}^{150}>-0.07$ (assuming a 5$\sigma$ upper limit at 60\,MHz). \label{ussfig}} 
\end{figure}

The spectrum of one intriguing USS source detected in the Lockman Hole is shown in Fig.~\ref{ussfig}. This object has a flux density of 100\,mJy in the LOFAR HBA image and exhibits an ultra-steep spectrum of $\alpha_{150}^{1400}=-1.6$ between 150\,MHz and 1.4\,GHz. However, the spectrum appears to turn over quite sharply below 150\,MHz, with the source undetected in the 60\,MHz LBA image. Using a conservative 5$\sigma$ upper limit of 100\,mJy at 60\,MHz we can place a lower limit of the spectral index between 60--150\,MHz of $\alpha_{60}^{150}>-0.07$. The ultra-steep spectral index, combined with the IR properties of this source ($K=19.9$ from UKIDSS DXS), indicate that this could be a high redshift radio galaxy ($z>4$). Observations at higher spatial resolution could also indicate if this source is a high-$z$ GPS or CSS source. 

We also note the similarities between this source and one of the peaked spectrum sources presented in Sec.~\ref{gps}. J104954+564507 also shows a very sharp spectral peak with a rising spectral index of $\alpha_{150}^{345}=0.36$ before turning over to exhibit an ultra-steep spectrum of $\alpha_{345}^{1400}=-1.99$. This also has a $K$ magnitude of 19.9 potentially placing it at a similar redshift. Another GPS source observed with the Murchinson Widefield Array (MWA; \citealt{mwa}) shows a similarly sharp turnover \citep{Callingham2015}, indicating that these could be part of a larger population of extreme peaked spectrum sources.

\subsection{Implications for larger LOFAR surveys} \label{tier1survey}

The LOFAR data presented here represents just a single nights observation of one field, a tiny fraction of the large-area sky survey currently being carried out with LOFAR (see Shimwell et al., 2016, submitted). As such, we can use the results found here to predict what we might expect to detect if similar observations were carried out across the whole sky above $\delta=0^{\circ}$. In the $\sim$7 square degrees covered by these observations (the Lockman--deep survey area) we have identified 85 ultra-steep spectrum sources and 7 sources with peaked spectrum at MHz frequencies. Extrapolating from these numbers we can therefore expect up to $\sim$300\,000 and $\sim$20\,000 USS, and peaked-spectrum sources respectively, assuming a survey carried out at the same depth and resolution presented here\footnote{The LOFAR all-sky survey aims to reach an rms of 0.1\,mJy with resolution of $\sim 5$\,arcsec, slightly deeper and at higher resolution than presented here.}.

However, these numbers assume that there is deep multi-frequency radio data available across large areas of sky, which is far from being the case. Limiting these numbers to just those sources that have been crossmatched with all-sky surveys such as NVSS, WENSS and VLSSr these numbers decrease to $\sim$13000 potential USS candidates and $\sim$2000 sources with peaked spectra (above a flux limit of $S_{150}=40$\,mJy). Nonetheless, this still represents an order of magnitude increase in the numbers of these source populations known, allowing for a better understanding of the evolution of these sources and how they relate to the general radio galaxy population. 

This highlights the need for complementary deep radio surveys at higher frequencies such as the continuum surveys planned at 1.4\,GHz with Apertif (APERture Tile in Focus - a focal plane array system installed on the WSRT; \citealt{apertif}) and the 3-GHz VLA Sky Survey (VLASS). This will enable us to push the spectral analysis presented here down to much deeper flux density limits and investigate how the source population changes in the sub-mJy regime. In addition to deeper multi-frequency radio data, ancillary data at other wavelengths would enable us to obtain a more comprehensive picture of the radio source population, providing more insight into the evolutionary stages of radio AGN.

\subsection{The Giant Radio Galaxy: HB13}

Rather than only selecting sources based on spectral information, we can also search for interesting sources based on morphology. One such example is HB13, a known giant radio galaxy, at the position 10h32m58.9s +56d44m53s, that was detected in early radio surveys but since then relatively neglected (but see tables with comparison to other giant radio galaxies discussed by \citet{Palma2000} and \citet{Saripalli1986}). This object was studied by \citet{Masson1979}  at 151~MHz (using data from the 6C survey) but originally selected, and named HB13, in the pencil beam survey of \citet{HanburyBrown1953}. The radio galaxy is identified with a galaxy at redshift $z=0.0450$. \citet{Masson1979} presents observations at 151 MHz and 1.4 GHz, although with very poor spatial resolution, reporting a total size of the radio source at low frequencies of 35\,arcmin.

\begin{figure}
\includegraphics[width=\linewidth]{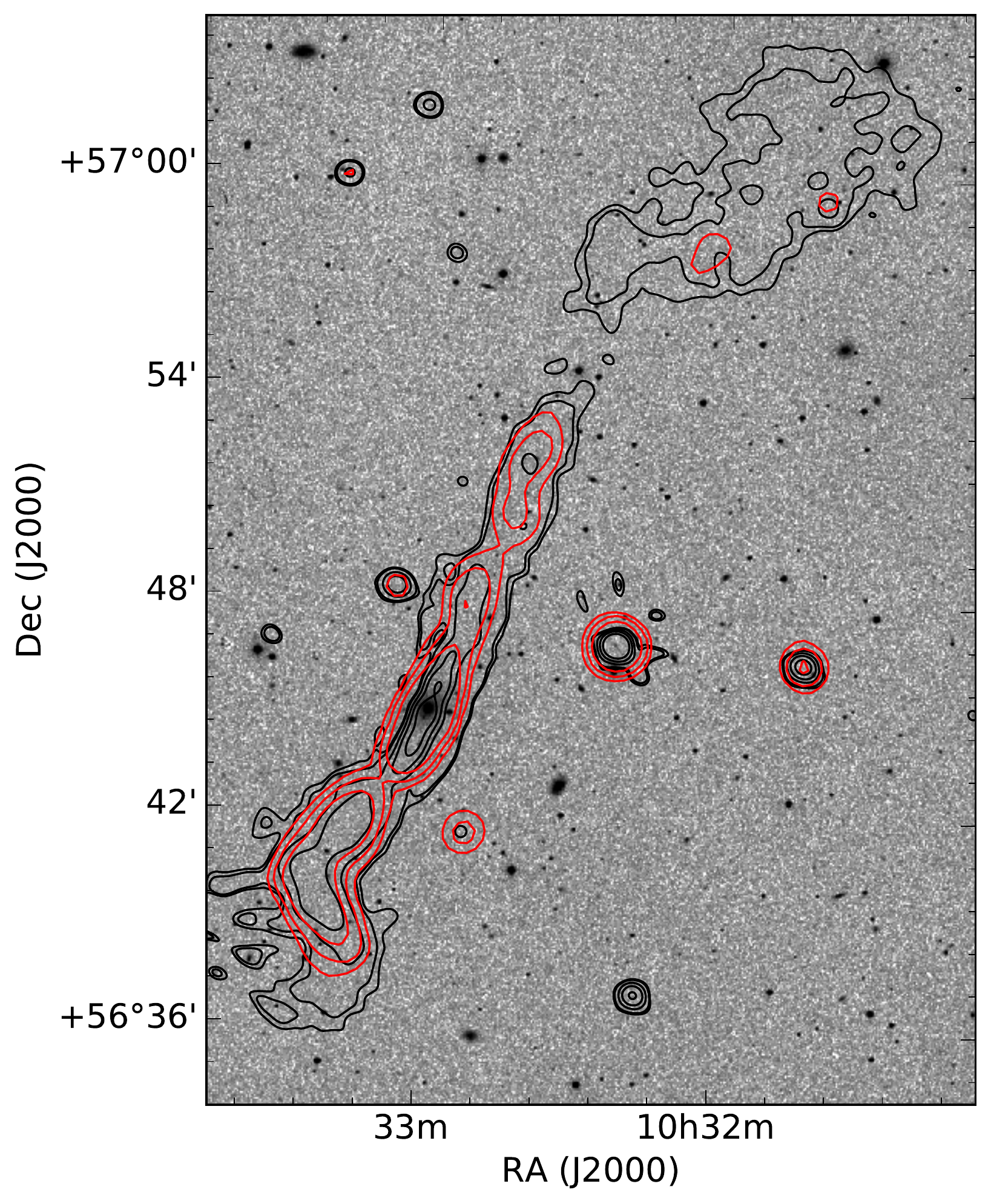}
\caption{The giant radio galaxy HB13. The greyscale image shows the optical image from DSS overlaid with NVSS 1.4-GHz contours in red (1, 2.5 and 5\,mJy\,beam$^{-1}$) and LOFAR 150-MHz contours in black (0.75, 1, 2.5, 5 and 10\,mJy\,beam$^{-1}$). \label{radiogal}} 
\end{figure}

Our image (see Fig.~\ref{radiogal}) shows radio emission extending about 30 arcmin, corresponding to about 1.6 Mpc (using the scale, 0.890 kpc/arcsec\footnote{assuming a cosmology with $H_0=69.6$\,km\,s$^{-1}$, $\Omega_0=0.29$ and $\Omega_\Lambda=0.71$.}. In the inner 2.5\,arcmin ($\sim$80 kpc) both sides of the radio jets appear to be relatively straight and symmetric while beyond that their structure seems to be less collimated and reminiscent of `plumes'. Large oscillations are also visible in the outer part of the jets. 

As illustrated by Fig.~\ref{radiogal}, the detailed structure of HB13 is seen in much more detail with LOFAR than in NVSS. Due to the low surface brightness sensitivity of the array, and the fact that these sources typically have steep spectrum, LOFAR is the ideal instrument to search for Giant Radio Galaxies (see also \citealt{Hardcastle2016}). 

\section{Conclusions}

We present 150-MHz LOFAR observations of the Lockman Hole field, allowing us to study the spectral index properties of low-frequency radio sources. The field was observed for 10~hrs using the HBA array covering the frequency range 110--172~MHz. The final 150~MHz image covers an area of 34.7 square degrees, has a resolution of 18.6$\times$14.7\,arcsec and reaches an rms of 160\,$\mu$Jy\,beam$^{-1}$ at the centre of the field. This results in a catalogue of 4882 sources within 3 degrees from the pointing centre. From these data we derive the 150-MHz source counts, which are in good agreement with other observations at 150\,MHz (e.g. \citealt{Williams2016}) as well as with extrapolations from observations of the Lockman Hole at other frequencies (Prandoni et al., 2016a,b, in preparation). 

By crossmatching this LOFAR 150-MHz catalogue with the NVSS, WENSS and VLSSr all-sky surveys, we form a sample with which to study the spectral index properties of low-frequency radio sources. Due to the different flux limits and resolutions of each of these surveys, we only include sources with deconvolved sizes less than 40\,arcsec and with $S_{150}>40$~mJy. This results in a sample of 385 LOFAR sources which we term the Lockman--wide sample. As expected for low-frequency selected sources, the sample is dominated by steep spectrum sources, with a median spectral index of $\alpha_{150}^{1400}=-0.82\pm0.02$ and an interquartile range of [$-0.94$, $-0.70$].

Although we reach sub-mJy flux densities in the LOFAR 150-MHz observations, the spectral analysis is severely limited by the much higher flux limits of other surveys that cover the entire LOFAR field of view (i.e. NVSS, WENSS, VLSSr). To investigate the spectral index properties of mJy sources selected at 150~MHz, we crossmatch the LOFAR catalogue with deeper 1.4~GHz observations from WSRT (Prandoni et al., 2016a, in preparation). This deep WSRT mosaic covers 7 square degrees with resolution 11$\times$9\,arcsec and reaches an rms of 11$\mu$Jy. This results in a sample of 1302 matched sources which we refer to as the Lockman--WSRT catalogue. 

Investigating the spectral index properties of this larger sample, again we find that the sample is dominated by steep-spectrum sources (82.1\,per\,cent), with 11.3\,per\,cent classified as flat-spectrum and 6.6\,per\,cent classified as USS. The median spectral index of this much larger sample is $\alpha_{150}^{1400}=-0.78\pm0.015$ with an interquartile range of [$-0.95$, $-0.65$]. The median spectral indices become slightly flatter with decreasing flux density, increasing from $\alpha_{150}^{1400}=-0.84$ for sources above 50\,mJy to $\alpha_{150}^{1400}=-0.76$ at 5--10\,mJy. At fainter flux densities the median spectral index stays relatively constant. 

Crossmatching this Lockman--WSRT catalogue with other deep, multi-frequency radio surveys available in the Lockman Hole we form the Lockman--deep sample with which to study the spectral properties of low-frequency radio sources above a flux density threshold of $S_{150}=9$\,mJy. This again shows that the majority of sources exhibit straight, power-law spectra over the full frequency range from 150\,MHz to 15\,GHz. Preliminary 60-MHz LOFAR observations of the Lockman Hole field shows tentative evidence that a large fraction of sources begin to flatten between 60\,MHz and 150\,MHz, but a more comprehensive analysis of the absolute flux calibration is needed to confirm this result.

Using this wide-frequency coverage we also search for sources with more extreme spectral properties such as Ultra-Steep Spectrum (USS) sources and peaked spectrum sources. We identify a total of 100 USS sources and 13 candidate peaked spectrum sources in the Lockman Hole. Using the $K$-$z$ relation we estimate that approximately 42\,per\,cent of these could be at $z>3$ and up to 21\,per\,cent at $z>4$ indicating that these could be candidate high-$z$ galaxies. However, further follow-up is required to confirm the physical properties of these objects. 

The depth reached in these observations closely matches that expected for the `Tier-1' all-sky LOFAR survey at 150~MHz. As such, the results obtained here provide insight into what we can expect to detect in the completed all-sky survey. However, this study also highlights the need for complementary, sensitive datasets across a wide range in frequency to maximise the scientific return of low-frequency radio surveys. 

\section*{Acknowledgements}

The authors thank the anonymous referee for useful comments that helped to improve the paper. LOFAR, the Low Frequency Array designed and constructed by ASTRON, has facilities in several countries, which are owned by various parties (each with their own funding sources), and that are collectively operated by the International LOFAR Telescope (ILT) foundation under a joint scientific policy. 

The research leading to these results has received funding from the
European Research Council under the European Union's Seventh Framework
Programme (FP/2007-2013) / ERC Advanced Grant RADIOLIFE-320745. Parts of this research were conducted by the Australian Research Council Centre of Excellence for All-sky Astrophysics (CAASTRO), through project number CE110001020. PNB is grateful for support from the UK STFC via grant ST/M001229/1, MJH and WLW acknowledge support from the UK Science and Technology Facilities Council [ST/M001008/1], TS acknowledges support from the ERC Advanced Investigator programme NewClusters 321271, and GJW gratefully acknowledges support from the The Leverhulme Trust.

This research made use of APLpy, an open-source plotting package for Python hosted at http://aplpy.github.com and Astropy, a community-developed core Python package for Astronomy (Astropy Collaboration, 2013).

\bibliographystyle{mnras}
\bibliography{lockmanhole} 




\bsp

\appendix
\section{Candidate Peaked Spectrum sources}   \label{appendixgps}

Table of candidate peaked spectrum sources detected in the Lockman Hole.
 
\begin{table*}
\caption{Catalogue of candidate GPS/CSS sources from both Lockman--deep and Lockman--wide samples. Flux densities are listed in mJy. The last three sources listed fall below the $S_{150}=9$\,mJy flux density threshold used to define the Lockman--deep sample so are not included in the analysis presented in this paper. We list them here for future reference. }
\scriptsize
\begin{tabular}{cccccccccc}
  \hline                       
Source ID & RA & DEC & $S_{60}$ & $S_{150}$ & $S_{345}$ & $S_{610}$ & $S_{1400}$ & $S_{15000}$ & Counterpart \\
& (h m s) & (d m s) & (mJy) &(mJy) & (mJy) & (mJy) & (mJy) & (mJy) & \\
  \hline
J110440+592450 & 11 04 40.1 & 59 24 51 & 1074.0$\pm$65{\textdagger} & 2230.3$\pm$172 & 2335.0$\pm$ 94{\textdaggerdbl} & --- & 644.2$\pm$19 & --- & \\
J110224+574727 & 11 02 24.0 & 57 47 27 & 1732.0$\pm$66{\textdagger}& 1946.3$\pm$149 & 1500.0$\pm$60{\textdaggerdbl} & --- & 478.0$\pm$14 & --- &\\
J105909+563249 & 10 59 09.0 & 56 32 49 & 549.0$\pm$73{\textdagger} & 795.2$\pm$196 & 212.0$\pm$10{\textdaggerdbl} & --- & 58.0$\pm$2 & --- & \\ 
J105826+565340 & 10 58 26.6 & 56 53 40 & 457.0$\pm$74{\textdagger} & 479.9$\pm$36 & 219.0$\pm$10{\textdaggerdbl} & ---& 54.7$\pm$2 & --- \\ 
J104833+600847 & 10 48 33.8 & 60 08 47 & $<$500.0{\textdagger} & 1242.5$\pm$95 & 1813.0$\pm$73{\textdaggerdbl} & --- & 1021.8$\pm$31 & --- & z=1.72 QSO{\text*}\\ 
J103559+583408 & 10 35 59.4 & 58 34 08 & $<$500.0{\textdagger} & 737.3$\pm$56 & 639.0$\pm$26{\textdaggerdbl} & --- & 397.7$\pm$12 & & $z_{\rm{ph}}=1.65$ QSO{\text**} \\
J105659+591145 & 10 56 59.9 & 59 11 45 & --- & 11.4$\pm$0.9 & 40.1$\pm$0.03 & 37.1$\pm$0.5 & 23.9$\pm$0.002 & ---&  \\
J105119+564018 & 10 51 19.6 & 56 40 19 & --- & 14.6$\pm$1.1 & 23.0$\pm$0.04 & 9.4$\pm$0.17 & 6.1$\pm$0.07 & --- & $K=17.6$ mag, $z=$2.378 QSO{\text*}\\ 
J104954+564507 & 10 49 54.6 & 56 45 07 & --- & 29.2$\pm$2.1 & 39.6$\pm$0.01 & --- & 2.4$\pm$0.06 & ---  &$K=19.9$ mag \\
J104435+585312 & 10 44 35.7 & 58 53 12 & --- & 14.8$\pm$1.1 & 22.3$\pm$0.02 & 21.3$\pm$0.25 & 8.8$\pm$0.01 & 0.48$\pm$0.06 & $K>21.0$ mag\\
J110105+581610 & 11 01 05.4 & 58 16 10 & 661.3$\pm$146 & 835.7$\pm$59 & 367.9$\pm$3 & 209.0$\pm$2 & 83.0$\pm$3 & ---  \\ %
J104930+590146 & 10 49 30.4 & 59 01 47 & $<$100 & 28.2$\pm$2.0 & 34.0$\pm$0.04 & 35.9$\pm$0.38 & 24.1$\pm$0.001 & 2.7$\pm$0.17 \\
J104917+583633 & 10 49 17.0 & 58 36 33 & 225.4$\pm$58.7 & 381.0$\pm$26.9 & 188.2$\pm$0.004 & --- & 52.3$\pm$0.0003 & $<$0.25 & $K=19.5$ mag\\
J105131+584950 & 10 51 31.4 & 58 49 50 & --- & 2.0 $\pm$0.44 & $<$4.0 & 17.4 $\pm$0.21 & 14.9 $\pm$0.001 & 0.78 $\pm$0.11 &$K>21.0$ mag\\ 
J105556+591654 & 10 55 55.9 & 59 16 54 & --- & 4.1 $\pm$0.57 & 8.1 $\pm$0.11 & 7.6 $\pm$0.24 & 5.6 $\pm$0.02 & ---  & \\ 
J105226+575508 & 10 52 25.5 & 57 55 08 & --- & 2.7 $\pm$0.48 & $<$4.0 & 12.3 $\pm$0.18 & 27.0 $\pm$0.0004 & 22.5 $\pm$1.3 &$K=19.1$ mag\\ 
  \hline 
\end{tabular}

   \begin{tablenotes}
    \item{\textdagger}{74\,MHz flux density from VLSSr}
	\item{\textdaggerdbl}{325\,MHz flux density from WENSS}
	\item{\text*}{Redshift from SDSS \citep{sdss}}
	\item{\text**}{Redshift from \citet{Richards2009}}
    \end{tablenotes}
\end{table*}

\section{LOFAR sources with excess emission at 150\,MHz}

Postage stamp images of some example LOFAR sources (including components) that do not have counterparts in the deep 1.4-GHz WSRT mosaic. This implies that the spectral index is steeper than $\alpha_{150}^{1400}=-1.2$.

\begin{figure*} 
\begin{minipage}{0.33\linewidth}
\includegraphics[width=\linewidth]{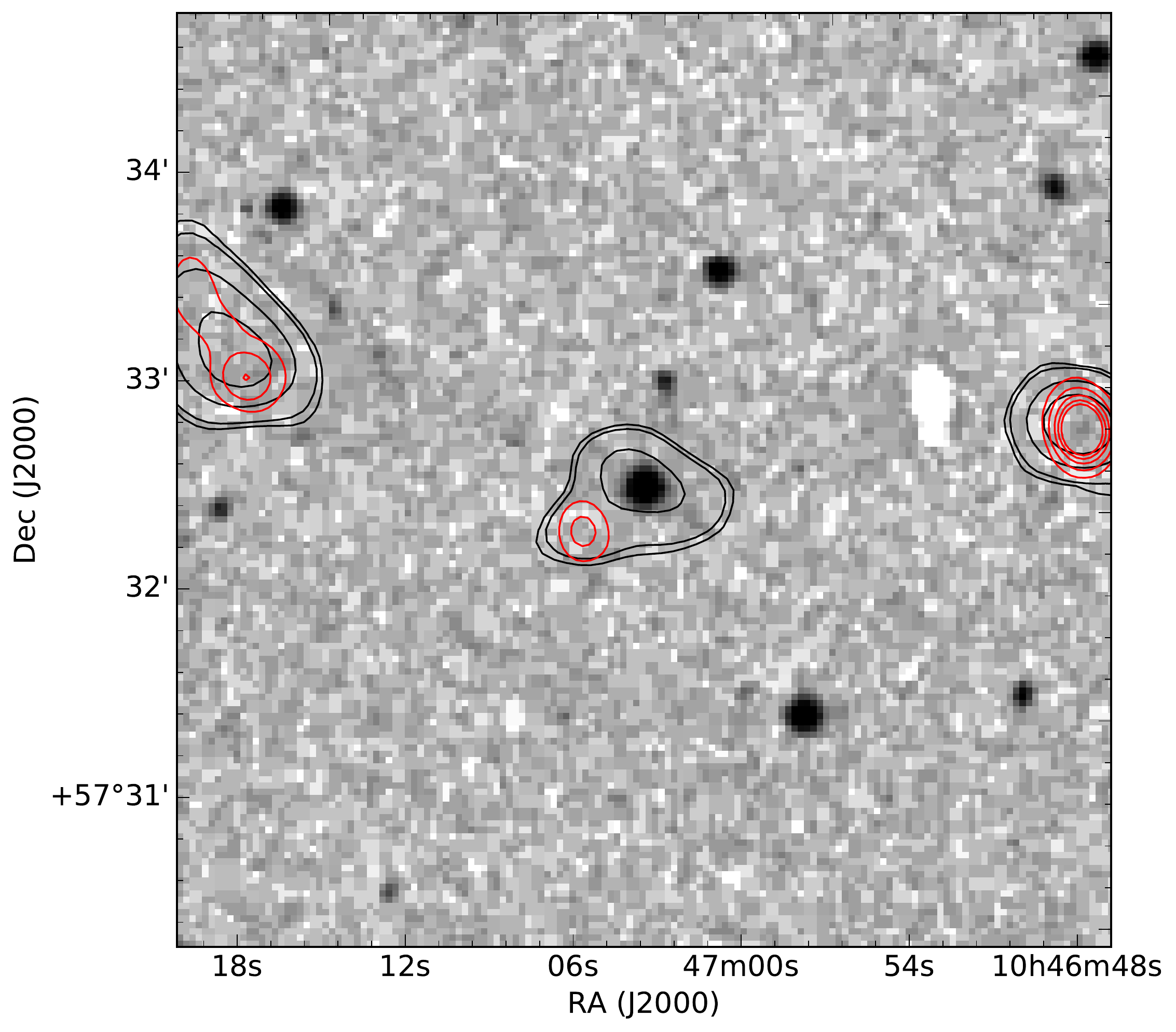}
\end{minipage}
\begin{minipage}{0.33\linewidth}
\includegraphics[width=\linewidth]{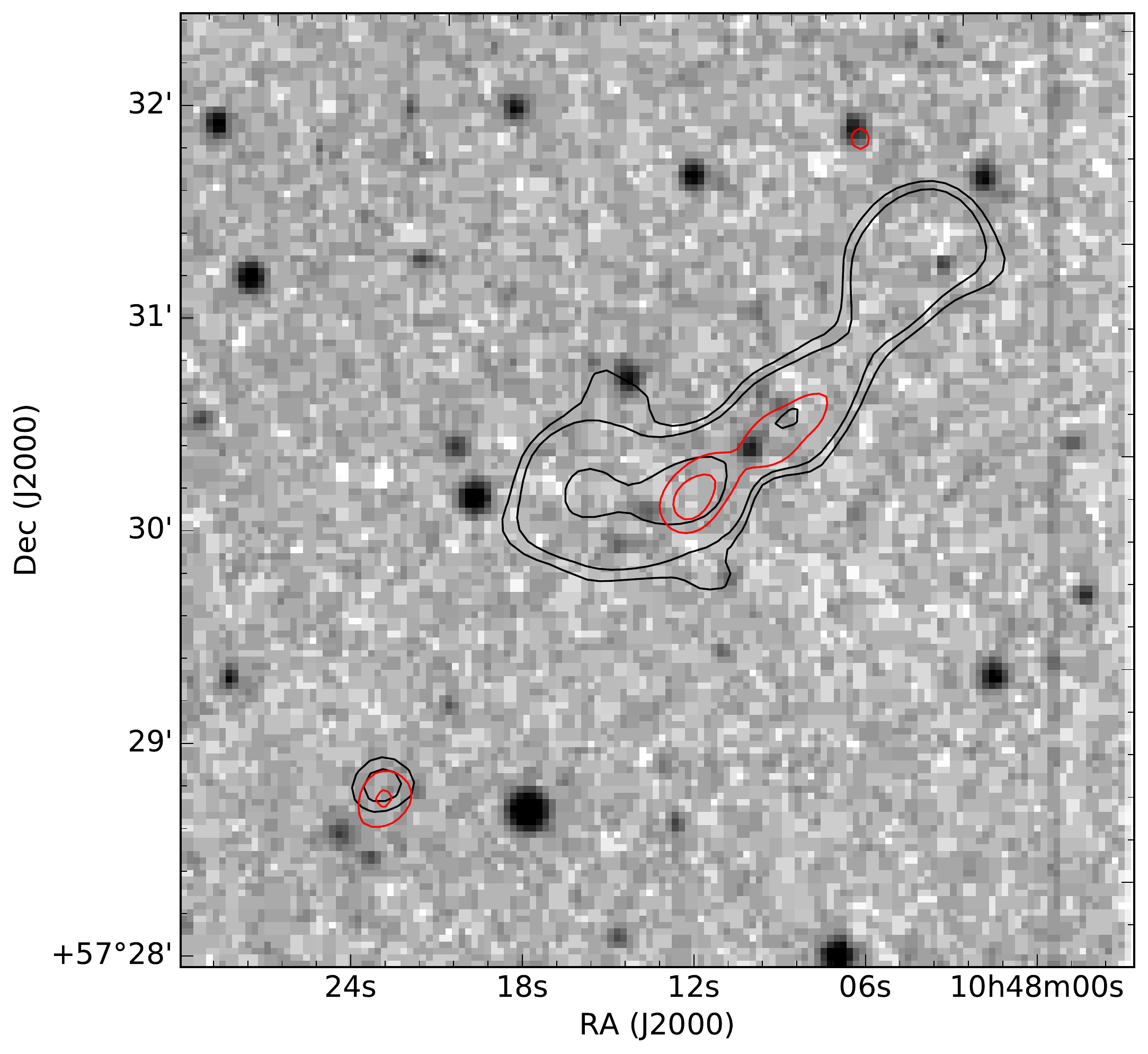}
\end{minipage}
\begin{minipage}{0.33\linewidth}
\includegraphics[width=\linewidth]{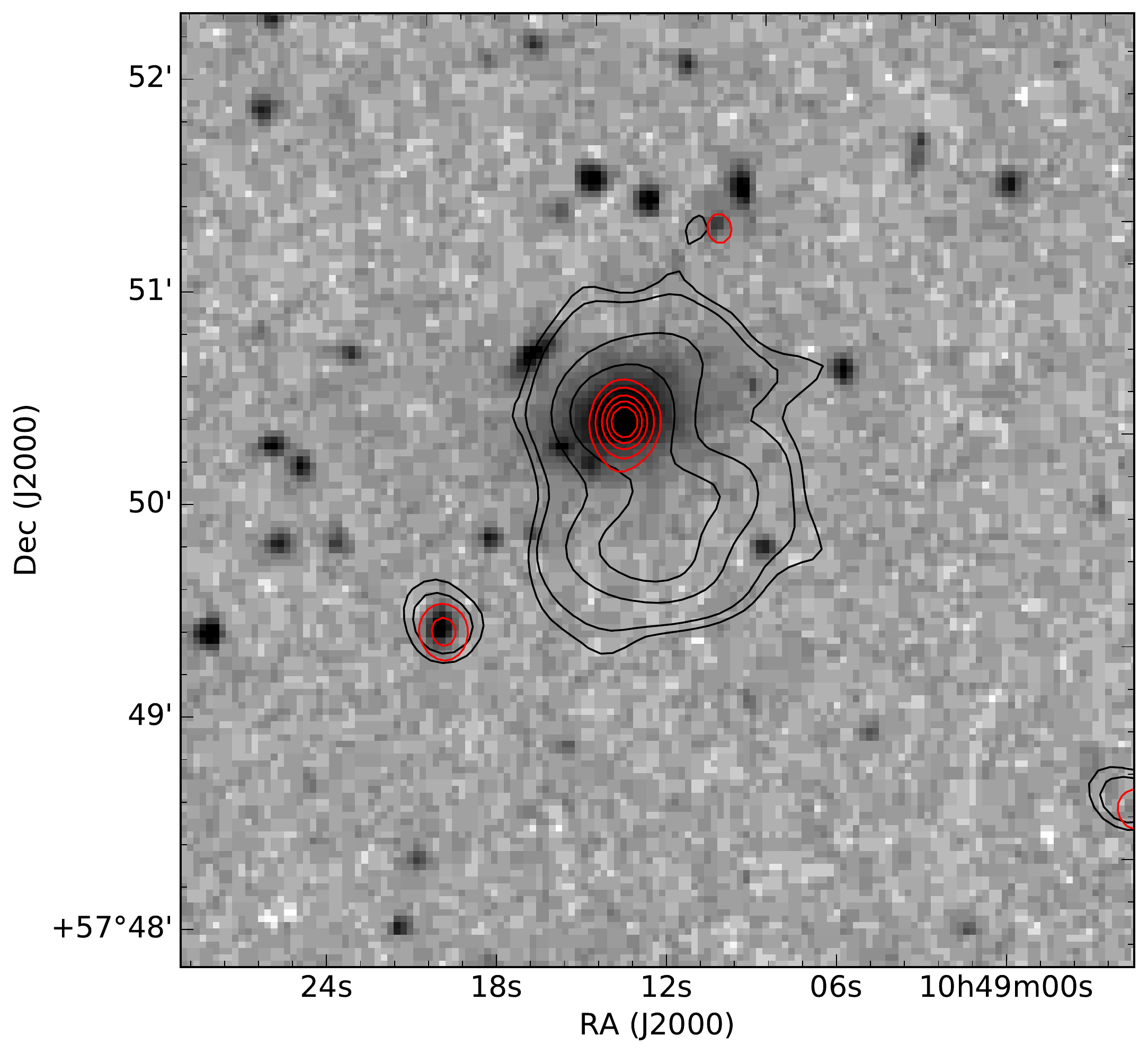}
\end{minipage}
\begin{minipage}{0.33\linewidth}
\includegraphics[width=\linewidth]{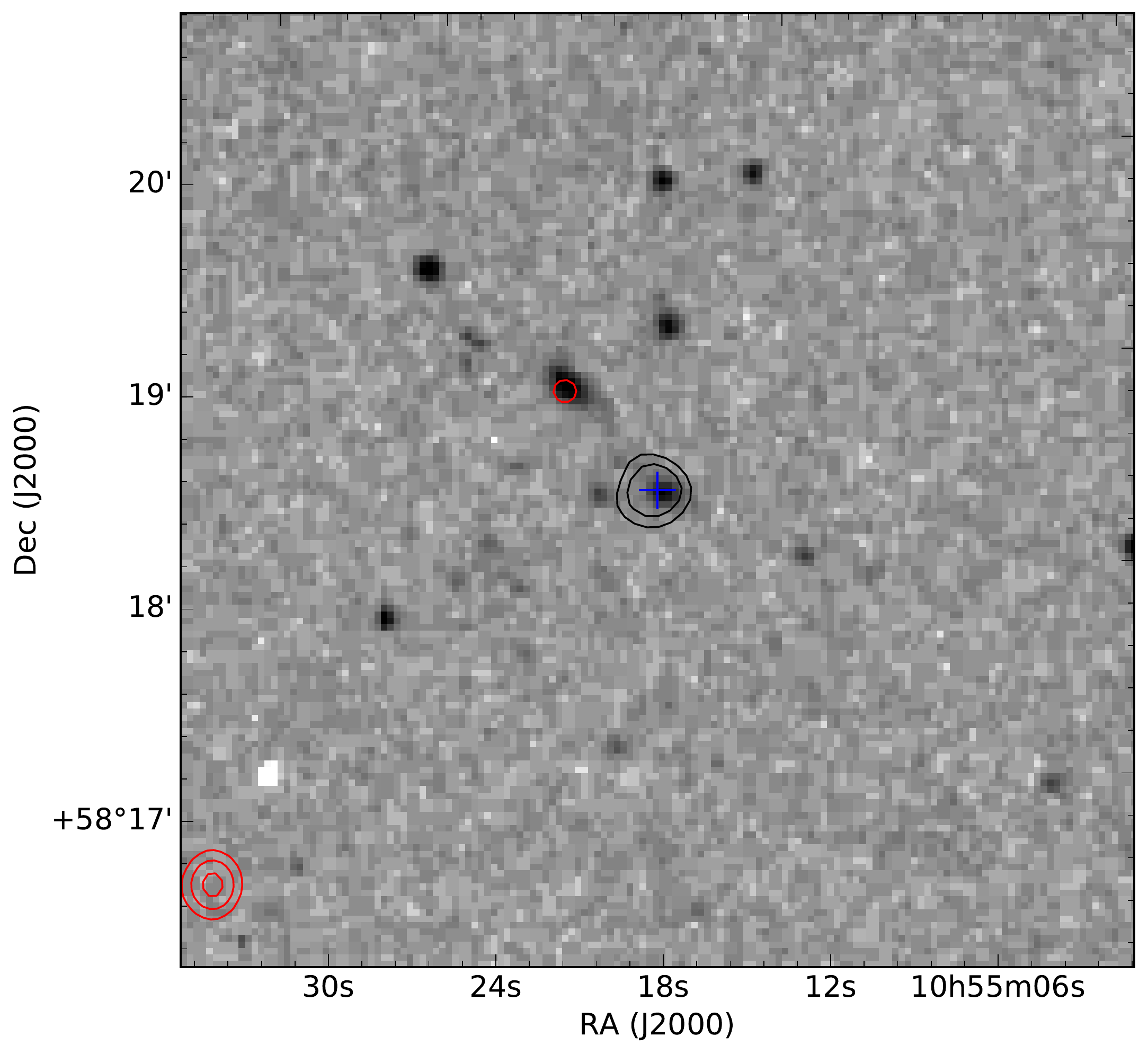}
\end{minipage}
\begin{minipage}{0.33\linewidth}
\includegraphics[width=\linewidth]{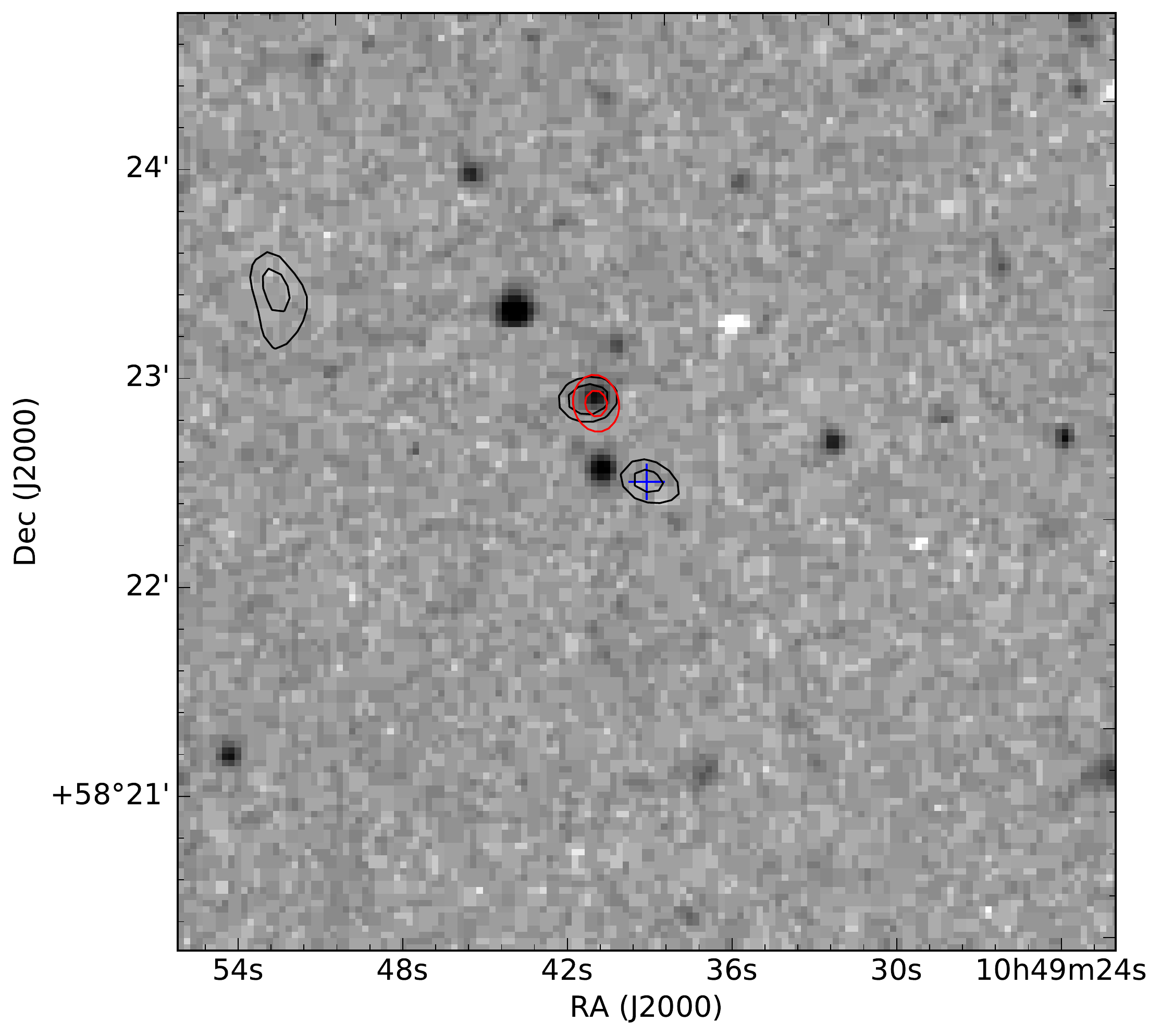}
\end{minipage}
\begin{minipage}{0.33\linewidth}
\includegraphics[width=\linewidth]{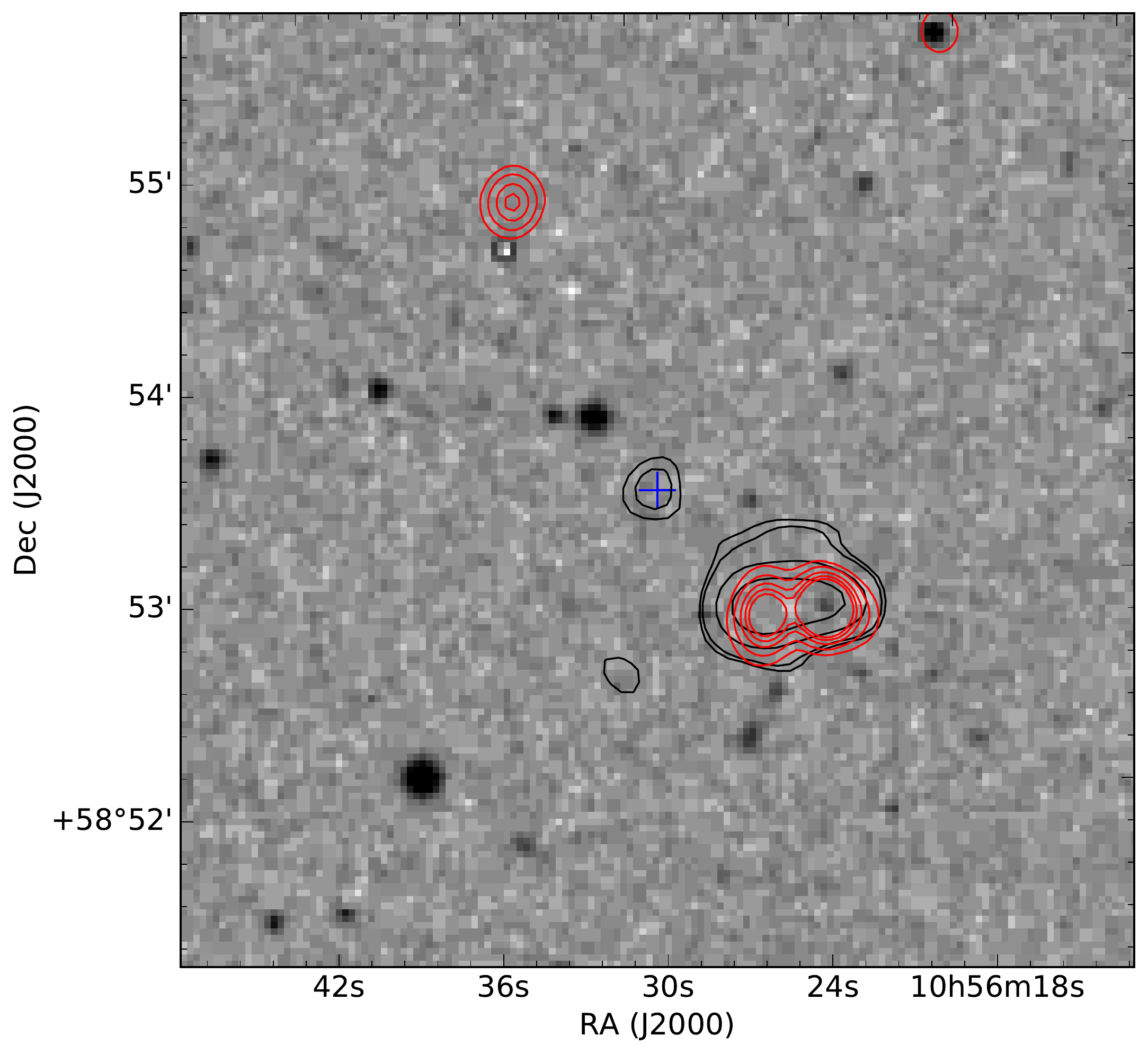}
\end{minipage}
\caption{LOFAR sources showing excess low-frequency emission not detected in the WSRT 1.4-GHz image. The flux limits of each survey mean that these sources must have spectral indices steeper than $\alpha_{150}^{1400}<-1.2$. The greyscale shows an optical image from DSS, the WSRT contours are shown in red at levels of 0.1, 0.25, 0.5, 0.75, 1.0 and 5.0\,mJy\,beam$^{-1}$ and the LOFAR contours are shown in black (0.75, 1.0, 2.5, 5.0 and 10.0\,mJy\,beam$^{-1}$). The bottom row shows sources shown that were completely undetected in the WSRT image (i.e. flag `L' in the Lockman--WSRT catalogue, the LOFAR positions are marked by the blue cross) and the top row shows objects where the LOFAR image detects additional components/emission that are not detected at 1.4\,GHz (flag `LC'). \label{lofnowsrt}}
\end{figure*}

\label{lastpage}

\end{document}